\documentclass{jfm}

\usepackage{graphicx}
\usepackage{subcaption}
\usepackage{xcolor}
\usepackage{pdflscape}
\usepackage{newtxtext}
\usepackage{newtxmath}
\usepackage{natbib}
\usepackage{hyperref}
\hypersetup{
    colorlinks = true,
    urlcolor   = blue,
    citecolor  = black,
}

\newcommand{\RomanNumeralCaps}[1]
\linenumbers

\title{Stabilisation of second Mack mode in hypersonic boundary layers through spanwise non-uniform surface temperature distribution}

\author{L. Boscagli\aff{1}
  \corresp{\email{l.boscagli@imperial.ac.uk}},
  G. Rigas\aff{1},
  O. Marxen\aff{2}  
 \and P. J. K. Bruce\aff{1}}

\affiliation{\aff{1}Department of Aeronautics, Imperial College London, London, SW7 2AZ, UK
\aff{2}School of Mechanical Engineering Sciences, University of Surrey, Guildford, GU2 7XH, UK}

\begin{document}
\maketitle

\begin{abstract}
The extreme heat fluxes characteristic of hypersonic flows significantly limit the flight envelope of hypersonic vehicles. The role of hydrodynamic instability and the onset of laminar to turbulent boundary layer transition is of notable importance. The effect of streaks on the suppression of planar (second Mack mode) instabilities has been previously investigated, but a potentially passive and non-intrusive control method has not been established yet. Recent work shows that streaks can be generated through a spanwise variation in surface temperature. %
This method exploits the aerothermodynamic characteristics of the flow, and therefore promises to be robust. This work uses direct numerical simulations to determine and quantify the effectiveness of this novel control method in the suppression of second Mack mode instability for a hypersonic boundary layer over a flat plate. The computational analyses cover a range of Mach numbers 4.8 to 6 and wall temperature ratios representative of both wind tunnel testing and flight scenarios. Among the range of configurations investigated the energy of the second Mack mode is reduced by up to approximately 60\% by the steady streaks. The streak wavelength parameter plays a significant role in the stabilisation benefits. For a Mach 6 configuration, for the most linearly amplified second Mack mode disturbance frequency, nearly optimum performance is achieved for a spanwise wavelength of approximately 8 to 10 times the local boundary layer thickness.    
These findings open new avenues for controlling hypersonic boundary layers and offer valuable guidance for future experimental campaigns aimed at validating this novel control strategy.   

\end{abstract}

\section{Introduction}
The development of aerospace technologies that travel with a flight Mach number ($M_{\infty}$) well above sonic is challenged by complex aerothermodynamic behaviours. This flight regime is typically referred to as hypersonic. Boundary layer instability and transition can significantly constrain the flight envelope and operation limits of hypersonic vehicles \citep{Lin2008}. The location point of laminar to turbulent transition in hypersonic boundary layers has a significant influence on viscous drag and aerodynamic heating of external surfaces of hypersonic vehicles, and is a dominant source of uncertainties during the design process \citep{Shea1992}. Relative to a laminar state, the heat flux for a turbulent boundary layer can be up to 8 times greater \citep{Leyva2017}. Thus, this motivates further research on transition control. 

For laminar hypersonic boundary layers, an important non-dimensional parameter is the relative Mach number $\overline{M}$, which is defined based on the velocity of the flow ($u$) relative to the phase speed ($c_{ph}$) of the hydrodynamic instability within the boundary layer. When $\overline{M}^2>1$, the compressible counterpart of the Rayleigh's equation admits multiple wave-like solutions, also referred to as higher Mack modes \citep{Mack1969}. For a flight Mach number ($M_{\infty}$) approximately between  4 and 6, and for a thermally insulated (adiabatic) wall, or under thermal equilibrium conditions (radiative-adiabatic, \citep{Anderson1989}), an important boundary layer instability mechanism is known to be two dimensional and dominated by high-frequency, $ \tilde{f} \in [10^5, 10^6]$Hz \citep{Laurence2016}, thermoacoustically driven \citep{Kuehl2018} waves trapped between the wall and the relative sonic line within the boundary layer \citep{Mack1975}. This instability mechanism is typically referred to as the second Mack mode. Although this is not a mode in a mathematical sense \citep{Fedorov2011}, the terminology is still generally accepted in the literature and therefore it is also used within the context of this work. The high-frequency dilatation work of the second Mack mode instability on the flow can also lead to significant local aerodynamic heating \citep{Zhu2018}, which can further reduce the aerothermal efficiency of hypersonic vehicles. 

The stability of compressible boundary layers is significantly affected by wall temperature \citep{Lees1946}. This is an important consideration for ground-testing. In high-enthalpy (flight representative) facilities, the wall temperature can be a small fraction of the freestream temperature ($\tilde{T}_w/\tilde{T}_{\infty}\approx0.1-0.3$, \citep{Bitter2015}), while this is not usually the case in wind tunnels operated at lower stagnation enthalpies. Based on Rayleigh's generalised inflection theorem \citep{Rayleigh1895}, inviscid inflectional instability modes can be stabilised by sufficient wall cooling for low-speed and supersonic flows \citep{Masad1992}. However, this is no longer true when higher Mack modes arise in hypersonic boundary layers. In particular, the second Mack mode is destabilised by wall cooling \citep{Mack1975,Bitter2015}. This effect is further exacerbated when the wall temperature is further reduced ($\tilde{T}_w/\tilde{T}_{\infty}<0.1$) and unstable supersonic modes also manifest \citep{Bitter2015,Chuvakhov2016,Saikia2022}. Wall heating instead tends to stabilise the second Mack mode \citep{Mack1975}. On the other hand, three dimensional, inflectional instabilities (e.g., first Mack mode) are stabilised by wall cooling \citep{Mack1969,Lysenko1984}. As a result of the significant impact of wall temperature on first and second Mack modes, several transition control strategies that exploit surface heat flux have been numerically attempted in the literature \citep{Zhao2018,Jahanbakhshi2021,Poulain2023thesis}. Although effective, active flow control techniques require careful energy input considerations \citep{Frohnapfel2012}. In addition, the practical and robust implementation of active flow control devices remains a challenge \citep{Gad2001}.

Passive control of hypersonic boundary layer transition has been experimentally and numerically attempted through the use of roughness elements \citep{Marxen2010,Fong2015,Taylor2016} or vortex generators \citep{Paredes2019}. \citet{Marxen2010} used high-order compressible DNS computations to investigate the growth rate of convective disturbances within a boundary layer at $M_{\infty}=4.8$ with two dimensional roughness elements. For high-frequency (second Mack mode type) disturbances, the spatial damping effect of the two dimensional, localised, roughness elements was significant. For a similar geometry configuration and for $M_{\infty}=5.92$, \citet{Duan2013} showed that the streamwise position of the roughness element is an important factor in the control of two and three dimensional (oblique) instabilities. For a cone configuration, \citet{Fong2015} showed that if the streamwise locations of the roughness elements is informed by numerical (linear) analysis of the boundary layer stability, it is possible to achieve stabilisation of both first and second Mack modes. However, these passive control devices present several implementation challenges at hypersonic speeds due to their long exposure to high heat flux. Thus, novel robust control methods are required for hypersonic regimes.

Effective transition delay for low-speed boundary layers using optimal streaks has been demonstrated by experimental \citep{Fransson2006} and numerical \citep{Cossu2002,Schlatter2010} studies. \citet{Bagheri2007} showed that Tollmien-Schlichting (TS) waves and oblique waves can be stabilised by finite amplitude streaks, that modify the mean flow distortion. It was also shown that the streak wavenumber for optimal growth of the streaks is not the most efficient to achieve TS-wave stabilisation. More recently, the theory and analysis has been extended to high-speed (compressible) boundary layers \citep{Paredes2016,Ren2016}. For thermally insulated conical bodies, \citet{Paredes2019} investigated the stabilisation of hypersonic boundary layers by optimally growing streaks through the parabolised stability equations (PSE). For a flight Mach number above 4, the generation of streaks was beneficial to reduce the amplification of second Mack mode and delay the onset of laminar to turbulent transition. \citet{Paredes2016} also showed that the theoretical benefits achievable by delaying the second Mack mode may be limited by potential adverse effects of the streaks on the first Mack mode. This is particularly true at lower flight Mach numbers ($M_{\infty}=3$), where, for an adiabatic flat plate configuration \citep{Paredes2017}, the interaction between the streaks sub-harmonics (spanwise wavelength, $\lambda$, greater than twice the fundamental wavelength, $\lambda_z$) and the first Mack mode can lead to earlier transition to turbulence in quiet, low external disturbance, environments. For a lower Mach number ($M_{\infty}=2$) boundary layer over an adiabatic flat plate, \citet{Sharma2019} and \citet{Kneer2022} conducted a set of parametric DNS studies and showed that streaks generated by a blowing and suction strip can successfully delay first mode oblique breakdown to turbulence. For a similar configuration, \citet{Celep2022} showed that uniform wall heating can reduce the useful range of control-streak amplitude that can successfully delay transition. For $M_{\infty}=4.5$, \citet{Zhou2023} showed that second mode oblique breakdown can also be successfully delayed through finite amplitude streaks. For low-speed (incompressible) flows, \citet{Andersson2001} showed that streaks can impart a spatial organisation to the supported instabilities, which manifests in the symmetric (varicose-type) or asymmetric (sinuous-type) characteristics of the eigenfunction of the instability mode relative to the streak structure. In compressible boundary layers, steady streaks typically undergo significant transient (non-modal) temporal \citep{Hanifi1996} and spatial \citep{Tumin2003} growth. Based upon this evidence, \citet{Caillaud2025} recently investigated through linearised Direct Numerical Simulations (DNS) the dynamics of non-modal instability for a hypersonic boundary layer ($M_{\infty}=6$) over an adiabatic flat plate with streaks generated through a volumetric momentum force. Several interaction mechanisms were determined based on the amplitude of the forcing streaks ($As_{u,0}$). For $As_{u,0} = 0.028$, the associated maximum amplitude of the streaks at the end of the domain was $As_{u}\approx 0.4$ and the symmetric, fundamental and first sub-harmonic second Mack mode were destabilised by the streaky flow. 
 
Recent computational \citep{Ozawa2025, Boscagli2025_SCITECH} and experimental \citep{Ozawa2025_SCITECH} studies showed that for a flat plate configuration it is possible to generate streaks within the boundary layer through a spanwise non-uniform wall temperature distribution. The method exploits the effect of heating and cooling on the mean velocity profile, which leads to thicker and thinner boundary layer profiles, respectively \citep{Anderson1989}. This can be passively attained through the use of alternate stripes of materials with different thermal properties, and by exploiting the high heat flux characteristics of the hypersonic regime. This non-intrusive, passive flow control technique has the potential to increase the aero-thermal-structural efficiency of hypersonic vehicles. Nevertheless, there is a need to determine the effectiveness of this control method due to conflicting mechanisms related to streaks and wall temperature effects on second Mack mode stabilisation. In addition, there is a need to determine the robustness of the control method for a range of operating conditions sufficiently representative of both wind tunnel and flight conditions due to the challenges associated to match flight representative conditions in low-enthalpy, quiet blow-down ground-test facilities. In particular, the effect of a (independent) change in Mach number and wall temperature ratios needs to be determined and quantified. 

For high-speed flows, high-temperature gas effects require some careful consideration \citep{Anderson1989}. Strong thermochemical non-equilibrium flows may be experienced by hypersonic vehicles operating under high specific total enthalpy conditions ($\tilde{h}_{0,\infty}>5\times 10^6$ J/kg) due to complex aerothermodynamics and chemical phenomena \citep{Leyva2017}, such as shock layer radiation, ablation,  etc.. The ratio of diffusion to reaction timescales, also known as Damk\"ohler number ($Da$), is an important non-dimensional parameter to characterise hypersonic flows, and boundary layer stability in particular. For the low-end spectrum of total enthalpies characteristics of hypersonic flight conditions ($M_{\infty}<10$, flight altitude $h<30000m$), \citet{Bitter2015} assessed thermal non-equilibrium effects on boundary layer stability. Vibrational excitation had a notable influence on base flow temperature while, for air, the effect of thermal non-equilibrium on maximum spatial growth rate for the second Mack mode was less than $8\%$ \citep{Bitter2015thesis}, and it did not affect the dominant aerodynamic mechanisms of the boundary layer stability. As such, for the working fluid and conditions of interest in this work, a vibrationally frozen ($Da \ll 1$) stability analysis is an acceptable assumption. Relative to chemical equilibrium, for a $M_{\infty}=10$ boundary layer over a flat plate, \citet{Marxen2014} showed that finite-rate chemistry leads to only slightly higher amplification factor for the second Mack mode. \citet{Passiatore2024} also reached similar conclusions relative to the effect of finite-rate chemistry on the linear amplification of the second Mack mode fundamental harmonic. However, it was also found that the transition point can be overall delayed by chemical non-equilibrium ($Da \sim \mathcal{O}(1)$) processes, which drain part of the modal energy from secondary instabilities that played a dominant role for the investigated breakdown to turbulence scenario. Overall, for the study of the evolution of small-amplitude disturbances for $M_{\infty} \leq 6$ and $\tilde{h}_{0,\infty}<2.0\times 10^6$ J/kg \citep{Anderson1989}, a calorically perfect gas modelling assumption is sufficiently valid. 

The novelty of this work is the assessment via Direct Numerical Simulations (DNS) of a high-speed boundary layer over a flat plate with zero pressure gradient of the effect of streaks generated through spanwise non-uniform surface temperature distributions on second Mack mode stabilisation, for a range of flight and wind tunnel-testing scenarios. The manuscript is structured as follows: section \ref{sec:methodology} presents the computational methods; results, discussion and synthesis of the computational assessment is presented in section \ref{sec:results}; conclusions and outlook are presented in section \ref{sec:conclusions}.

\section{Methodology}\label{sec:methodology}
The development of small-amplitude disturbances within a high-speed boundary layer over a flat plate with uniform and non-uniform surface temperature distributions is investigated by means of 3D Direct Numerical Simulations (DNS). Linear Stability Theory (LST) analyses are used to inform the selection of some of the boundary conditions for the DNS computations, and for some a-posteriori verification and characterisation of the triggered instability. In the sections below a brief description of the numerical methods and notation, and the formulation of the wall boundary conditions used is provided.

\subsection{Direct numerical simulations}\label{sec:DNS}
\subsubsection{Governing equations and numerical method}
The three dimensional, time-dependent, compressible formulation of the Navier Stokes equations is solved for a calorically perfect gas (air). The non-dimensional equations for the conservation of mass, balance of momentum and energy conservation are expressed as in \citet{Marxen2010}, and these are also included in appendix \ref{appendix:gov_eq} for completeness.  The non-dimensionalisation is mostly based on the free-stream conditions \citep{Marxen2010}, which are indicated with subscript $(\cdot)_{\infty}$. The dimensional variables are marked with the symbol $\tilde{(\cdot)}$, whereas the latter is omitted for the non-dimensional form. Sutherland's law, with Sutherland's temperature $\tilde{T}_s=110.4 $ K \citep{Anderson1989}, is used to compute viscosity. From the used non-dimensionalisation of the Navier-Stokes equations, the Reynolds number ($Re_{\infty}$) and Prandtl number ($Pr_{\infty}$) formulation is as follows,

\begin{equation}
Re_{\infty} = \tilde{\rho}_{\infty} \tilde{c}_{\infty} \tilde{L}_{ref} / \tilde{\mu}_{\infty}
\end{equation} 

\begin{equation}
Pr_{\infty} = \tilde{\mu}_{\infty} \tilde{c}_{p} / \tilde{k}_{\infty} ,
\end{equation} 
where $\tilde{\rho}_{\infty}$, $\tilde{c}_{\infty}$, $\tilde{\mu}_{\infty}$ and $\tilde{k}_{\infty}$ are the freestream density, speed of sound, dynamic viscosity and thermal conductivity, respectively, $\tilde{L}_{ref}$ is the reference lengthscale, and $\tilde{c}_{p}$ is the specific heat at constant pressure. The three-dimensional velocity vector is indicated as $\left[u_1 \: u_2 \: u_3 \right]^T = \left[u \: v \: w \right]^T$, and it is a function of the spatial coordinates $\left[x_1 \: x_2 \: x_3 \right]^T = \left[x \: y \: z \right]^T$. 

In all the figures below, velocity and temperature scales are normalised with the freestream velocity, $\tilde{u}_{\infty}$, and static temperature, $\tilde{T}_{\infty}$, respectively. In place of the non-dimensional streamwise coordinate, $x$, a local Reynolds number, $Re_x=\sqrt{xRe_{\infty}M_{\infty}}$, is sometimes also used. The ratio of the specific heats ($\gamma$) is set to $\gamma=1.4$, and $Pr_{\infty}=0.71$. 

The structure and methods used for the DNS solver closely follow the algorithm described by \citet{Nagarajan2003} and \citet{Nagarajan2007}. The equations are discretised on a spatially structured, curvilinear grid with a staggered approach for the conservative variables. A time-accurate solution is achieved through a sixth-order compact finite difference scheme within the interior nodes of the domain and with an explicit $3^{rd}$ order Runge-Kutta time-stepping method \citep{Marxen2010}. The compressible DNS solver has been extensively used and verified for the computation of linear (small-amplitude) and non-linear evolution of boundary layer disturbances \citep{Marxen2011}, with \citep{Marxen2013} and without \citep{Marxen2010} high-temperature gas effects.

\subsubsection{Computational domain and boundary conditions}\label{sec:domain}
\begin{figure}
  \centering 
  \subfloat[]{\includegraphics[width=0.5\textwidth]{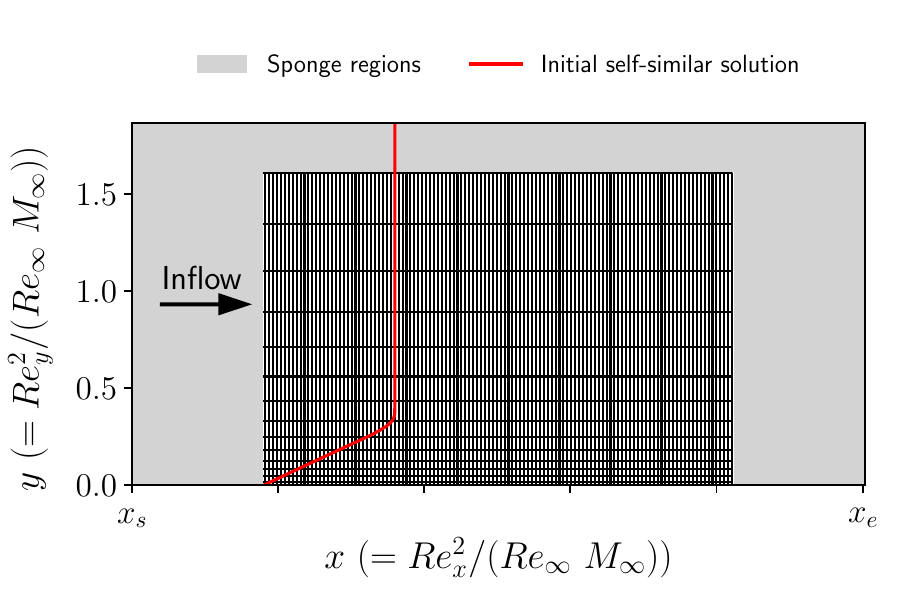}\label{fig:DNS_domain_XY}} 
  \hfill 
  \subfloat[]{\includegraphics[width=0.5\textwidth]{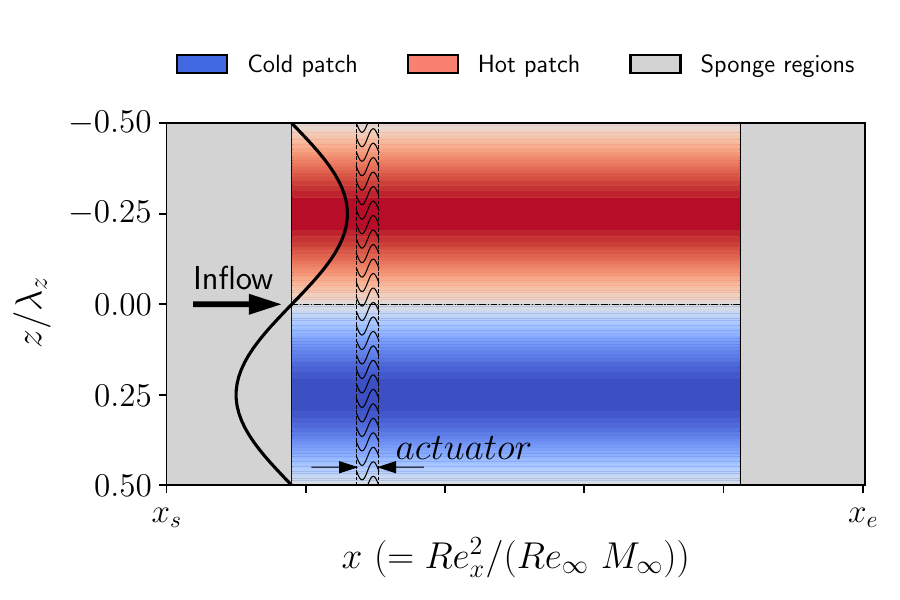}\label{fig:DNS_domain_XZ}} 
  \caption{(\textit{a}) Streamwise, $x$, and (\textit{b}) spanwise, $z$, 2D schematics of the computational domain, boundary conditions and initial solution. Streamwise and wall-normal, y, grid refinement displayed every $10^{th}$ and $15^{th}$ point, respectively. Flow is left to right, and the domain is periodic in the spanwise direction.}
  \label{fig:DNS_domain}
\end{figure}

The computational domain for the DNS (figure \ref{fig:DNS_domain}) includes the viscous wall, where a laminar self-similar solution develops, and inflow, outflow and upper boundaries, where sponge regions are used to damp the solution towards a self-similar laminar state and prevent spurious reflection of pressure waves (figure \ref{fig:DNS_domain_XY}). Periodic boundary conditions are applied in the spanwise direction at both sides of the domain (figure \ref{fig:DNS_domain_XZ}). 

In the streamwise $x$, and spanwise $z$, directions, the grid nodes are uniformly distributed. For each of the computations the number of grid nodes is adjusted such that approximately 22 nodes per second Mack mode streamwise wavelength are used. Based on previous studies for an adiabatic flat plate with \citep{Passiatore2024} and without \citep{Ma2003} high-temperature gas effects, this guarantees sufficient streamwise resolution to capture two-dimensional instability waves. In the wall-normal direction 211 nodes ($n_y$) are used, with the grid stretching toward the wall \citep{Marxen2010}, such that for each of the computations the boundary layer at is resolved with at least 30 points near the domain inflow, where the boundary layer is thinner. In the spanwise direction, 13 points per spanwise wavelength of the streaks ($\lambda_z$) are used. A grid refinement study showed that the discretisation error on second Mack mode amplification factor due to spanwise grid resolution is within $6\%$ (further details are in Appendix \ref{appendix:grid_refinement}). The spanwise extent of the computational domain ($\lambda_{z,domain}$) corresponds to the fundamental harmonic of the streaks, $\lambda_z$. For these investigations, streaks subharmonics are not modeled as an early assessment showed that they have no influence on the linear amplification of the second Mack mode (further details of the assessment are in Appendix \ref{appendix:sub_harmonic}).

The computational time step is adjusted so that 600 time steps are used within each fundamental period ($\tau=2\pi/\omega$). The latter is defined based on the angular frequency ($\omega$) of the blowing and suction method used to trigger second Mack mode instability within the domain, as further described in the following section (section \ref{sec:actuator}). The choice of the computational time step is based on previous studies \citep{Marxen2010}, and it guarantees sufficient temporal resolution to capture the second Mack mode instabilities.

\subsubsection{Disturbance forcing}\label{sec:actuator}
To trigger boundary layer instabilities and promote transition to turbulence, a wall-normal momentum perturbation is introduced downstream of the domain inflow and upstream of the region of interest. The formulation (equation \ref{eq:actuator_eq}) is similar to that used by \citet{Pagella2002} and \citet{Marxen2010}, 
\begin{equation}
\begin{cases}
    \frac{(\tilde{\rho}\tilde{v})_{wall}}{(\tilde{\rho}\tilde{c})_{\infty}} = (\rho v)_{wall} = A_v \cos \left( k\frac{2 \pi}{\lambda_z} z \right) \sin(\omega t) \sin(n \xi)\exp(-\frac{1}{\sqrt{2}} 	\xi^2)\\
    \xi = \frac{x - x_{c,strip}}{L_{strip}}
\end{cases}
\label{eq:actuator_eq}
\end{equation}
For a more concise notation, in the rest of the text, this boundary condition will be referred to as actuator. The mathematical formulation is similar to the one used by \citet{Pagella2002}. The streamwise location of the center of the actuator ($x_{c,strip}$) and its length ($L_{strip}$) are determined based on linear stability analyses as described in section \ref{sec:LST}. The amplitude of the perturbation introduced by the actuator ($A_v$) is set to $A_v=0.0006M_{\infty}$. This choice is based on previous studies in the literature \citep{Egorov2006,Unnikrishnan2020}, and it is sufficiently small to avoid bypass of the linear instability regime. In equation \ref{eq:actuator_eq}, the parameter $n$ control the number of actuators used to trigger the instability. A preliminary assessment showed that $n=4$ provided a sufficiently computationally efficient way to trigger boundary layer instability. For two dimensional perturbations, such as those used to trigger second Mack mode instabilities, $k$ is set to $0$. The streamwise distribution of the blowing and suction forcing law resemble a dipole, and therefore vortical disturbances are mostly excited \citep{Harris1997}. 

\subsubsection{Wall temperature boundary condition}\label{sec:tw}
The wall temperature boundary condition is  
\begin{equation}
    T_w = T_{w,base} \left(1 + A_{T_{w}} \sin \left( \frac{2 \pi}{\lambda_z} z \right) \right)\\
\label{eq:Tw_eq}
\end{equation}
where $A_{T_w}$ sets the amplitude of the wall temperature variation  relative to the baseline (uniform) wall temperature. The wall temperature is imposed as a modification to the internal energy, and a five-cell stencil linear interpolation is used to get the value at the cell centre where the conservative variables are stored. Further details about the arrangement of conservative and thermodynamic flow variables as well as the use of interpolation schemes for non-periodic boundaries are in \citet{Nagarajan2003}. A linear temporal ramp-up of $A_{T_w}$ is used as part of the convergence strategy. Within that period, data are discarded as part of the initial numerical transient and not taken into account within the analysis.  A blending function along the streamwise direction similar to the one imposed at the sponge regions \citep{Franko2013} is also used to ensure smooth transition from uniform to non-uniform wall temperature and avoid numerical discontinuities. 

\subsection{Linear stability theory}\label{sec:LST}
Parallel, Linear Stability Theory (LST) analysis is used to inform the selection of the computational domain size ($[x_s,x_e]$, figure \ref{fig:DNS_domain_XY}) for the DNS, as well as the choice of the temporospatial frequencies of the blowing and suction actuation region used to trigger boundary layer instabilities. The ansatz formulation for the solution of the linearised Navier Stokes equations ($q'$) is expressed as follows,

\begin{equation}
q'(x,y,z;t) = \hat{q}(y) e^{i\left( \alpha x + \beta z -\omega t \right)}
\end{equation}
where $\alpha$ and $\beta$ are the streamwise and spanwise wavenumbers, respectively, $\omega$ is the angular frequency and $\hat{q}$ is the wall normal distribution of the eigenfunction. Further details about the numerical implementation of the LST code are in \citet{Mack1976}.
The LST is used within a spatial framework, and therefore $\alpha$ is complex, while $\beta$ and $\omega$ are real numbers. The spatial growth rate is expressed by $\alpha_i$ and the laminar boundary layer is linearly unstable for $-\alpha_i>0$. The LST results presented in this work were benchmarked with existing data in the literature. For $M_{\infty}>4$, the difference in the spatial growth rate for the second Mack mode was below $10 \%$ (further details are in appendix \ref{appendix:LST_verification}). The agreement is deemed satisfactory for the purpose of this work, which is focused on the assessment, via DNS, of the effect of non-uniform surface temperature distribution on the second Mack mode stabilisation fora hypersonic boundary layer. 

\subsection{Data analysis methods}
The computations were advanced in time for about $250$ to $300$ times the fundamental period ($\tau=2\pi/\omega$). An initial numerical transient was discarded to allow the initial pressure disturbance due to the actuator to be convected outside the domain. Data were collected at a sampling rate $300/\tau$ for approximately $10\tau$, which provided sufficient spectral resolutions and statistical convergence of the amplification factor and growth rate. The streamwise evolution of the streak amplitude ($As_u(x)$) was determined based on the following definition, 
\begin{equation}
As_{u}(x) = \frac{1}{2} \left[ \max_{y,z}{\left( U(x) - U_b(x) \right)} - \min_{y,z}{\left( U(x) - U_b(x) \right)} \right]
\label{eq:Asu}
\end{equation}
In equation \ref{eq:Asu}, $U_b$ is the non-dimensional streamwise velocity for the base flow with spanwise uniform surface temperature distribution. This definition was initially introduced for low speed flows \citep{Andersson2001}, and adopted in most of the recent literature for supersonic and hypersonic flows \citep{Paredes2019, Caillaud2025}. 

The flow field is homogeneous in the spanwise direction, and therefore a frequency ($f$) and spanwise wavenumber ($k$) Fourier decomposition of the primitive variables is used to determine the amplitude of the perturbations due to the steady streaks, $(f,k)=(0,\pm 1)$, second Mack mode, $(f,k)=(1,0)$, and non-linear interactions, $(f,k)=(1,\pm1)$. In the rest of the text, the $\pm$ symbol is dropped for a more concise notation. The Chu's energy ($E_{Chu}^{fk}$, \citep{Chu1965}) is used to track the evolution of the boundary layer instabilities and it is defined as follows, 

\begin{equation}
\begin{split}
E_{Chu}^{fk}(x) = \frac{1}{2} \int_{0}^{L_y} \Biggl[ \overline{\rho}\left( \hat{u}\hat{u}^* + \hat{v}\hat{v}^* + \hat{w}\hat{w}^* \right) +\\
                  + \frac{\overline{T}}{\gamma M_{\infty}^2 \overline{\rho}} \hat{\rho}\hat{\rho}^* 
                  + \frac{\overline{\rho}}{\gamma \left( \gamma - 1 \right) M_{\infty}^2 \overline{T}}\hat{T}\hat{T}^* \Biggr] dy           
\end{split}
\label{eq:Echu}
\end{equation}
In equation \ref{eq:Echu}, $\overline{(\cdot)}$, $(\cdot)'$ and $\hat{(\cdot})$ indicate the mean flow deformation, the amplitude of the fluctuations and the Fourier coefficient, respectively, and $(\cdot)^*$ indicates the complex conjugate. $L_y$ indicates the wall-normal extent of the computational domain. The Chu's energy is chosen as a metric to quantify the modal energy as this takes into account both kinetic and thermodynamic energy contributions \citep{Unnikrishnan2020,Guo2023}, which are both relevant in the present study where streaks are generated through manipulation of the surface temperature. In addition, the Chu's energy it is also a commonly used metric in compressible linear input/output analysis \citep{Bugeat2019}, for the study of modal and non-modal boundary layer linear stability.

For the uncontrolled case, where explicitly indicated in the figure caption, the streamwise growth rate ($\sigma(x)$) and phase speed ($c_{ph}(x)$) of boundary layer hydrodynamic instabilities are computed as follows,

\begin{equation}
\sigma(x) = \frac{d}{d x} \ln \left( |\hat{p}_w| \right) 
\end{equation}

\begin{equation}
c_{ph}(x) = Re_{\infty}M_{\infty}F \left( \frac{d \Phi}{d x}  \right)
\end{equation}
In the preceding equations, $\hat{p}_w$ is the temporal Fourier coefficient of the wall static pressure fluctuations ($p'_w$), $F$ ($=\omega/\left( M_{\infty}^2 Re_{\infty} \right)$) is the non-dimensional forcing frequency usually used in linear stability theory, and $\Phi$ is the phase of the Fourier coefficient $\hat{p}_w$. In the uncontrolled case where only the second Mack mode is triggered, the flow remains two dimensional, $(x,y)$, and therefore $p'_w$ are spanwise averaged and only the amplitude of the fundamental harmonic ($\omega$) is used for the computation of $\sigma(x)$ and $c_{ph}(x)$. This data processing approach closely follows the methodology used by \citet{Egorov2006} and \citet{Marxen2010}. In addition, \citet{Mayer2011} shows that the use of static pressure fluctuations for the computation of the growth rate is likely less affected by non-parallel effects compared to streamwise velocity fluctuations. Thus, this is an appropriate metric for comparing DNS with parallel, LST results.

\section{Results}\label{sec:results}
In the following sections, the effect of a spanwise non-uniform surface temperature variation on second Mack mode amplification is determined and quantified. The operating conditions for the initial uncontrolled (section \ref{sec:results_uncontrolled}) and controlled (section \ref{sec:results_controlled}) case study  are based upon previous work in the literature \citep{Ozawa2023} with $M_{\infty}=6$, $\tilde{T}_{\infty}=216.7$K and unit Reynolds number based on the free stream speed $Re_{unit} \approx 11 \times 10^6$1/m. The effectiveness of the control method is then verified through a parametric assessment (section \ref{sec:results_parametric}) for a range of operating conditions representative of wind-tunnel and flight scenarios. 

\subsection{Baseline configuration}\label{sec:results_uncontrolled}
\begin{figure}
  \centering 
  \includegraphics[trim={0cm 0cm 0cm 0cm},clip,width=0.5\textwidth]{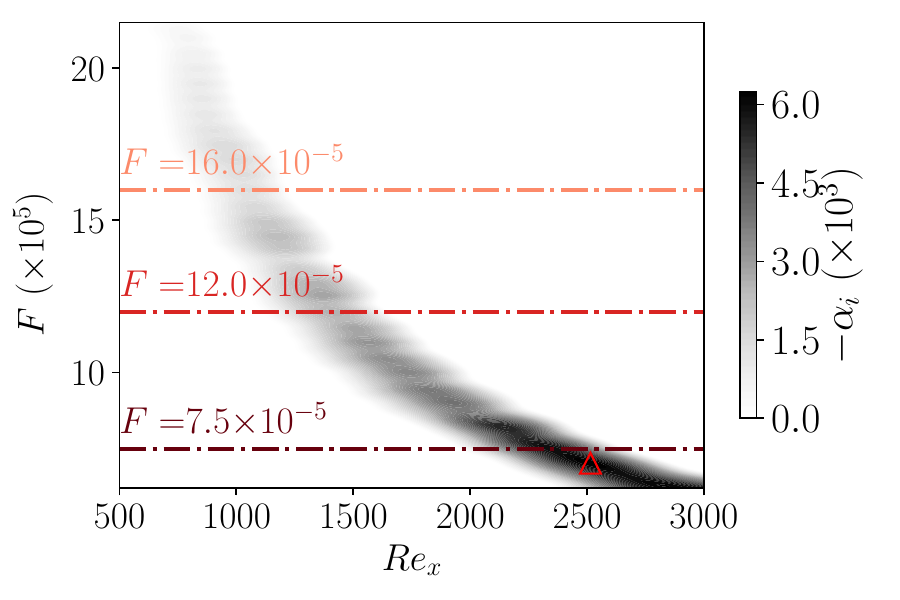} 
  \caption{Second Mack mode growth rate based on linear stability analysis for a laminar, self-similar base flow with $\tilde{T}_{\infty}=216.7 \textnormal{ K}$, $\tilde{p}_{\infty}=5475 \textnormal{ Pa}$ and $T_{w,base}=3$.}
  \label{fig:LST_contour}
\end{figure}

A cold flat plate is used as a baseline (uncontrolled, $A_{T_w}=0$) case, with $T_{w,base}=3$, which corresponds to approximately 42\% of the adiabatic wall temperature and it is sufficiently representative of flight conditions \citep{Schneider1999}. The non-dimensional forcing frequency ($F=\omega/(M_{\infty}^2 Re_{\infty})$) is $F=7.5\times10^{-5}$, and it is close to the most linearly amplified one based on LST analysis for a laminar, self-similar base flow (figure \ref{fig:LST_contour}). This choice is similar to previous work on DNS studies of boundary layer stability \citep{Pagella2002,Egorov2006} and transition \citep{Ryu2015}. An assessment of the sensitivity of control effectiveness to forcing frequency is also discussed in section \ref{sec:F_effect}. 

\begin{figure}
  \centering 
  \subfloat[]{\includegraphics[width=0.5\textwidth]{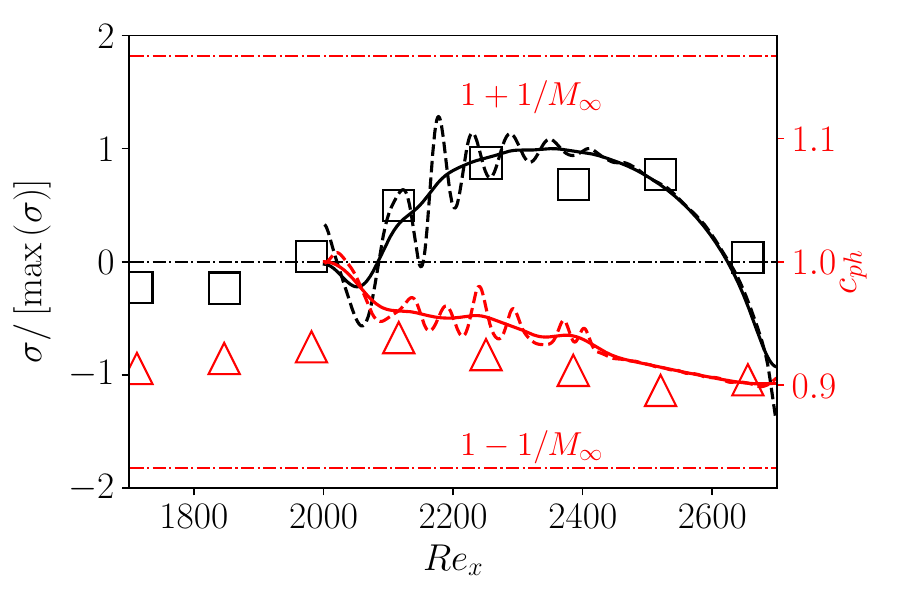}\label{fig:growth_rate_no_streaks}} 
  \hfill 
  \subfloat[]{\includegraphics[width=0.5\textwidth]{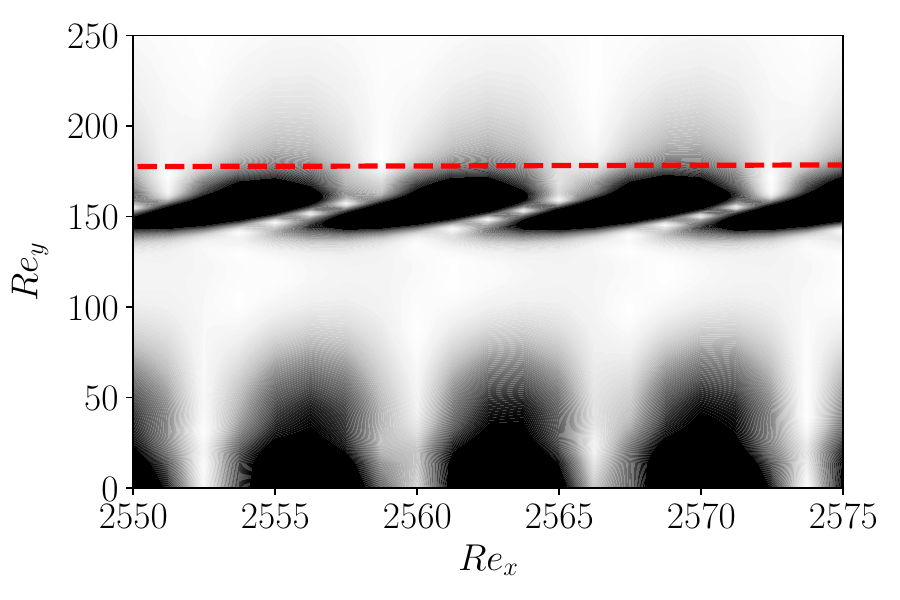}\label{fig:drhodx_no_streaks}} 
  \caption{(\textit{a}) Second Mack mode growth rate ($\sigma$, black) and non-dimensional phase speed ($c_{ph}$, red) based on (uncontrolled) DNS (lines) and LST (markers); filtered (dashed line) and unfiltered (solid line) DNS data computed from wall static pressure fluctuations. Black dot-dashed line demarcates second Mack mode stable ($\sigma<0$) and unstable($\sigma>0$) regions, respectively; red dot-dashed lines mark the phase speed of slow ($1-1/M_{\infty}$) and fast ($1+1/M_{\infty}$) acoustic waves. (\textit{b}) DNS time snapshot of streamwise density gradient fluctuations; red dashed line: $u=0.999$. Uniform ($T_w=3$) case.}
  \label{fig:uncontrolled_2nd_mode}
\end{figure}

For this case, the maximum growth rate ($\sigma$) of the second Mack mode occurs at approximately $Re_x\approx 2500$ (figure \ref{fig:growth_rate_no_streaks}), which is equivalent to a Reynolds number based on local boundary layer thickness ($\delta_{99}$) and freestream velocity $Re_{\delta_{99}} \approx 30940$. The instability manifests with a typical phase speed $c_{ph}\approx 0.9$, and rope-like signature in the fluctuations of the streamwise density gradient (figure \ref{fig:drhodx_no_streaks}). 
Both the growth rate and the phase speed show oscillation with a streamwise varying wavelength. These were also identified in previous numerical work \citep{Sivasubramanian2014,Ryu2015} that used high-order, spatial discretization schemes, and likely attributed to shock-ripples due to the actuator strip that was used to promote laminar to turbulence transition. In \citet{Mayer2011}, the phase speed of the instability mechanism for a Mach 3 boundary layer also shows similar oscillations, although only the decay phase of the instability is reported, and therefore the source remains unknown. More recently, \citet{Hader2024} reported the presence of similar oscillations in the envelope of wall static pressure fluctuations for a Mach 6 transitional boundary layer over a cone. The broadband forcing introduced to emulate natural transition was identified as the source of the oscillations. Within the context of this work, which is focused on the stabilisation of the second Mack mode via streaks, a gaussian filter is used to remove the spurious oscillation from the second Mack mode growth rate profile and enable a more quantitative comparison between the DNS and the LST. Relative to the LST results, the difference in the integrated area underneath the unstable region ($-\alpha_i \ge 0$) for the DNS simulations is approximately less than 1\%. The agreement between LST and DNS (figure \ref{fig:growth_rate_no_streaks}) confirms the appropriate selection of the time-space characteristics of the wall-normal momentum perturbation to trigger the second Mack mode. 

\subsubsection{Effect of uniform heating and cooling}
\begin{figure}
  \centering 
  \subfloat[]{\includegraphics[width=0.5\textwidth]{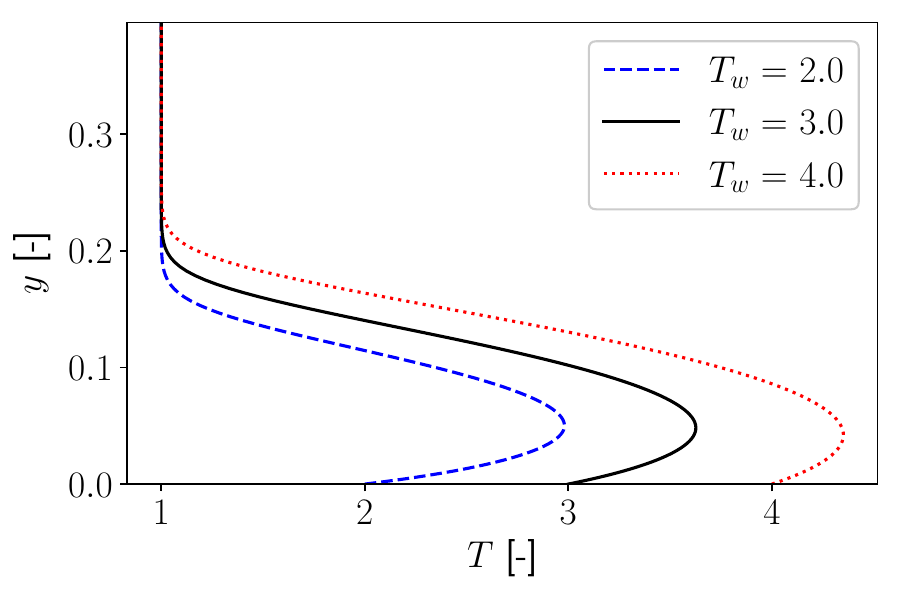}\label{fig:T}}
  \hfill 
  \subfloat[]{\includegraphics[width=0.5\textwidth]{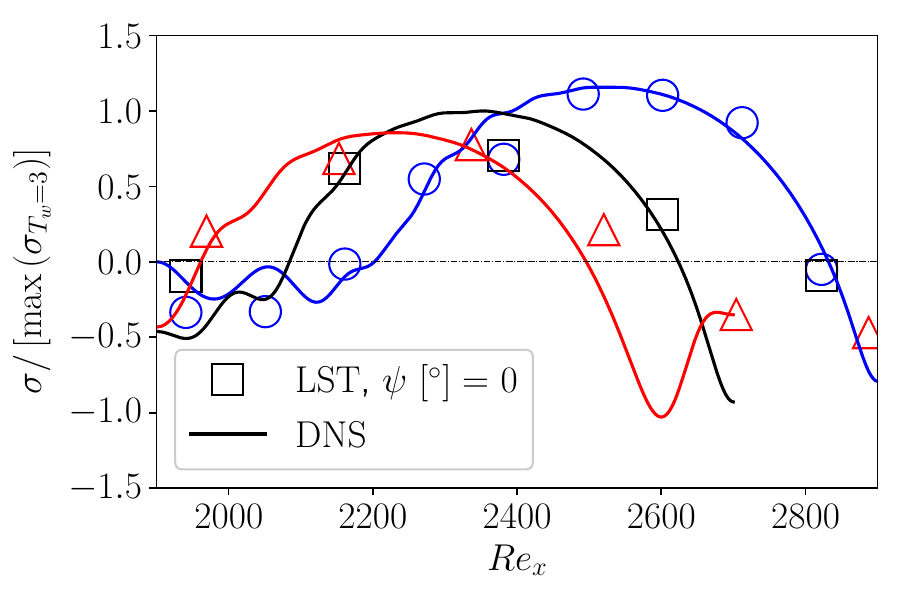}\label{fig:LST_DNS}} 
  \caption{(\textit{a}) Self-similar temperature profiles and (\textit{b}) second Mack mode growth rate based on (uncontrolled) DNS (lines) and LST (markers). In (\textit{b}), the DNS data are computed from the spanwise averaged wall static pressure fluctuations; the black dashed line demarcates second Mack mode stable ($\sigma<0$) and unstable($\sigma>0$) regions, respectively.}
  \label{fig:uncontrolled_cooling_heating_2nd_mode}
\end{figure}

To further verify that the second Mack mode instability is successfully triggered within the DNS domain, the wall temperature was uniformly increased ($T_w=4$) and decreased ($T_w=2$) relative to the baseline computation (figure \ref{fig:T}) and the DNS results are compared with the LST results. For this case study, the growth rate of the instability ($\sigma$) in the DNS is computed based on the spanwise averaged wall static pressure fluctuations and the results are normalised relative to the maximum growth rate for the baseline case ($\max\left(\sigma_{T_w=3}\right)$). To ease figure readability, only the spatially filtered growth rates are reported for the DNS, although oscillations due to the actuator are also present at different wall temperatures .As expected based on previous research \citep{Mack1975}, cooling and heating destabilises and stabilises the second Mack mode, respectively (figure \ref{fig:LST_DNS}). Both DNS and LST were able to capture these effects, and this provides confidence that second Mack mode instability was triggered in the DNS computations, despite some differences in the decay rate between LST and DNS.   

\subsection{Effect of streaks on second Mack mode stabilisation}\label{sec:results_controlled}
\begin{figure}
  \centering 
  \includegraphics[trim={0cm 0cm 0cm 0cm},clip,width=0.5\textwidth]{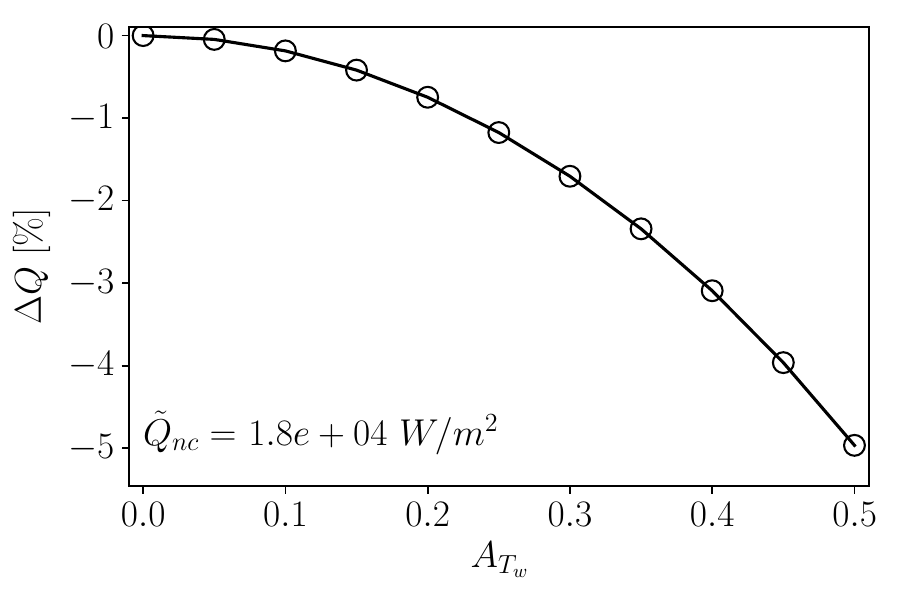} 
  \caption{Effect of non-uniform wall temperature on integrated surface heat flux. Numerically estimated based on a compressible, self-similar laminar boundary layer for a Mach 6 flat plate configuration with $T_{\infty}=216.7 \textnormal{ K}$, $\tilde{p}_{\infty}=5475 \textnormal{ Pa}$ and $T_{w,base}=3$.}
  \label{fig:power_input}
\end{figure}

For the controlled configuration, the amplitude of the spanwise temperature variation is set to $A_{T_w}=0.3$ for both the hot and cold patch. Thus, the surface temperature distribution is anti-symmetric relative to the x axis, and the base flow surface temperature for both the controlled and uncontrolled case remains the same. This is to mimic a passive flow control method configuration, for which a practical implementation has been proposed by \citet{Ozawa2025} using appropriately selected materials with different thermal characteristics. In the DNS studies, as a results of the base flow wall temperature being held constant, the integrated surface heat flux ($\tilde{Q}$) slightly reduces for the controlled configurations. Relative to the uncontrolled configuration, the reduction in $\tilde{Q}$ for the controlled cases is due to the non-linear relationship between surface temperature and heat transfer. The boundary layer in the DNS computations remains laminar, and therefore $\tilde{Q}$ can be estimated apriori for both the controlled ($\tilde{Q}_{c}$) and uncontrolled ($\tilde{Q}_{nc}$) configurations using the wall heat transfer ($\tilde{q}$) relationship for a compressible, self-similar, laminar boundary layer over a flat plate with zero pressure gradient \citep{White2006}, which is expressed as follows,

\begin{equation}
\tilde{q}(\tilde{x},\tilde{z}) = 0.332 \tilde{\rho}_{\infty}  \tilde{u}_{\infty} \tilde{c}_p \sqrt{\frac{\tilde{\mu}_w(\tilde{x},\tilde{z})}{\tilde{\rho}_{\infty}  \tilde{u}_{\infty} \tilde{x}}} (\tilde{T}_{aw} - \tilde{T}_w(\tilde{x},\tilde{z}))
\label{eq:heat_transfer_laminar}
\end{equation}
where $\tilde{c}_p$ is the isobaric specific heat for air, $\tilde{\mu}_w$ is the molecular viscosity at the wall, and $\tilde{T}_{aw}$ is the adiabatic wall temperature. Numerical integration of equation \ref{eq:heat_transfer_laminar} along the streamwise ($x$) and spanwise ($z$) directions for various spanwise wall temperature perturbation, $A_{T_w}$, provides an estimate of the difference in the energy balance for controlled and uncontrolled configurations ($\Delta Q = (\tilde{Q}_c - \tilde{Q}_{nc})/\tilde{Q}_{nc}$). For a Mach 6 boundary layer at $20000 \textnormal{ m}$ altitude conditions and with $T_{w,base}=3$, increasing $A_{T_w}$ from 0 to 0.5 approximately leads to a 5\% reduction in integrated surface heat flux relative to the uncontrolled ($A_{T_w}=0$) configuration (figure \ref{fig:power_input}). This outcome arises from the modelling choice to regulate temperature in the DNS simulations.  Thus, the control method can be classified as active \citep{Gad-el-Hak_2000} as implemented in the computational model, while for a flight-relevant practical implementation this can also be regarded as a passive flow control management concept \citep{Fiedler1990}, in that a non-uniform surface temperature distribution may be achieved without an external power device by tailoring the surface thermal properties and thickness so that the required temperature distribution is driven by the local aerothermodynamic heat transfer environment as recently proposed by \citet{Ozawa2025}.

For the controlled case, the maximum ($T_w=4$) and minimum ($T_w=2$) wall temperature is approximately $58\%$ and $29\%$ the adiabatic wall temperature, respectively. Both the controlled and the uncontrolled configurations are initialised with a self-similar laminar solution for an isothermal flat plate boundary layer corresponding to the uncontrolled (uniform) baseline wall temperature, $T_{w,base}=3$. For the controlled case, a spanwise non-uniform surface temperature is then used. As anticipated in the introduction, both streaks and heating and cooling affect the second Mack mode stabilisation. Thus, several controlled configurations with different streak amplitude are used to provide an assessment of the influence of the streaks on second Mack mode amplification (table \ref{tab:Asu_effect}). The resulting streak amplitude is varied either through a change in the streamwise location  where the spanwise non-uniform surface temperature is enforced ($x_{T_w,s}$) relative to the end of the blowing and suction region ($x_{bs,e}$), or through a change of the fundamental spanwise wavelength of the streaks ($\lambda_z$).

\begin{table}
  \begin{center}
\def~{\hphantom{0}}
  \begin{tabular}{lcccccc}
      Case & $T_{w,base}$  & $\lambda_z$   &   $A_{T_w}$ & $x_{T_w,s}-x_{bs,e}$ & $Re_{x_{T_w,s}}$\\[3pt]
       C0  & 3 & 1.2 & 0.3 & -10 & 1580\\
       C1a  & 3 & 1.2 & 0.3 & 0 & 1870\\
       C2  & 3 & 1.2 & 0.3 & 5 & 2000\\
       C3  & 3 & 1.2 & 0.3 & 10 & 2120\\
       C1b  & 3 & 0.9 & 0.3 & 0 & 1870\\
       C1c  & 3 & 1.5 & 0.3 & 0 & 1870\\
       C1d  & 3 & 2.4 & 0.3 & 0 & 1870\\
       C1e  & 3 & 4.8 & 0.3 & 0 & 1870\\
  \end{tabular}
  \caption{Summary of controlled configurations investigated for the initial case study; $M_{\infty}=6$, $(Re_{\infty}M_{\infty})=1.0\times 10^5$, $\tilde{h}_{0,\infty}=1.8\times 10^6$J/kg.}
  \label{tab:Asu_effect}
  \end{center}
\end{table}

\begin{figure}
  \centering 
  \includegraphics[trim={0cm 0cm 0cm 0cm},clip,width=\textwidth]{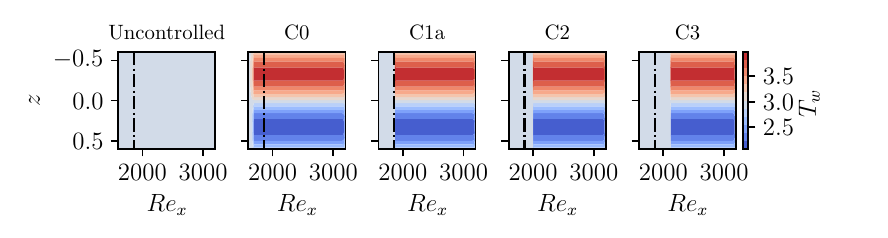} 
  \caption{DNS results showing wall temperature distribution for the uncontrolled and controlled configurations under investigation, relative to the end of the diturbance forcing region (dot-dashed line).}
  \label{fig:twall_overlap_effect}
\end{figure}

The range of streamwise locations investigated spanned from a case with overlap between the disturbance forcing actuator and the spanwise non-uniform surface temperature distribution (case C0, figure \ref{fig:twall_overlap_effect}), to configurations where the spanwise non-uniform surface temperature boundary condition is  enforced progressively closer to the onset of the second Mack mode (cases C1 to C3). The case with overlap (case C0) is not further investigated, as an initial assessment showed that it is important to avoid overlap between the disturbance forcing and the control method to consistently determine and quantify the effect of streaks on second Mack mode linear amplification (further details are in appendix \ref{appendix:overlap_effect}).  

\begin{figure}
  \centering 
  \subfloat[]{\includegraphics[width=0.5\textwidth]{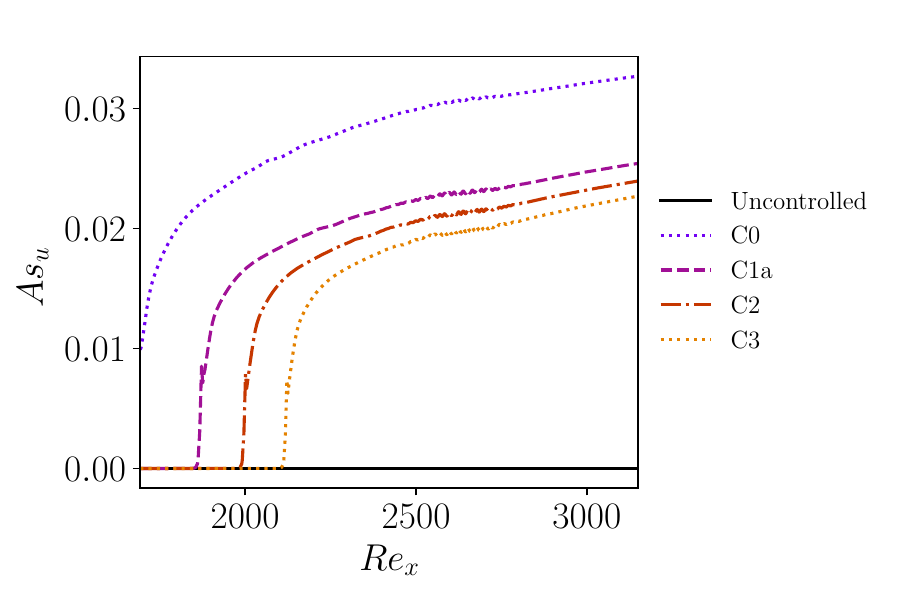}\label{fig:Asu_xtw_effect}} 
  \hfill 
  \subfloat[]{\includegraphics[width=0.5\textwidth]{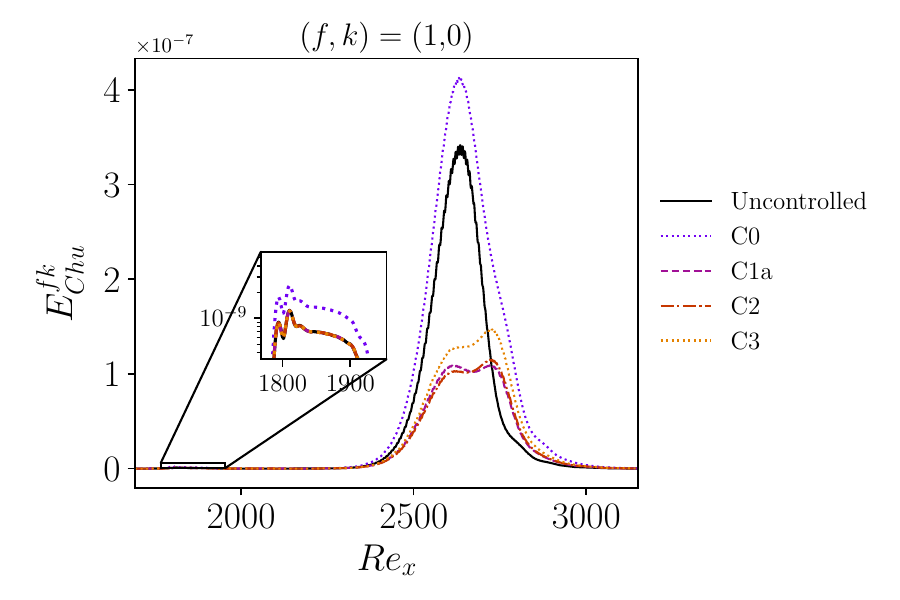}\label{fig:EChu_xtw_effect}}  
  \hfill 
  \subfloat[]{\includegraphics[width=0.5\textwidth]{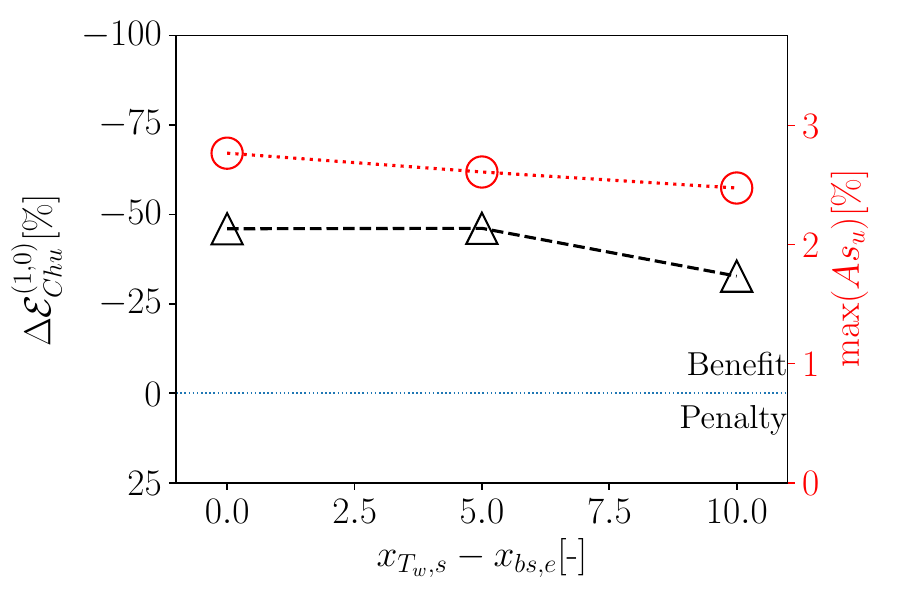}\label{fig:summary_xtw_effect}}
  \caption{DNS results showing the effect of actuator/control overlap on (\textit{a}) streak amplitude and (\textit{b}) second Mack mode energy. (\textit{c}) Influence of $x_{T_w,s}$ on second Mack mode stabilisation (left y-axis) and maximum streak amplitude (right y-axis).}
\end{figure}

For the cases C1a, C2 and C3, the streak amplitude undergoes a noticeable growth from the start of the non-uniform wall temperature distribution to the end of the computational domain (figure \ref{fig:Asu_xtw_effect}). The streaks reduce the  energy of the second Mack mode (figure \ref{fig:EChu_xtw_effect}), which is stabilised by the spanwise non-uniform surface temperature distribution. The inset in figure \ref{fig:EChu_xtw_effect} depicts the modal energy associated to the forcing disturbance via blowing and suction, which is the same for the controlled and uncontrolled configurations. Thus, the stabilisation of the second Mack mode due to the spanwise non-uniform surface temperature is quantified based on the percentage ratio $\Delta \mathcal{E}_{Chu}^{(1,0)}$[\%], which is defined as follows, 

\begin{equation}
\Delta \mathcal{E}_{Chu}^{(1,0)} = \frac{\left( \int_{x_s}^{x_e} E_{Chu,c}^{(1,0)}dx - \int_{x_s}^{x_e} E_{Chu,nc}^{(1,0)}dx \right)}{\int_{x_s}^{x_e} E_{Chu,nc}^{(1,0)}dx}100
\label{eq:second_mode_stabilisation}
\end{equation}
Where $E_{Chu,nc}^{(1,0)}$ and $E_{Chu,c}^{(1,0)}$ are the second Mack mode energies for the uncontrolled and controlled case, respectively. The metric $\Delta \mathcal{E}_{Chu}^{(1,0)}$ quantifies the stabilisation of the planar second Mack mode induced by the control streaks. The contribution of non-linear components, $(f,k)=(1,1)$, arising from spanwise non-uniformity remains small to approximately $ 15\%$, as discussed in further detail in Appendix~\ref{appendix:overlap_effect}. The choice of this energy metric, as opposed to logarithmic growth rate, is also motivated by recent laminar to turbulence transition studies \citep{Boscagli2025_SCITECH} showed transition delay via low amplitude  ($As_u<0.05 \tilde{u}_\infty$) control streaks, due to a significant reduction of high-frequency shear-stresses associated to the second Mack mode planar wave. An energy norm conveys the mean level of fluctuations in small amplitude disturbances \citep{Chu1965}, and therefore it is an appropriate metric within the context of this work. 

As the control method is activated closer to the onset of the second Mack mode, the amplitude of the streaks slightly reduces and the method also becomes less effective (figure \ref{fig:summary_xtw_effect}). For example, relative to case C1a, the amplitude of the streaks reduces by approximately $0.3\%$ for case C3, and $\Delta \mathcal{E}_{Chu}^{(1,0)}$ also reduces by approximately $10\%$. This is an indication that the streaks generated through the spanwise non-uniform surface temperature distribution contribute to the stabilisation of the second Mack mode. This is further investigated by changing the streak amplitude through a change in the streaks fundamental spanwise wavelength.

\begin{figure}
  \centering 
  \subfloat[]{\includegraphics[trim={12cm 4cm 12cm 5cm},clip,width=0.5\textwidth]{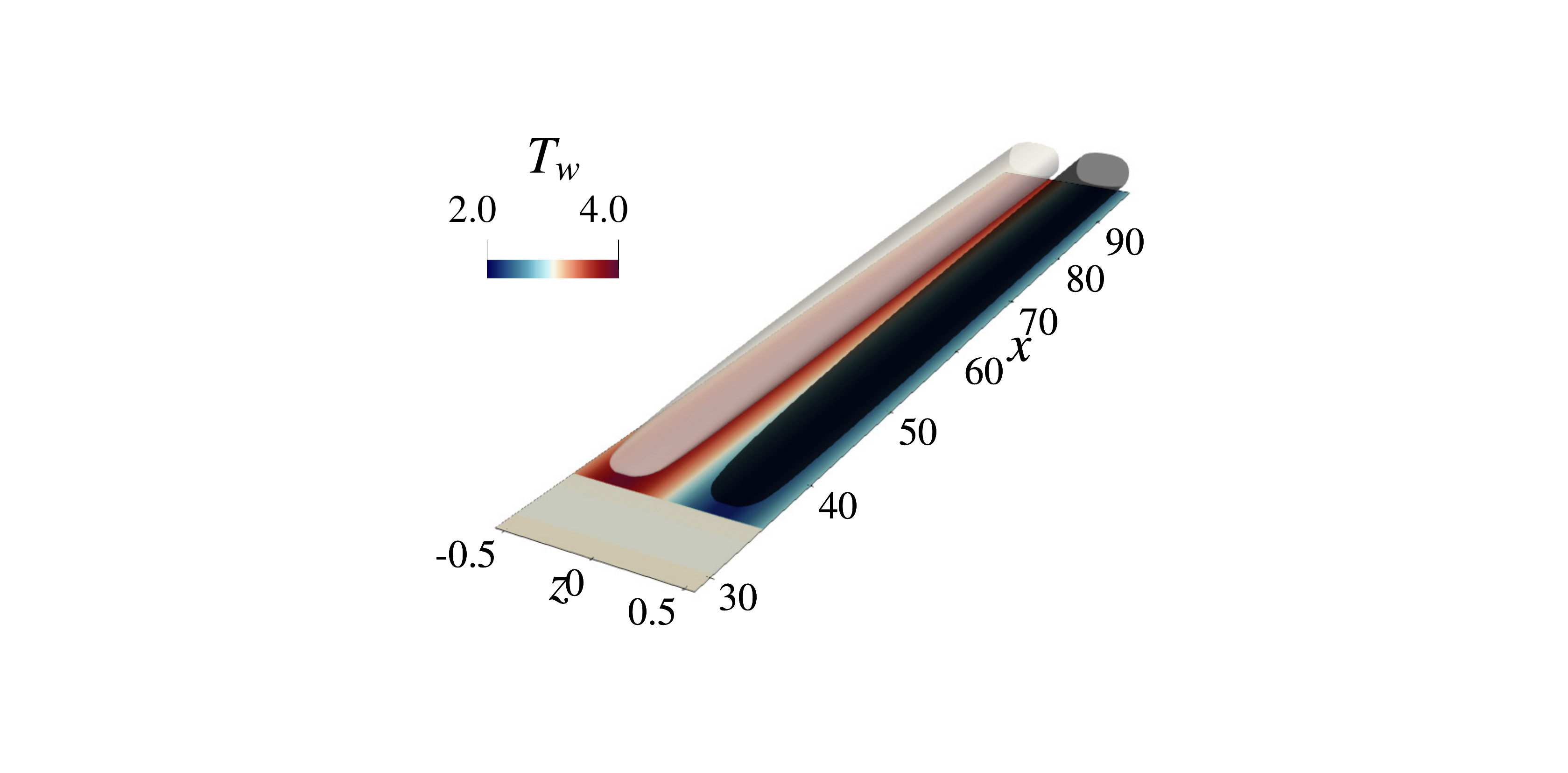}\label{fig:streaks_lambda_z_1p2}} 
  \hfill 
  \subfloat[]{\includegraphics[trim={12cm 4cm 12cm 5cm},clip,width=0.5\textwidth]{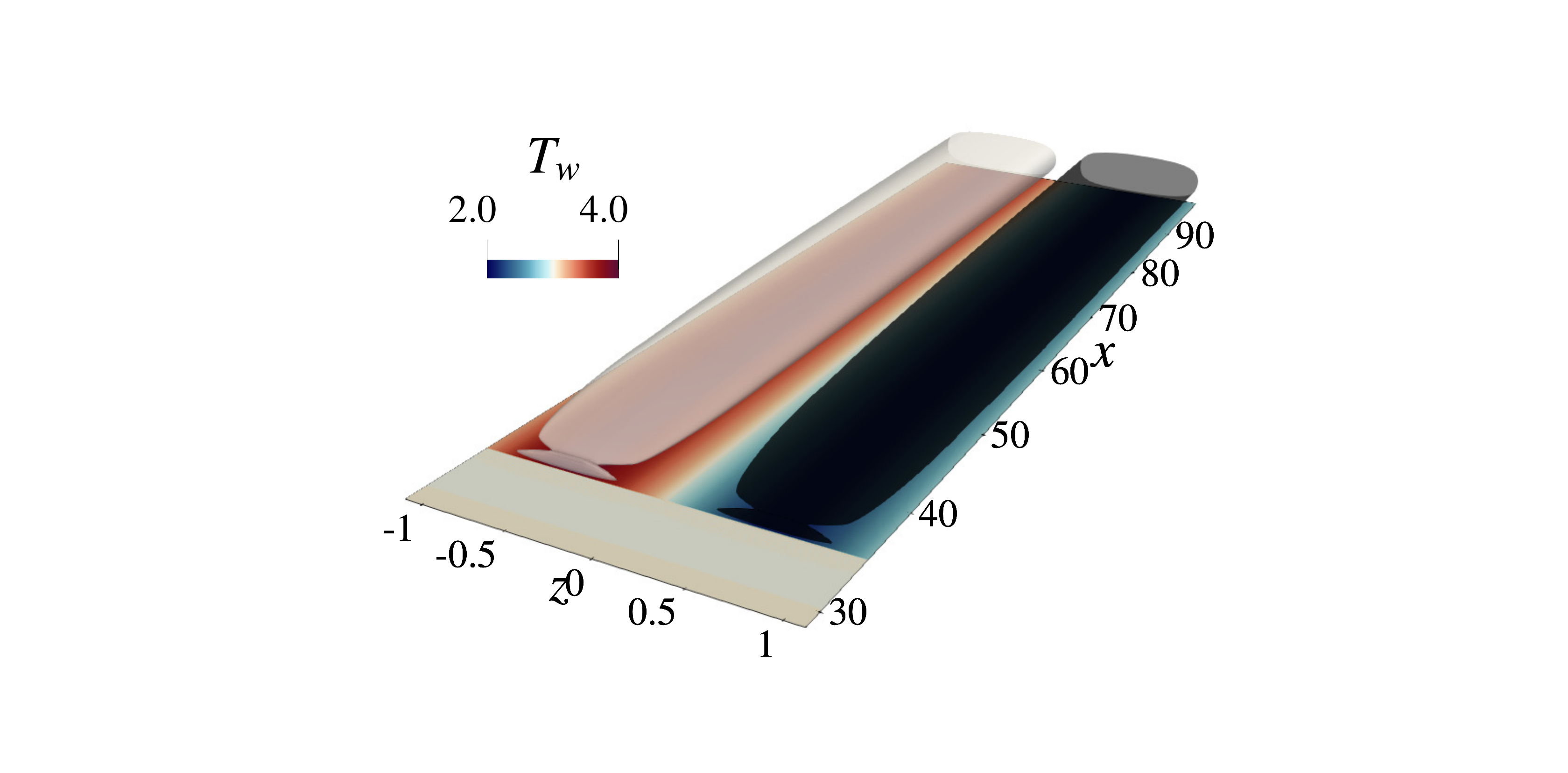}\label{fig:streaks_lambda_z_2p4}}  
  \caption{DNS results showing the effect of streak wavelength ($\lambda_z$) on  streaks streamwise growth. (\textit{a}) $\lambda_z=1.2$ and (\textit{b}) $\lambda_z=2.4$. Isosurfaces show streamwise velocity fluctuations of the streak fundamental harmonic $(f,k)=(0,1)$, with positive ($+0.01$, black) and negative ($-0.01$, white) values.}
\end{figure}

\begin{figure}
  \centering 
  \subfloat[]{\includegraphics[width=0.5\textwidth]{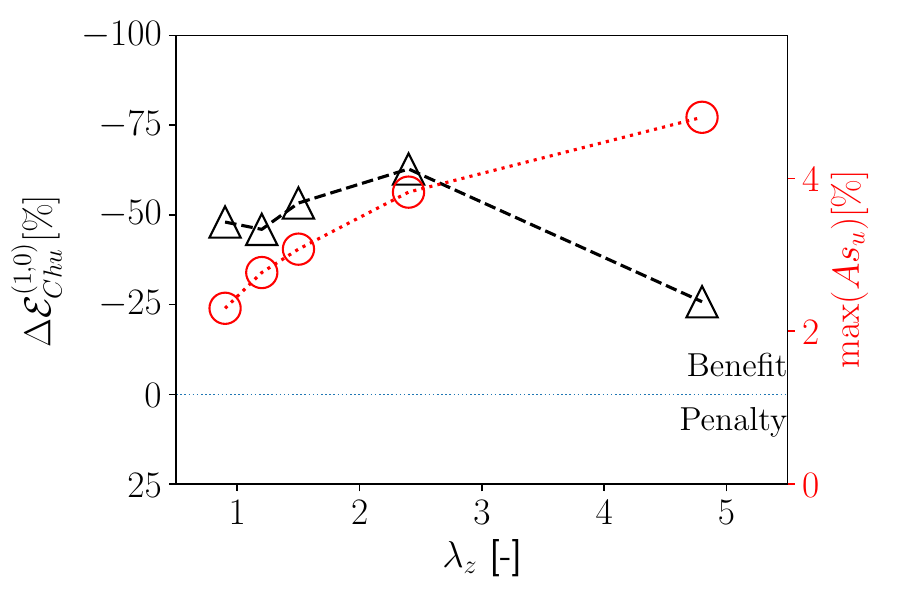}\label{fig:summary_lambda_z_effect}} 
  \hfill 
  \subfloat[]{\includegraphics[width=0.5\textwidth]{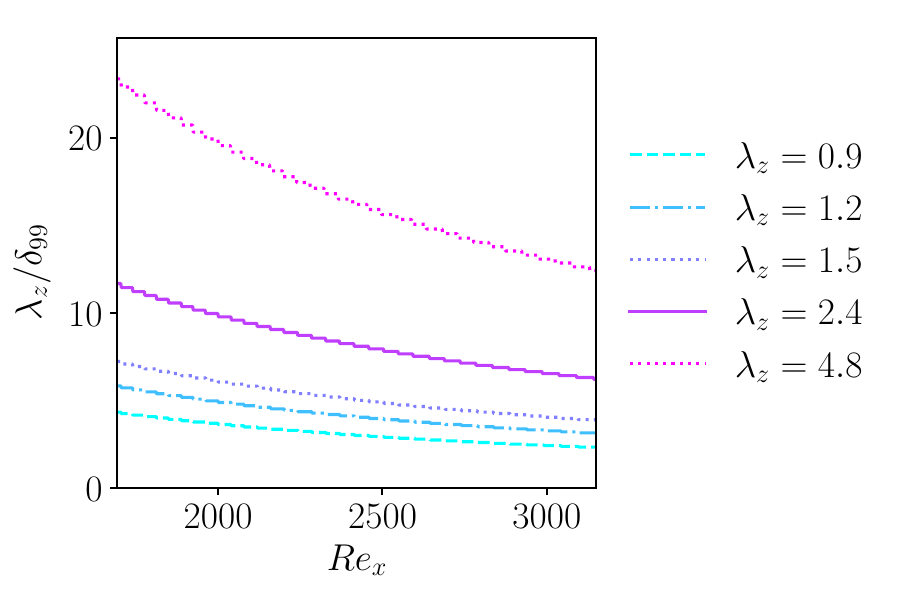}\label{fig:lambda_z_effect_delta_99}}  
  \caption{DNS results showing the (\textit{a}) influence of $\lambda_z$ on second Mack mode stabilisation (left y-axis) and maximum streak amplitude (right y-axis); (\textit{b}) non-dimensional streamwise distribution of the ratio of the base flow boundary layer thickness ($\delta_{99}$) to the fundamental spanwise wavelength of the streaks ($\lambda_z$).}
\end{figure}

The spanwise extent of the computational domain $\lambda_z$ is varied between $\lambda_z=0.9$ and $4.8$ (cases C1a, C1b, C1c and C1d in table \ref{tab:Asu_effect}), to further assess the combined effect of streak amplitude and spanwise wavelength on second Mack mode stabilisation. As $\lambda_z$ is increased from 1.2 (figure \ref{fig:streaks_lambda_z_1p2}) to 2.4 (figure \ref{fig:streaks_lambda_z_2p4}) the maximum streak amplitude increases by approximately 1\%, and, relative to the uncontrolled case, the stabilisation effect on second Mack mode also increases from approximately 45 to 62\%(figure \ref{fig:summary_lambda_z_effect}). However, doubling the streak wavelength from $\lambda_z=2.4$ to $4.8$, produces a noticeable loss in control performance with a reduction in $\Delta \mathcal{E}_{Chu}^{(1,0)}$ from 62\% to 25\%. Relative to the local boundary layer thickness ($\delta_{99}$), the investigated wavelengths of the streaks range from approximately $\lambda_z\approx4\delta_{99}$ to $15\delta_{99}$ at the location of maximum amplification of the second Mack mode (figure \ref{fig:lambda_z_effect_delta_99}). The analyses at $M_{\infty}=6$ indicate that nearly optimum stabilisation is achieved for $\lambda_z\approx10\delta_{99}$, with a noticeable loss in performance for larger spanwise wavelengths. Overall, an increase in streak amplitude either via a change in $x_{T_w,s}$ or in $\lambda_z$ leads to an increase in the second Mack mode stabilisation effect of the control method. Thus, this indicates that the reduction of the linear amplification of the second Mack mode may not be caused by the surface temperature, rather by the streaks. In the next section, the physical mechanism underlying the control effects are further investigated.

\subsubsection{Mechanisms of stabilisation}\label{sec:stab_mechanisms}
\begin{figure}
  \centering 
  \includegraphics[width=0.75\textwidth]{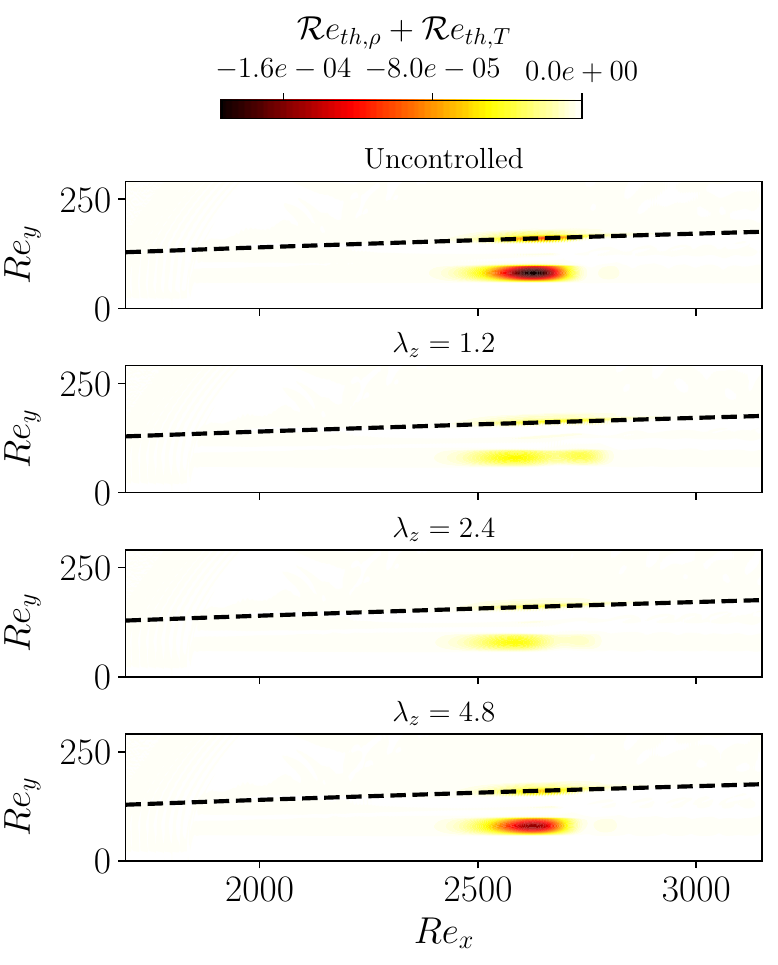}   
  \caption{Spatial (x-y) distribution of the thermoacoustic Reynolds stresses for uncontrolled and controlled configurations based on DNS data. The black dashed line indicates the outer edge of the boundary layer ($u\approx0.999$) for the base flow.}
\label{fig:lambda_z_effect_Re_stresses}  
\end{figure}

\begin{figure}
  \centering 
\includegraphics[width=0.5\textwidth]{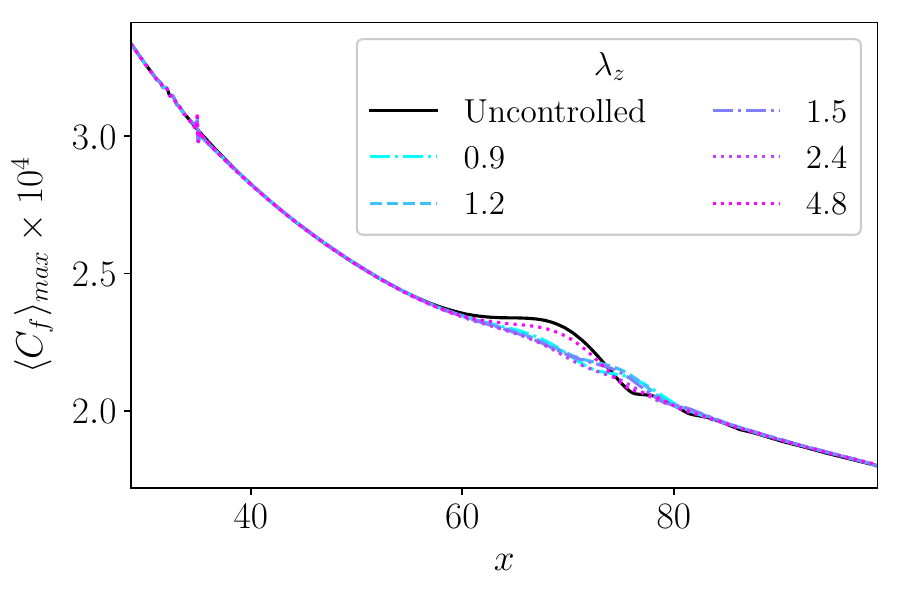}
  \caption{DNS results showing the effect of streak wavelength on the streamwise distribution of the envelope of the instantaneous, spanwise-averaged skin friction coefficient.}
\label{fig:lambda_z_effect_Cf}   
\end{figure}

The thermoacoustic Reynolds stresses ($\mathcal{R}e_{th}$), that represent a driving source of second Mack mode instability \citep{Kuehl2018,Chen2023}, are also investigated as a further confirmation of the stabilisation effect of the streaks. Based upon the inviscid, parallel derivation in \citet{Kuehl2018}, $\mathcal{R}e_{th}$ related to the second Mack mode acoustic energy are defined as follows,
\begin{equation}
\mathcal{R}e_{th,\rho} = \frac{d}{dy}\left(\overline{T} |\hat{\rho}|^{(1,0)} |\hat{v}|^{(1,0)} \right)
\end{equation}
\begin{equation}
\mathcal{R}e_{th,T} = \frac{d}{dy}\left(\overline{\rho} |\hat{T}|^{(1,0)} |\hat{v}|^{(1,0)} \right)
\end{equation}    
When the sum of the two terms in the equations above is negative ($\mathcal{R}e_{th,\rho}+\mathcal{R}e_{th,T} < 0$), the energy of the disturbance is amplified \citep{Kuehl2018}. Relative to the uncontrolled configuration, for these operating conditions the streaks always reduce the magnitude of the negative thermoacoustic Reynolds stresses (figure \ref{fig:lambda_z_effect_Re_stresses}). For the configuration with $\lambda_z=4.8$, the damping effect reduces and the stabilisation benefit is eroded as already discussed in figure \ref{fig:summary_lambda_z_effect}. A similar trend is also identified in the envelope of the instantaneous, spanwise-averaged skin friction coefficient ($\langle C_f \rangle_{max}$, figure \ref{fig:lambda_z_effect_Cf}). Relative to the uncontrolled case, the streaks always reduces the amplitude of the high-frequency, peak stresses, although for this case study where only small-amplitude disturbances are investigated, the benefit remains marginal. 

\begin{figure}
  \centering 
  \subfloat[]{\includegraphics[trim={0cm 0cm 0cm 0cm},clip,width=\textwidth]{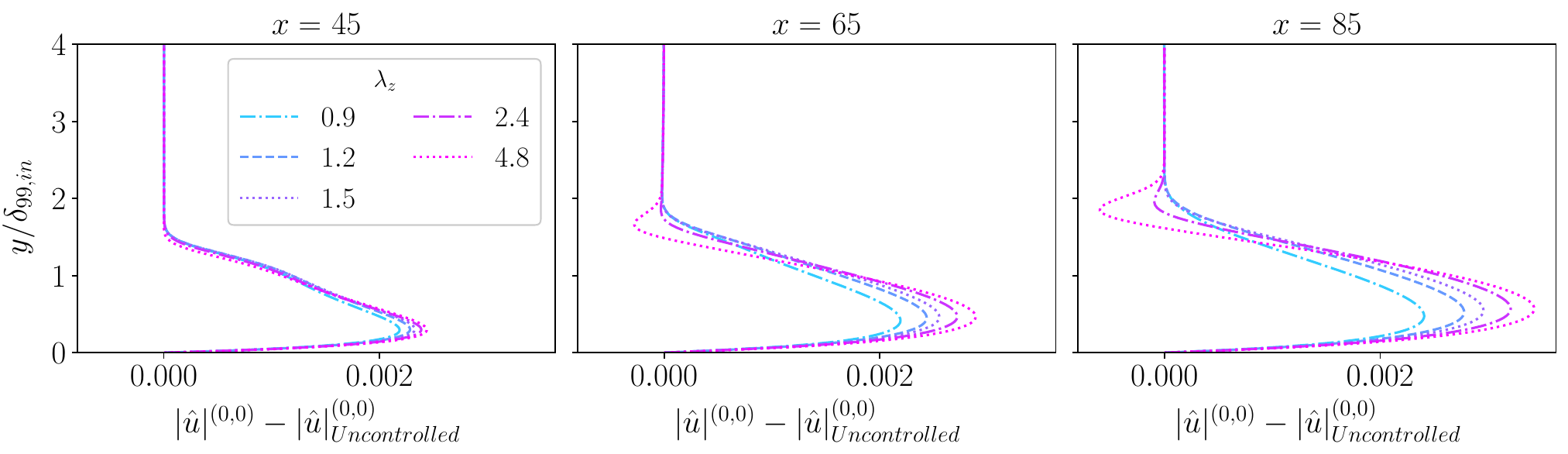}\label{fig:MFD_Velocity}} 
  \hfill 
  \subfloat[]{\includegraphics[trim={0cm 0cm 0cm 0cm},clip,width=\textwidth]{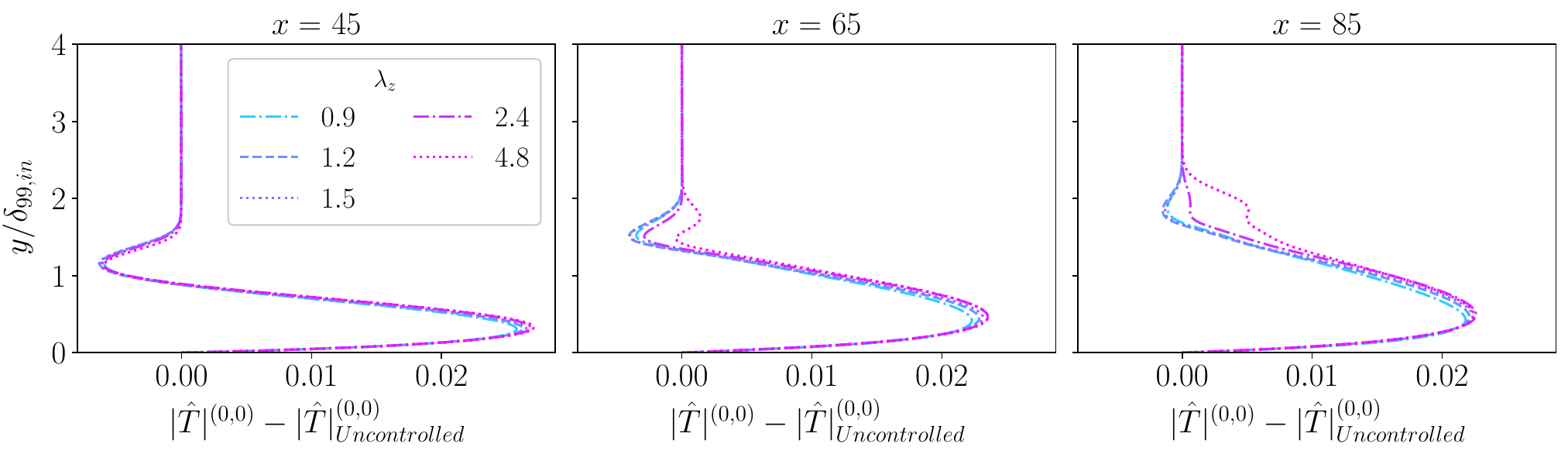}\label{fig:MFD_Temperature}}  
  \caption{DNS results showing the effect of streak wavelength ($\lambda_z$) on  base flow, $(f,k)=(0,0)$, deformation. Perturbation (\textit{a}) streamwise velocity and (\textit{b}) static temperature profiles at various streamwise locations ahead ($x=45$), across ($x=65$) and downstream ($x=85$) of the second Mack mode.}
  \label{fig:MFD}
\end{figure}

\begin{figure}
  \centering 
  \includegraphics[trim={0cm 0cm 0cm 0cm},clip,width=\textwidth]{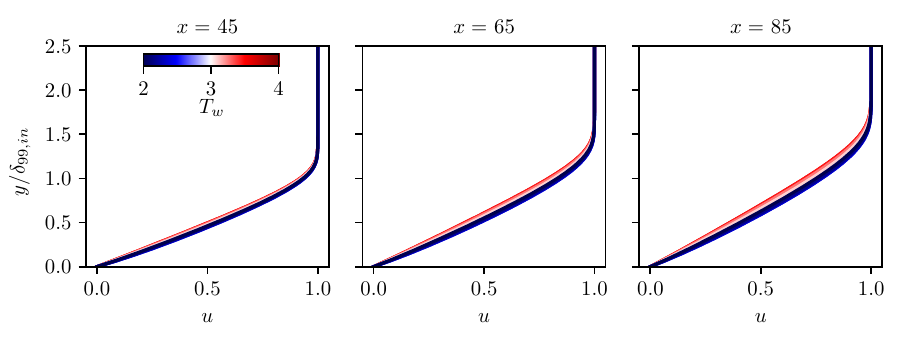}
  \caption{DNS results showing the effect of wall temperature on streamwise velocity profiles at various streamwise locations ahead ($x=45$), across ($x=65$) and downstream ($x=85$) of the second Mack mode. Configuration with $\lambda_z=4.8$.}
  \label{fig:MFD_lambda_z_4p8}
\end{figure}

Previous work \citep{Ren2016} for a Mach 6 configuration and for similar amplitude of the streaks ($As_u\in[1,5]\%$), has identified the base flow deformation due to non-linear interaction of the control streaks as the dominant mechanism of stabilisation of both first and second Mack modes. Figures \ref{fig:MFD_Velocity} and \ref{fig:MFD_Temperature} depict the perturbation base flow profiles for the streamwise velocity and static temperature, respectively, at various streamwise locations ahead ($x=45$), across ($x=65$) and downstream ($x=85$) of the second Mack mode. The wall normal coordinate is scaled with the boundary layer thickness at the inlet of the domain ($\delta_{99,in}$). The amplitude of the streak increases with the streak wavelength (figure \ref{fig:summary_lambda_z_effect}) and the amplitude of the base flow modification becomes greater, for both the velocity and temperature perturbation fields. This effect is more prominent downstream of the second Mack mode (figure \ref{fig:MFD}, $x=85$), where the boundary layer is mostly affected by the spanwise non-uniform wall temperature (figure \ref{fig:MFD_lambda_z_4p8}). Notably, this base flow modification due to the streaks leads to fuller velocity profiles near the wall, which may be beneficial for transition delay as also indicated in previous studies \citep{Cossu2002,Wassermann2002,Ren2016,Paredes2017}. An increase in $\lambda_z$ from 2.4 to 4.8 leads to the onset of an inflection point in the perturbation base flow velocity field, both across (figure \ref{fig:MFD_Velocity}, $x=65$) and downstream of the second Mack mode (figure \ref{fig:MFD_Velocity}, $x=85$). This is likely to produce secondary, inflectional instabilities, although these may not be fully supported by the low amplitude streaks \cite{Cossu2002}. Overall, this is consistent with previous work for a Mach 6 configuration \citep{Ren2016} that used optimal perturbations to generate the control streaks. \citet{Ren2016} shows that for $As_{u} < 0.05\tilde{u}_{\infty}$ the amplitude of both the first and second Mack mode can be reduced through control-streaks. Overall, this provides a plausible explanation of the reduction in the stabilising effect of the control method for $\lambda_z>2.4$ (figure \ref{fig:summary_lambda_z_effect}).

\begin{figure}
  \centering 
  \subfloat[]{\includegraphics[trim={0cm 0cm 0cm 0cm},clip,width=0.5\textwidth]{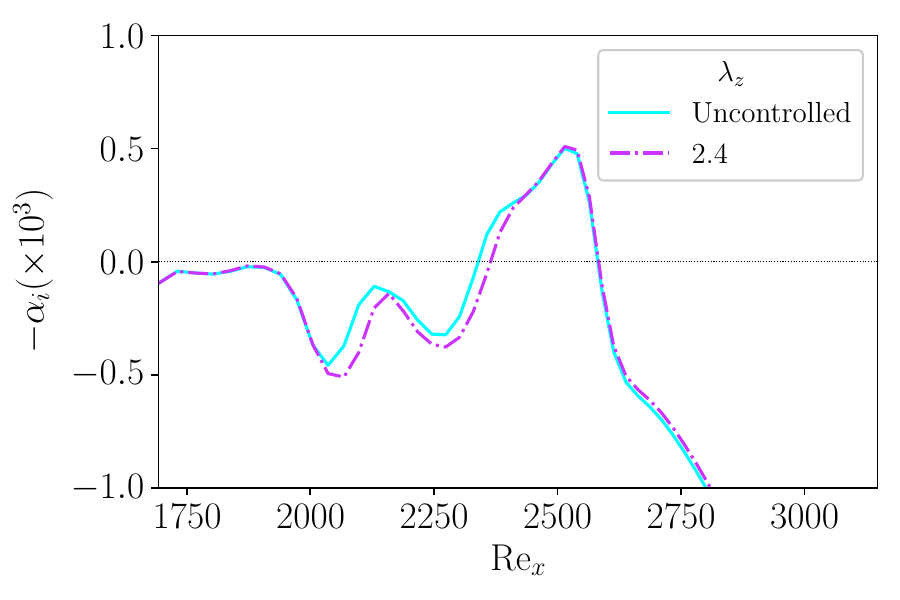}\label{fig:LST_alpha_7p5}} 
  \hfill 
  \subfloat[]{\includegraphics[trim={0cm 0cm 0cm 0cm},clip,width=0.5\textwidth]{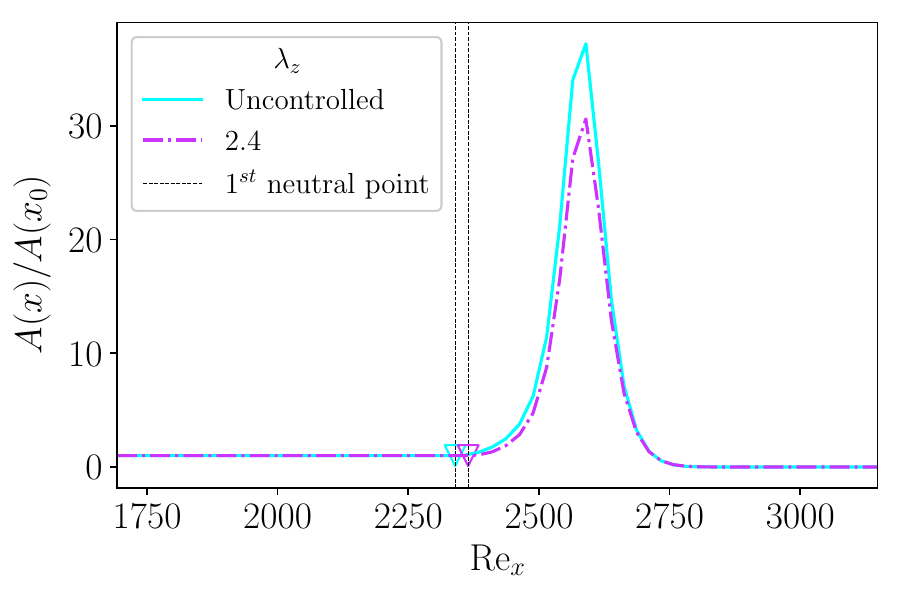}\label{fig:LST_A_7p5}}
  \hfill
  \subfloat[]{\includegraphics[trim={0cm 0cm 0cm 0cm},clip,width=0.5\textwidth]{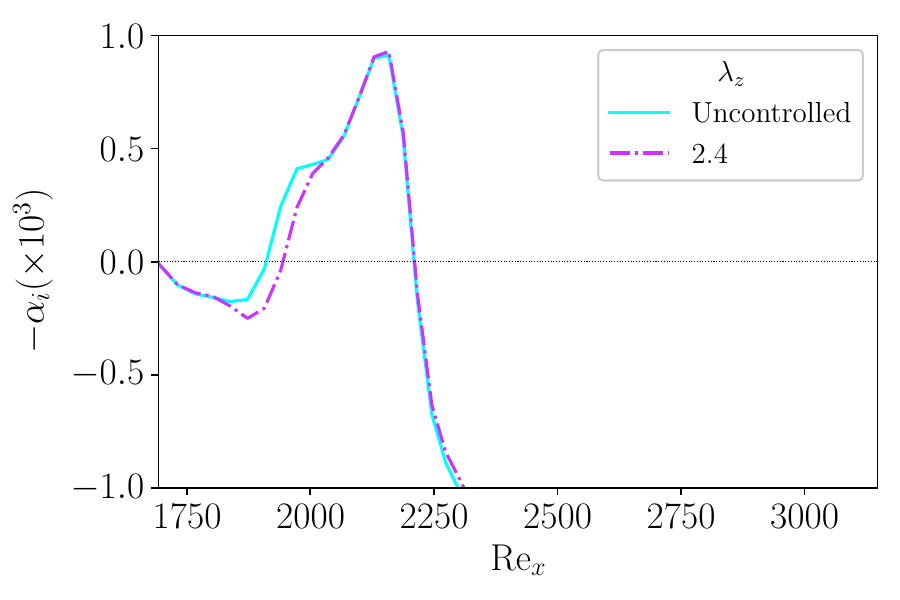}\label{fig:LST_alpha_12}} 
  \hfill 
  \subfloat[]{\includegraphics[trim={0cm 0cm 0cm 0cm},clip,width=0.5\textwidth]{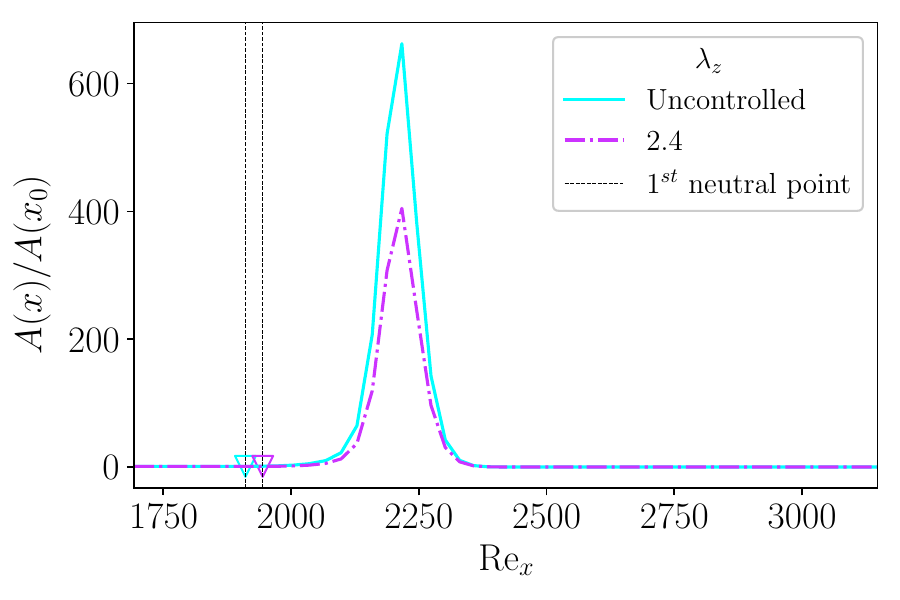}\label{fig:LST_A_12}}  
  \caption{Local, parallel LST of the DNS base flow showing the effect of control streaks on growth rate, (\textit{a}), (\textit{c}), and normalised disturbance amplitude, (\textit{b}), (\textit{d}). (\textit{a}),(\textit{b}): $F=7.5\times10^{-5}$; (\textit{c}),(\textit{d}): $F=12\times10^{-5}$.}
  \label{fig:MFD_LST}
\end{figure}

Local, parallel LST analysis of the DNS base flow, $(f,k)=(0,0)$, is used to further confirm the driving role of the base flow modification due to the streaks on the stabilisation of the second Mack mode. Two configurations are investigated, uncontrolled and controlled with $\lambda_z=2.4$. Figure \ref{fig:MFD_LST} depicts growth rate, $-\alpha_i$, and amplitude evolution, $A(x)=\exp{\left( \int_{x_0}^{x} -\alpha_i(x^{\prime}) \, dx^{\prime}\right)}$, normalised relative to the position of the first neutral point ($x_0$) for two disturbance frequencies ($F=[7.5, 12.0]\times10^{-5}$). In both scenarios, the modification of the base flow due to the streaks shifts $x_0$ further downstream, while the position of the second neutral point remains unchanged. In agreement with the DNS results, the amplitude of the two-dimensional, second mack mode is reduced by the control streaks (figures \ref{fig:LST_A_7p5}, \ref{fig:LST_A_12}). It is acknowledged that bi-global stability would be required \citep{Groskopf2016} for a comprehensive assessment of the effect of three-dimensional mean flow modification. However, as identified in previous studies \citep{Paredes2019}, for weak control streaks ($As_u<5\%$) the three dimensional mean flow deformation is likely to play a secondary role on the stabilisation of the second Mack mode. 

Overall, these investigations show that spanwise non-uniform surface temperature is unlikely to be able to generate large amplitude ($As_u>0.1\tilde{u}_{\infty}$), narrowly spaced streaks with a wavelength similar to the one for optimally growing streaks. These would be needed to delay laminar to turbulence transition under both first \citep{Sharma2019} and second \citep{Zhou2023} Mack mode dominated scenarios. Previous research \citep{Paredes2016} has shown that intrusive devices such as vortex generator and roughness elements can be used for this purpose. However, as the amplitude of the streaks increases, streak instability can also occur as previously identified for supersonic boundary layers \citep{Paredes2017}, and therefore this requires to iterate through the design process to identify an optimal configuration of the passive control devices \citep{Klauss2022}.

\begin{figure}
  \centering 
  \includegraphics[width=0.5\textwidth]{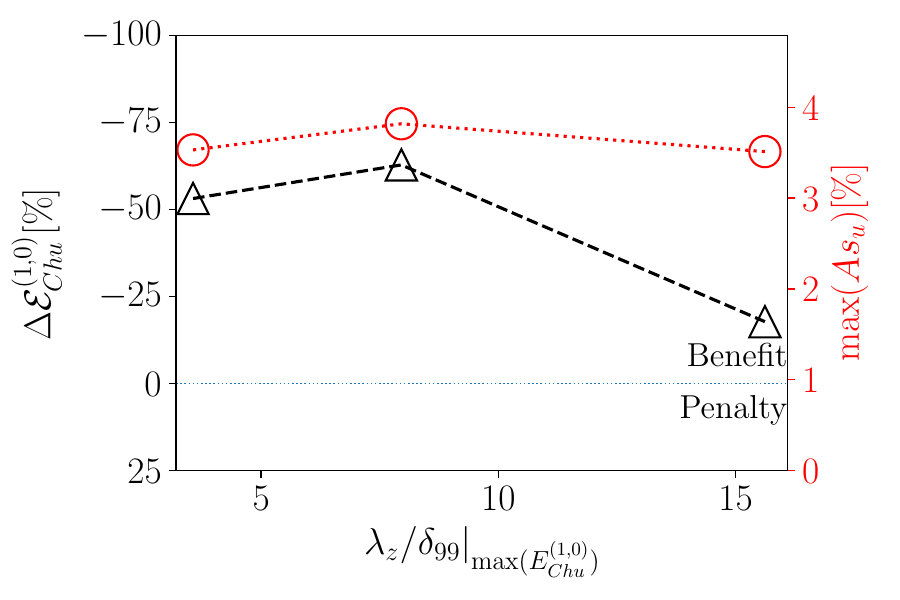}
  \caption{DNS results showing the influence of $\lambda_z$ on second Mack mode stabilisation (left y-axis) at (nearly) constant maximum streak amplitude (right y-axis).}
  \label{fig:summary_lambda_z_effect_const_Asu}
\end{figure}

To confirm the role of streak wavelength on the stabilisation of the second Mack mode, the case study presented in figure \ref{fig:summary_lambda_z_effect} is also assessed adjusting the amplitude of the surface temperature variation $A_{T_w}$ to keep the amplitude of the streaks constant. Table \ref{tab:lambda_z_effect_Asu_constant} summarizes operating and boundary conditions for this assessment. The analysis indicates that for a constant maximum amplitude of the streaks ($\max(As_u)$) and for the most linearly amplified forcing frequency, the maximum stabilisation is achieved for $\lambda_z$ approximately 8 times the local boundary layer thickness ($\delta_{99}|_{\max(E_{Chu}^{(1,0)})}$) at the maximum amplitude of the second Mack mode (figure \ref{fig:summary_lambda_z_effect_const_Asu}), therefore confirming the important role of streak wavelength on the stabilisation mechanism of this control method. However, the amplitude and wavelength of the control streaks remain intrinsically coupled, and it is not possible to fully establish the dominance of one parameter over the other.

\begin{table}
  \begin{center}
\def~{\hphantom{0}}
  \begin{tabular}{cccccccccc}
     $\tilde{T}_{w,\infty}$ & $Re_{unit}$ & $T_{w,base}$ & $Re_{x_{s}}$ & $Re_{x_{e}}$ & $Re_{x_{c,strip}}$ & $L_{strip}$ & $\lambda_{z}$ & $A_{T_w}$\\ [3pt]
     $216.17$K & $10.9\times10^6$1/m & $3.0$ & $1600$  & $3200$ & $1800$ & $4.45$ & $1.2$ & $0.38$\\  
     $216.17$K & $10.9\times10^6$1/m & $3.0$ & $1600$  & $3200$ & $1800$ & $4.45$ & $2.4$ & $0.3$\\ 
     $216.17$K & $10.9\times10^6$1/m & $3.0$ & $1600$  & $3200$ & $1800$ & $4.45$ & $4.8$ & $0.22$\\ 
  \end{tabular} 
  \caption{Summary of operating and boundary conditions for the the assessment of streak wavelength variation at (nearly) constant streak amplitude; $M_{\infty}=6$, $(Re_{\infty}M_{\infty})=1.0\times 10^5$, $F=\omega/(M_{\infty}^2 Re_{\infty})=7.5\times10^{-5}$. }
  \label{tab:lambda_z_effect_Asu_constant}
  \end{center}
\end{table}

\subsubsection{Sensitivity of control effectiveness to disturbance frequency}\label{sec:F_effect}

The sensitivity of control effectiveness to changes in forcing frequency and streak wavelength is assessed, and the configurations are summarized in table \ref{tab:F_effect}. Relative to the baseline configuration with a forcing frequency ($F=7.5\times10^{-5}$) close to the most linearly amplified one, two more configurations are assessed with $F=12\times10^{-5}$ and $16\times10^{-5}$ (figure \ref{fig:LST_contour}). The streak wavelength is varied relative to the base flow boundary layer thickness at the maximum energy of the second Mack mode ($\delta_{99}|_{\max(E_{Chu}^{(1,0)})}$), and the streak amplitude is also quantified at the same location ($As_u|_{\max(E_{Chu}^{(1,0)})}$). 

\begin{table}
  \begin{center}
\def~{\hphantom{0}}
  \begin{tabular}{cccccccccc}
     $\tilde{T}_{w,\infty}$ & $Re_{unit}$ & $T_{w,base}$ & $Re_{x_{s}}$ & $Re_{x_{e}}$ & $F=\omega/(M_{\infty}^2 Re_{\infty})$ & $Re_{x_{c,strip}}$ & $L_{strip}$ & $A_{T_w}$\\ [3pt]
     $216.17$K & $10.9\times10^6$1/m & $3.0$ & $1600$  & $3200$  & $7.5\times10^{-5}$  & $1800$ & $4.45$ & $0.3$\\ 
     $216.17$K & $10.9\times10^6$1/m & $3.0$ & $600$  & $2200$  & $12.0\times10^{-5}$  & $800$ & $2.09$ & $0.3$\\ 
     $216.17$K & $10.9\times10^6$1/m & $3.0$ & $100$  & $1800$  & $16.0\times10^{-5}$  & $350$ & $0.97$ & $[0.2,0.3]$\\ 
  \end{tabular} 
  \caption{Summary of operating and boundary conditions for the disturbance frequency assessment; $M_{\infty}=6$, $(Re_{\infty}M_{\infty})=1.0\times 10^5$. }
  \label{tab:F_effect}
  \end{center}
\end{table}

\begin{figure}
  \centering 
  \subfloat[]{\includegraphics[trim={0cm 0cm 0cm 0cm},clip,width=0.5\textwidth]{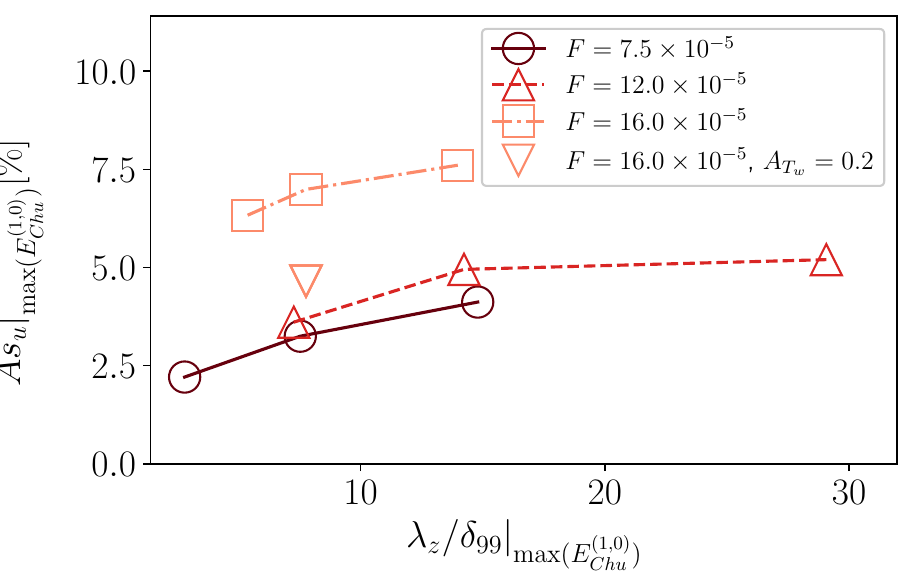}\label{fig:F_effect_streaks}} 
  \hfill 
  \subfloat[]{\includegraphics[trim={0cm 0cm 0cm 0cm},clip,width=0.5\textwidth]{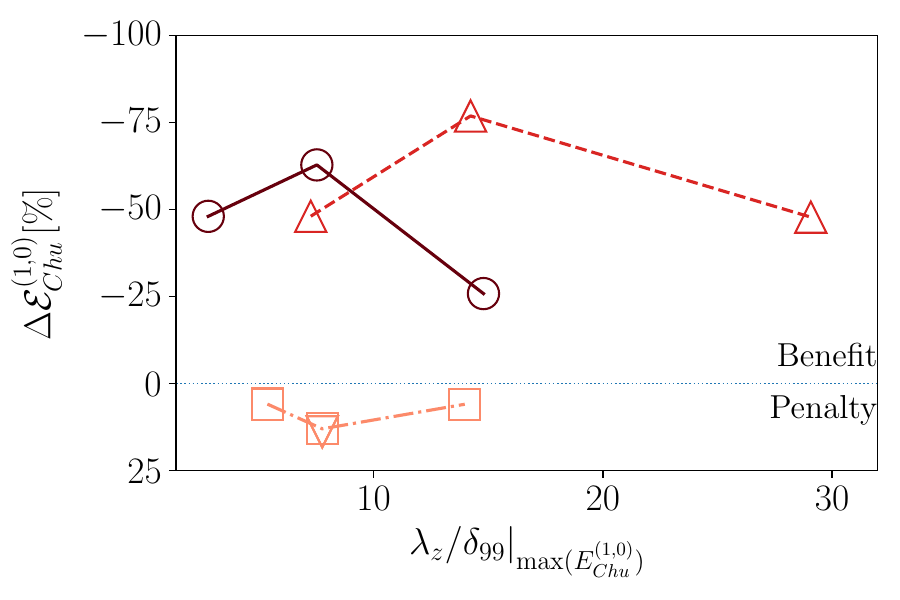}\label{fig:F_effect_stabilisation}} 
  \caption{DNS results showing the sensitivity of control effectiveness to changes in disturbance frequency. (\textit{a}) Streak amplitude and (\textit{b}) control effectiveness for various streak wavelength to local boundary layer thickness ratio. $M_{\infty}=6$, $T_{w,base}=3$, $T_{\infty}=216.7$K.}
\end{figure}

Firstly, the spanwise temperature variation is held constant to $A_{T_w}=0.3$ for all the configurations. As the disturbance forcing frequency is increased, $As_u|_{\max(E_{Chu}^{(1,0)})}$ increases for similar streak wavelength to boundary layer thickness ratio. This stems from a reduction in $\delta_{99}|_{\max(E_{Chu}^{(1,0)})}$, and therefore a greater heat flux per boundary layer height. 
For the case with $F=12\times10^{-5}$, the streak wavelength to local boundary layer thickness ratio is varied between approximately 8 to 30, and the streak amplitude  varies between $0.03\tilde{u}_{\infty}$ to $0.05\tilde{u}_{\infty}$ (figure \ref{fig:F_effect_streaks}), respectively. Compared to the case with $F=7.5\times10^{-5}$, for $F=12\times10^{-5}$ the peak stabilisation is achieved for slightly greater $\lambda_z/\delta_{99}|_{\max(E_{Chu}^{(1,0)})}\approx15$ (figure \ref{fig:F_effect_stabilisation}), and  the maximum stabilisation is also greater, as a result of the lower second Mack mode amplification and greater streak amplitude. For the range of streak amplitude and wavelength investigated (figure \ref{fig:F_effect_streaks}), when the disturbance frequency is further increased to $F=16\times10^{-5}$, the control streaks have nearly no effect on the stabilisation of the second Mack mode (figure \ref{fig:F_effect_stabilisation}).

Previous work \citep{Kuehl2016} assessed the effect of low amplitude ($As_u <5\%$) G\"ortler and second Mack mode instability interactions using 2D and 3D PSE for a Mach 6 boundary layer over a cone, at low-stagnation temperature conditions ($\tilde{T}_{\infty}=300$K). Based on a (local) effectiveness metric for the interaction of the two modes, the 2D and 3D PSE identified opposite trend, with frequency dependent and independent effectiveness, respectively. \citet{Kuehl2016} also suggested ineffectiveness of vortex-like modes to control second Mack mode dominated transition. However, the effectiveness metric only included velocity perturbations, therefore neglecting the effect on density and temperature. In addition to the differences in the methodology and configuration, in the current study the thermoacoustic effect of the streaks on the second Mack mode are taken into account through the Chu's energy based effectiveness (integral, Fourier-based) metric, and therefore the results are not directly comparable. However, the effectiveness of the streaks in second Mack mode stabilisation has been proved effective by the present and previous other studies \citep{Ren2016,Paredes2019,Kneer2022}, therefore confirming the role of streaks as a valid transition control strategy for hypersonic wall-bounded flows.

\begin{figure}
  \centering 
  \subfloat[]{\includegraphics[trim={0cm 0cm 0cm 0cm},clip,width=\textwidth]{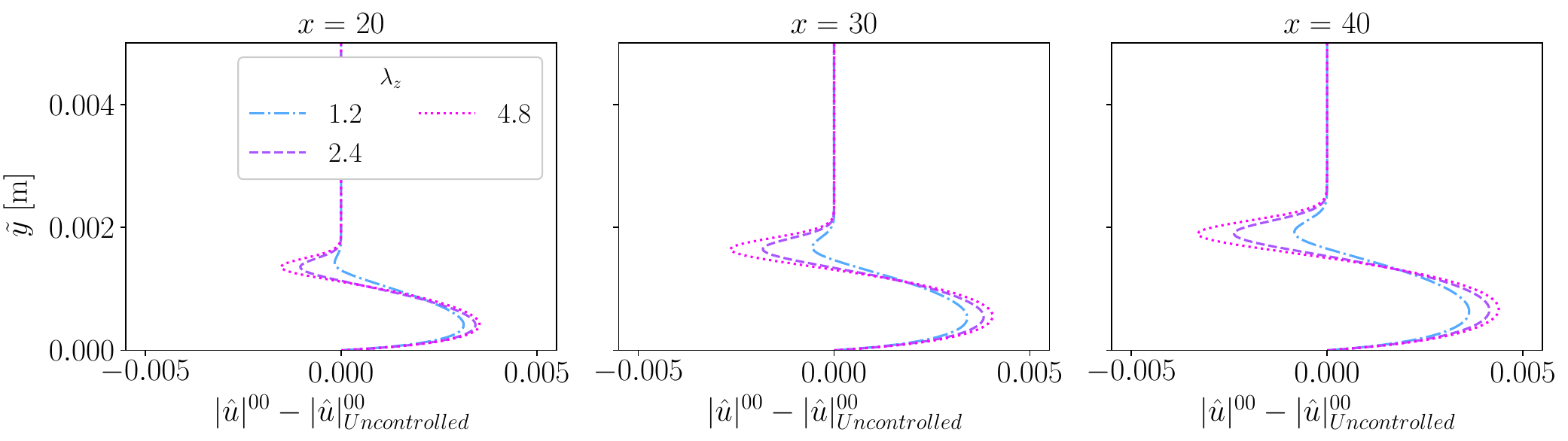}\label{fig:F1p2_U00}} 
  \hfill 
  \subfloat[]{\includegraphics[trim={0cm 0cm 0cm 0cm},clip,width=\textwidth]{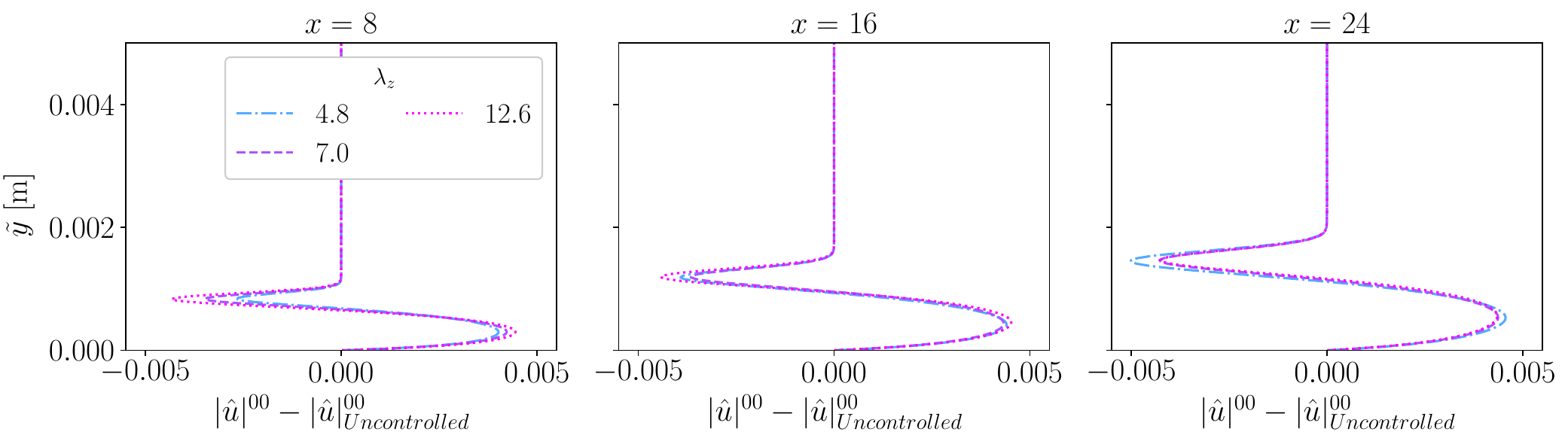}\label{fig:F1p6_U00}}
  \caption{DNS results showing the effect of streak wavelength ($\lambda_z$) on  base flow, $(f,k)=(0,0)$, deformation. Perturbation streamwise velocity profiles at various streamwise locations ahead, across and downstream of the second Mack mode with disturbance forcing frequency (\textit{a}) $F=12\times10^{-5}$ and (\textit{b}) $F=16\times10^{-5}$. $M_{\infty}=6$, $T_{w,base}=3$, $T_{\infty}=216.7$K.}
  \label{fig:F_effect_U00} 
\end{figure}

In the previous section (section \ref{sec:stab_mechanisms}), it is shown that the base flow deformation due to non-linear interaction of the control streaks is the dominant stabilisation mechanism for the linearly most amplified disturbance frequency. As the frequency is increased, $\delta_{99}|_{\max(E_{Chu}^{(1,0)})}$ reduces, and the effect of the streaks on base flow deformation becomes prominent away from the wall (figure \ref{fig:F_effect_U00}), where the streaks generate an inflection point in the streamwise velocity profile. This is further exacerbated for the case with the greatest frequency investigated ($F=16.0\times10^{-5}$, figure \ref{fig:F1p6_U00}), where the streak effect away from the wall dominates the mean flow deformation closer to the wall. Overall, this provides an explanation for the lack of control method effectiveness. For the configuration with $F=16.0\times10^{-5}$, the streak amplitude is considerably increased to 6-8\%. This is a consequence of the temperature variation being held constant, with a thinner the boundary layer in the region of second Mack mode amplification compared to the cases with $F=7.5\times10^{-5}$ and $12.0\times10^{-5}$. Thus, to ensure that this effect is not dominated by the increase in streak amplitude, for the case with $\lambda_z=7.0$ and $16.0\times10^{-5}$ the temperature variation is reduced from $A_{T_w}=0.3$  to $0.2$, such that the streak amplitude also reduces from approximately 7.5\% to 5\% (figure \ref{fig:F_effect_streaks}). Despite the reduction in amplitude, the control streaks are not able to significantly affect the second Mack mode energy (figure \ref{fig:F_effect_stabilisation}), therefore confirming an important role of the streak wavelength.

In the next sections, the effectiveness of the control method on second Mack mode stabilisation is parametrically investigated through a change in specific total enthalpy and Mach number, as well as base flow wall temperature, to determine the robustness of the method to a change in operating conditions. The disturbance forcing frequency is the nearly most linearly amplified one, and therefore the results provide a quantitative assessment of the maximum second Mack mode stabilisation that is achievable through control streaks that are passively generated through a spanwise temperature variation.

\subsection{Parametric studies}\label{sec:results_parametric}
The influence of freestream specific total enthalpy ($\tilde{h}_{0,\infty}$) and Mach number ($M_{\infty}$) on the effectiveness of the control method is independently assessed (table \ref{tab:summary_table_DNS}). All the computations start with a uniform surface temperature ($T_{w,base}$) and $\lambda_z=1.2$. For the controlled configurations, the spanwise non-uniform wall temperature boundary conditions is only enforced downstream of the actuator region ($x_{T_w,s}-x_{bs,e} \approx 0$), and $A_{T_w}=0.3$. A list of additional control parameters and the motivation for the choice are provided in each of the following subsections.

\begin{table}
\centering
\caption{Overview of operating conditions and computational domain size for DNS parametric studies}
\label{tab:summary_table_DNS}
\begin{tabular}{cccccccc}
\multicolumn{1}{c}{\textbf{$M_{\infty} \textnormal{ [-]}$}} &
  \multicolumn{1}{c}{\textbf{$Re_{\infty} \textnormal{ [-]}$}} &
  \multicolumn{1}{c}{\textbf{$p_{\infty} \textnormal{ [Pa]}$}} &
  \multicolumn{1}{c}{\textbf{$T_{\infty} \textnormal{ [K]}$}} &
  \multicolumn{1}{c}{\textbf{$L_{ref} \textnormal{ [m]}$}} &
  \multicolumn{1}{c}{\textbf{$(x_s,x_e) \textnormal{ [-]}$}} &
  \multicolumn{1}{c}{\textbf{$(L_y,L_z) \textnormal{ [-]}$}} &
  \multicolumn{1}{c}{\textbf{$(N_x,N_y,N_z) \textnormal{ [-]}$}} \\
$4.8$ & $20833$ & $1197$ & $124.8$ & $24\times10^{-3}$ & $(7.23,44.1)$ & $(1.575,1.2)$ & $(1200,211,13]$ \\
$5.4$ & $18518$ & $599.9$ & $102.5$ & $32\times10^{-3}$ & $(7.23,57.6)$ & $(1.575,1.2)$ & $(1200,211,13)$ \\
$6.0$ & $16666$ & $369$ & $37.9$ & $9\times10^{-3}$ & $(10.0,67.6)$ & $(1.575,1.2)$ & $(1200,211,13)$ \\
$6.0$ & $16666$ & $1420$ & $85.4$ & $9\times10^{-3}$ & $(19.6,90.0)$ & $(1.575,1.2)$ & $(1200,211,13)$ \\
$6.0$ & $16666$ & $5475$ & $216.7$ & $9\times10^{-3}$ & $(25.6,102.4)$ & $(1.575,1.2)$ & $(1200,211,13)$ \\
\end{tabular}
\end{table}

\subsubsection{Effect of specific total enthalpy}\label{sec:h0_effect}
The effect of freestream specific total enthalpy on the effectiveness of the control method is assessed for a fixed Mach number $M_{\infty}=6$ and Reynolds number $Re_{u_{\infty}}=(Re_{\infty}M_{\infty})=10^5$.  Due to the ideal gas law assumption, this assessment is not intended to investigate thermochemical non-equilibrium resulting from changes in stagnation enthalpy. Instead, it aims to cover a range of ground- and flight-representative conditions that would lead to an equivalent variation in wall temperature ratio. The range of values of $\tilde{h}_{0,\infty}$ used is listed in table \ref{tab:h0_effect}, and it is motivated by the need to assess flight representative (flight altitude of approximately 20000m) condition ($\tilde{h}_{0,\infty}=1.8 \times 10^6$J/kg) corresponding to a freestream static temperature $\tilde{T}_{\infty}=216.7$K, as well as ground-testing conditions. For the latter, the freestream static temperature is a result of the operating total temperature of the tunnel and the tested Mach number. As a result of a change in the freestream static temperature, the total pressure is adjusted to hold the unit Reynolds number ($Re_{unit}$) constant to $Re_{unit}=10.9\times 10^6$1/m, which is common to both flight test \citep{Schneider1999} as well as ground testing \citep{Ceruzzi2024} conditions. In addition, while the amplitude and angular frequency of the forcing disturbance is kept constant, the bounds of the computational domain and the position and extent of the blowing and suction region ($Re_{x,strip}$) are offset in the streamwise direction to accommodate the changes in onset, growth and decay of the second Mack mode due to the changes in freestream total temperature. This was informed by an assessment of the shift of the neutral curves for the second Mack mode through LST analyses, and it enabled DNS studies with the same streamwise resolution for second Mack mode fundamental wavelength with no increase in computational cost. To also reflect these changes, the amplitude of the streaks ($As_u$) is quantified at the streamwise position of the peak energy for the second Mack mode, and it is indicated by the addition of the subscript $|_{max(E_{Chu}^{\left(1,0\right)})}$. It is acknowledged that under flight and ground test representative conditions a wide range of time and length scales of the forcing disturbance may be encountered. This assessment is not within the scope of this work and therefore it is not captured in these studies. The decision to hold the forcing frequency constant across the range of stagnation enthalpies investigated is a modelling choice than an attempt to capture realistic conditions.

\begin{table}
  \begin{center}
\def~{\hphantom{0}}
  \begin{tabular}{ccccccccccc}
     $\tilde{h}_{0,\infty}$ & $Re_{unit}$ & $T_{w,base}$ & $Re_{x_{s}}$ & $Re_{x_{e}}$ & $\omega/(M_{\infty}^2 Re_{\infty})$ & $Re_{x_{c,strip}}$ & $L_{strip}$ & $\lambda_z$ & $A_{T_w}$\\ [3pt]
     $0.3\times10^6$J/kg & $10.9\times10^6$1/m & $3.0$ & $1200$  & $2800$  & $7.5\times10^{-5}$  & $1400$ & $3.47$ & $1.2$ & $0.3$\\ 
     $0.7\times10^6$J/kg & $10.9\times10^6$1/m & $3.0$ & $1400$  & $3000$  & $7.5\times10^{-5}$  & $1600$ & $3.96$ & $1.2$ & $0.3$\\ 
     $1.8\times10^6$J/kg & $10.9\times10^6$1/m & $3.0$ & $1600$  & $3200$  & $7.5\times10^{-5}$  & $1800$ & $4.45$ & $1.2$ & $0.3$\\ 
  \end{tabular} 
  \caption{Summary of operating and boundary conditions for the specific total enthalpy assessment; $M_{\infty}=6$, $(Re_{\infty}M_{\infty})=1.0\times 10^5$. }
  \label{tab:h0_effect}
  \end{center}
\end{table}

\begin{figure}
  \centering 
\includegraphics[width=0.5\textwidth]{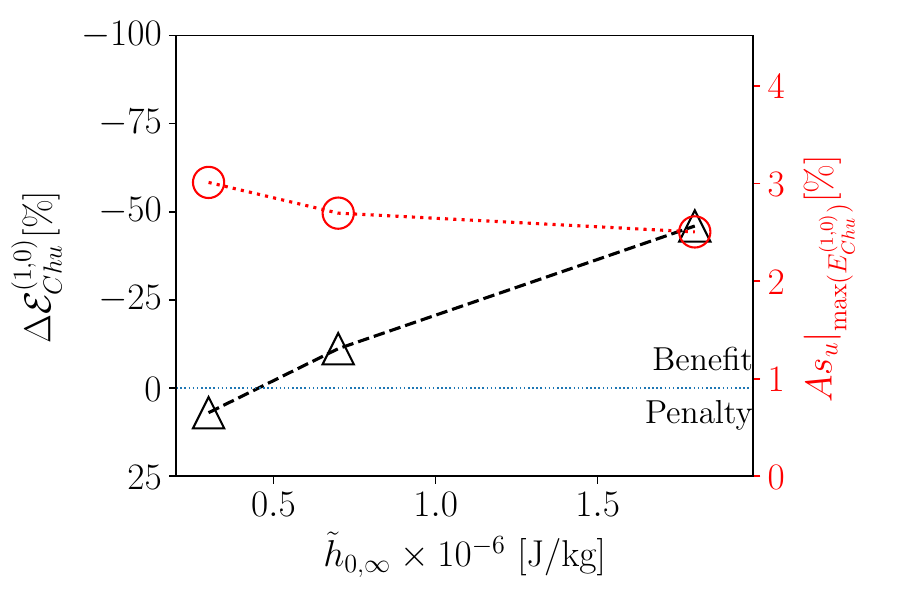}
  \caption{DNS results showing the influence of $\tilde{h}_{0,\infty}$ on second Mack mode stabilisation (left y-axis) and streak amplitude at the streamwise location of maximum amplification of the second Mack mode (right y-axis).}
\label{fig:h0_effect}   
\end{figure}

\begin{figure}
  \centering 
\includegraphics[width=0.75\textwidth]{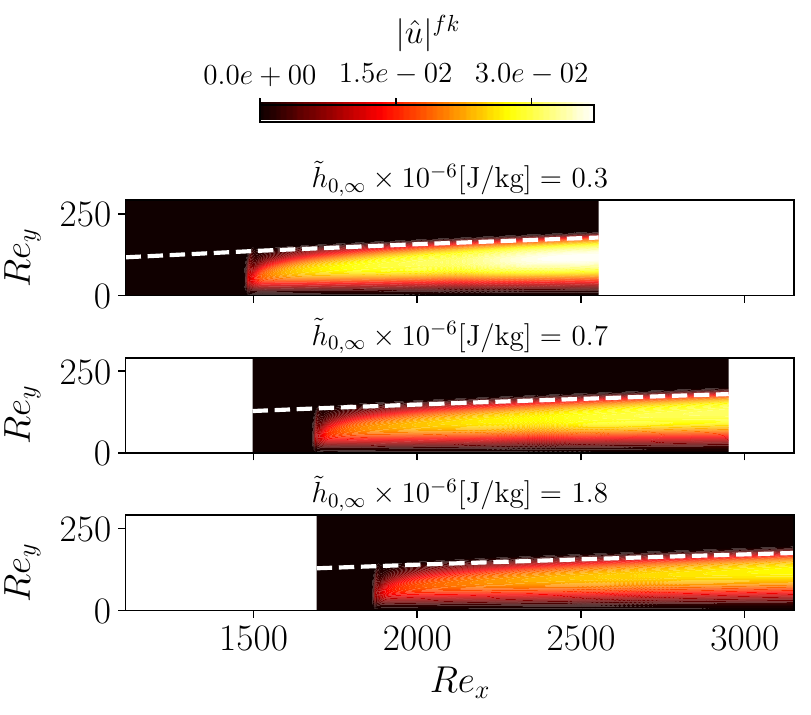} 
  \caption{DNS results showing the effect of freestream total enthalpy on the spatial (x-y) distribution of the amplitude of the Fourier mode corresponding to the fundamental harmonic of the streaks, $(f,k)=(0,1)$. The white dashed line indicates the outer edge of the boundary layer ($u\approx0.999$) for the base flow.}
\label{fig:h0_effect_streaks}  
\end{figure}

The freestream specific total enthalpy is progressively increased from $\tilde{h}_{0,\infty}=0.3 \times 10^6$J/kg to $1.8 \times 10^6$J/kg and the stabilisation effect of the streaks is quantified using the quantity $\Delta \mathcal{E}_{Chu}^{(1,0)}$ introduced in the previous section. It is found that while at $\tilde{h}_{0,\infty}=0.3 \times 10^6$J/kg the streaks slightly destabilise the second Mack mode (figure \ref{fig:h0_effect}), the polarity of the control method effectiveness reverses as $\tilde{h}_{0,\infty}$ increases and the beneficial effect of the streaks on second Mack mode stabilisation is recovered already at $\tilde{h}_{0,\infty}=0.7 \times 10^6$J/kg with $\Delta \mathcal{E}_{Chu}^{(1,0)} \approx 10\%$, which further increases to approximately $50\%$ at flight conditions ($\tilde{h}_{0,\infty}=1.8 \times 10^6$J/kg). The analysis overall indicates that the control mechanism is likely to be more effective at flight conditions, despite a more comprehensive assessment to changes in operating conditions would be required to generalize these results. It is also shown that the streak amplitude at the maximum amplification of the second Mack mode slightly changes (figure \ref{fig:h0_effect}), and it increases with a reduction in total enthalpy. This is further evident from the streamwise and wall-normal distribution of the amplitude of the $(f,k)=(0,1)$ Fourier coefficients for the streamwise velocity depicted in figure \ref{fig:h0_effect_streaks}.  The increase in streak amplitude with a reduction in stagnation enthalpy is driven by the modelling choice to hold the non-dimensional wall temperature distribution constant across the range of conditions investigated, and the physical reduction of boundary layer thickness at lower stagnation enthalpies as a result of colder wall temperature. Overall, this leads to an increase in surface heat flux per boundary layer thickness for the lower stagnation enthalpy conditions. Further investigations on appropriate scaling parameters to tune the amplitude of the control-streaks are discussed in section \ref{sec:Mach_effect}, where the effect of Mach number on the control-streaks effectiveness is assessed. 

\begin{figure}
  \centering 
\includegraphics[width=0.5\textwidth]{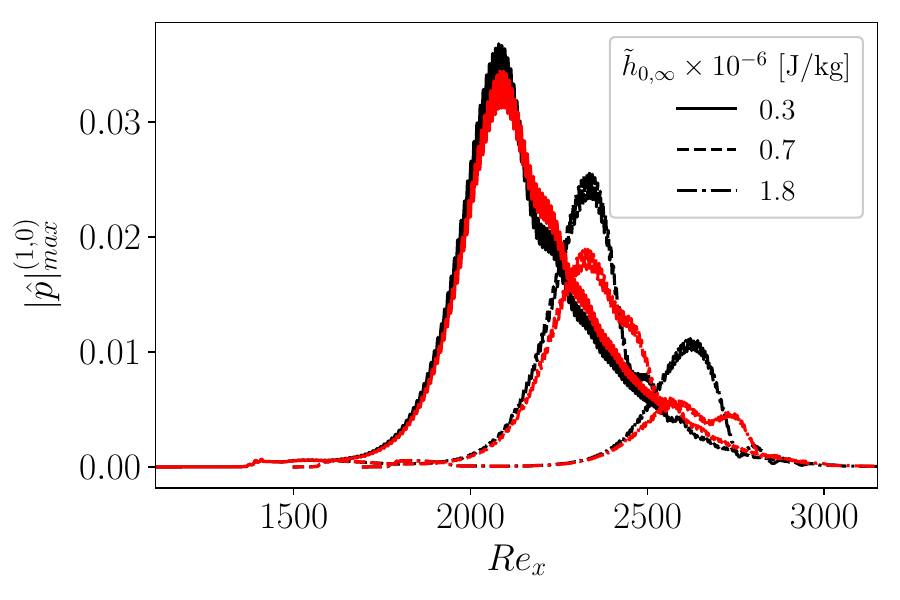}
  \caption{DNS results showing the influence of $\tilde{h}_{0,\infty}$ on the streamwise distribution of the wall-normal, maximum amplitude of the second Mack mode static pressure fluctuations for the the uncontrolled (black lines) and controlled (red lines) configurations.}
\label{fig:h0_effect_pmax_second_mode}   
\end{figure}

Further inspection of the streamwise distribution of the wall-normal maximum amplitude of the $(f,k)=(1,0)$ static pressure fluctuations (figure \ref{fig:h0_effect_pmax_second_mode}) shows that the second Mack mode planar wave is significantly destabilised by a reduction in freestream total enthalpy for both the controlled and uncontrolled configurations. As $\tilde{h}_{0,\infty}$ is reduced, the static temperature ($\tilde{T}_{\infty}$) also reduces and the wall gets significantly colder in absolute terms, although its non-dimensional ratio relative to the freestream static temperature is held constant. Overall, this results in a cooling effect which destabilises the second Mack mode, which is in agreement with a previous study in the literature \citep{Bitter2015} looking at the effect of freestream total enthalpy on second Mack mode growth rate via linear stability analyses. The effect of wall temperature on the effectiveness of control-streaks on the transition to turbulence via oblique breakdown was previously investigated for supersonic ($M_{\infty}=2$) conditions by \citep{Celep2022}. The effect of wall cooling on the second Mack mode stabilisation effect of control-streaks with nearly constant non-dimensional amplitude and spanwise wavelength is a novel contribution of this work. 

\subsubsection{Effect of Mach number}\label{sec:Mach_effect}
The effect of Mach number on the second Mack mode stabilisation performance of the control method is assessed for a fixed freestream specific total enthalpy $\tilde{h}_{0,\infty}\approx0.7\times10^6$ J/kg and Reynolds number $Re_{u_{\infty}}=(Re_{\infty}M_{\infty})=10^5$. Relative to the initial case study at $M_{\infty}=6$, two more configurations at $M_{\infty}=4.8$ and $5.4$ are investigated. The choice of the selected Mach number range is motivated by the need to assess operating conditions relevant for second Mack mode instability, while at the same time avoiding increasing Mach beyond $6$. Under flight scenario would lead to total enthalpies for which high-temperature gas effects would likely become relevant. As the freestream total temperature is held constant across the range of Mach numbers, the freestream static temperature changes accordingly. For the case with $M_{\infty}=4.8$ and $5.4$, two separate configurations are assessed where either the ratio of the base flow wall temperature to the freestream static temperature or to the laminar, adiabatic wall temperature ($\tilde{T}_{aw}$) is held constant, relative to the operating condition for the case study at $M_{\infty}=6$. The bounds for the computational domain and forcing frequencies are determined based on linear stability analyses.

\begin{table}
  \begin{center}
\def~{\hphantom{0}}
  \begin{tabular}{cccccccccc}
     $M_{\infty}$ & $\tilde{T}_{w,base}/\tilde{T}_{\infty}$ & $\tilde{T}_{w,base}/\tilde{T}_{aw}$ & $Re_{x_{s}}$ & $Re_{x_{e}}$ & $\omega/(M_{\infty}^2 Re_{\infty})$ & $Re_{x_{c,strip}}$ & $L_{strip}$ & $\lambda_z$ & $A_{T_w}$\\ [3pt]
     $4.8$ & [$2.1$, $3.0$] & [$0.42$, $0.61$] & $850$  & $2100$  & $16.0\times10^{-5}$  & $1000$ & $1.91$ & $1.2$ & $0.3$\\ 
     $5.4$ & [$2.5$, $3.0$] & [$0.42$, $0.51$] & $850$  & $2400$  & $12.0\times10^{-5}$  & $1000$ & $2.38$ & $1.2$ & $0.3$ \\ 
     $6.0$ & $3.0$ & $0.42$ & $1400$  & $3000$  & $7.5\times10^{-5}$  & $1600$ & $3.96$ & $1.2$ & $0.3$ \\ 
  \end{tabular} 
  \caption{Summary of operating and boundary conditions for the Mach number assessment; $\tilde{h}_{0,\infty}=0.7\times10^6$ J/kg, $(Re_{\infty}M_{\infty})=1.0\times 10^5$. }
  \label{tab:M_effect}
  \end{center}
\end{table}

\begin{figure}
  \centering 
  \subfloat[]{\includegraphics[width=0.5\textwidth]{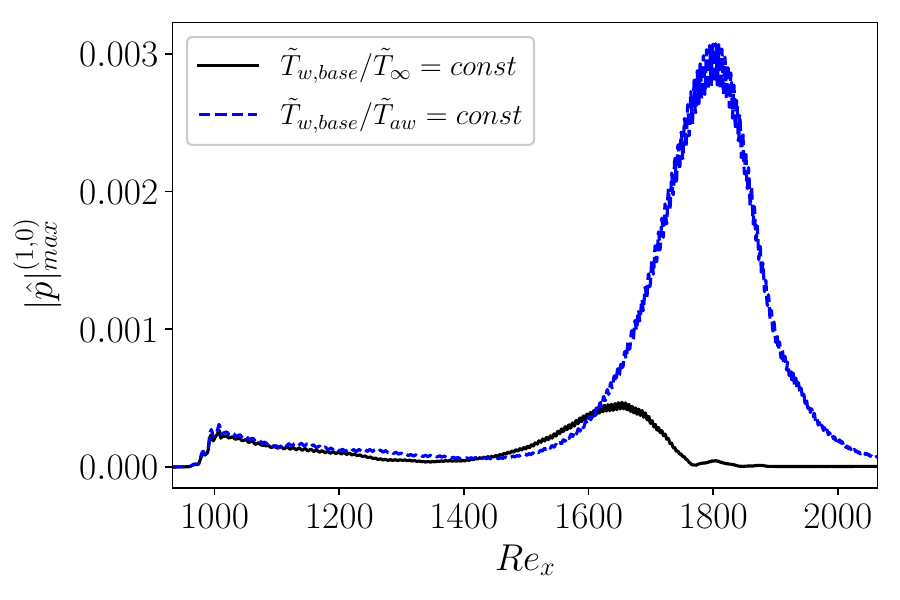}\label{fig:M4p8_effect_second_mode}} 
  \hfill 
  \subfloat[]{\includegraphics[width=0.5\textwidth]{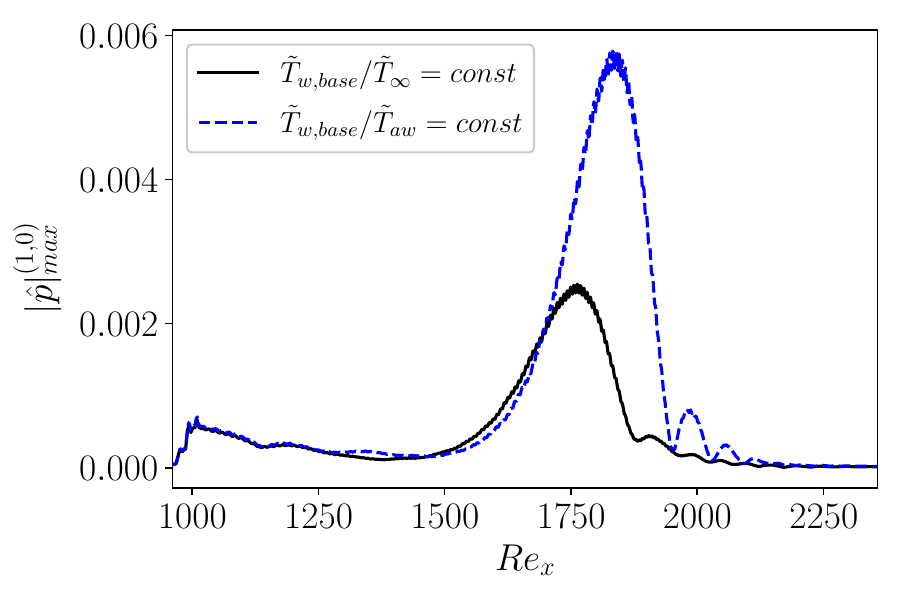}\label{fig:M5p4_effect_second_mode}}
  \caption{DNS results showing the effect of base flow wall temperature on the streamwise distribution of the wall-normal, maximum amplitude of the second Mack mode static pressure fluctuations for the uncontrolled configurations. (\textit{a}) $M_{\infty}=4.8$; (\textit{b}) $M_{\infty}=5.4$.}
  \label{fig:M_effect_second_mode_uncontrolled} 
\end{figure}

\begin{figure}
  \centering 
  \subfloat[]{\includegraphics[width=0.5\textwidth]{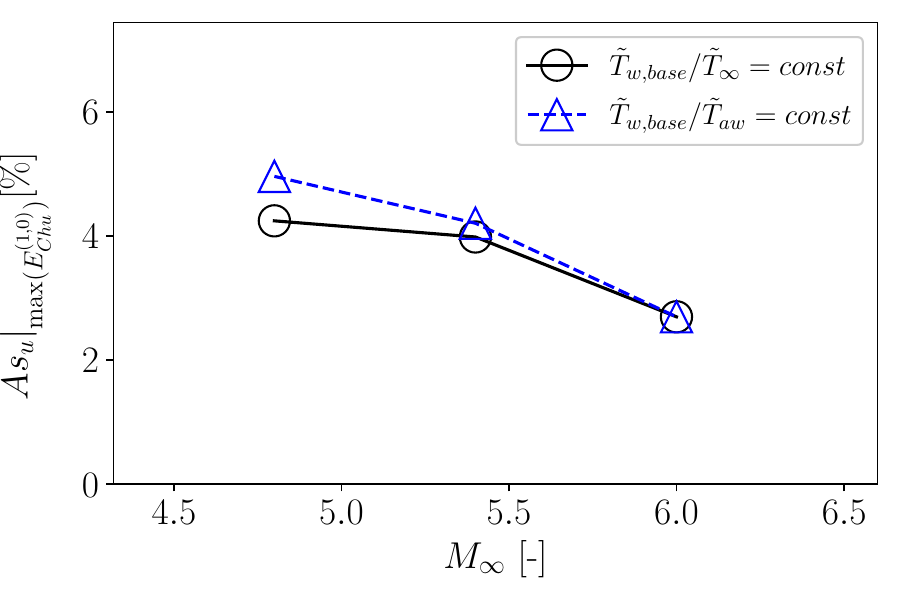}\label{fig:M_effect_streaks_amplitude}} 
  \hfill 
  \subfloat[]{\includegraphics[width=0.5\textwidth]{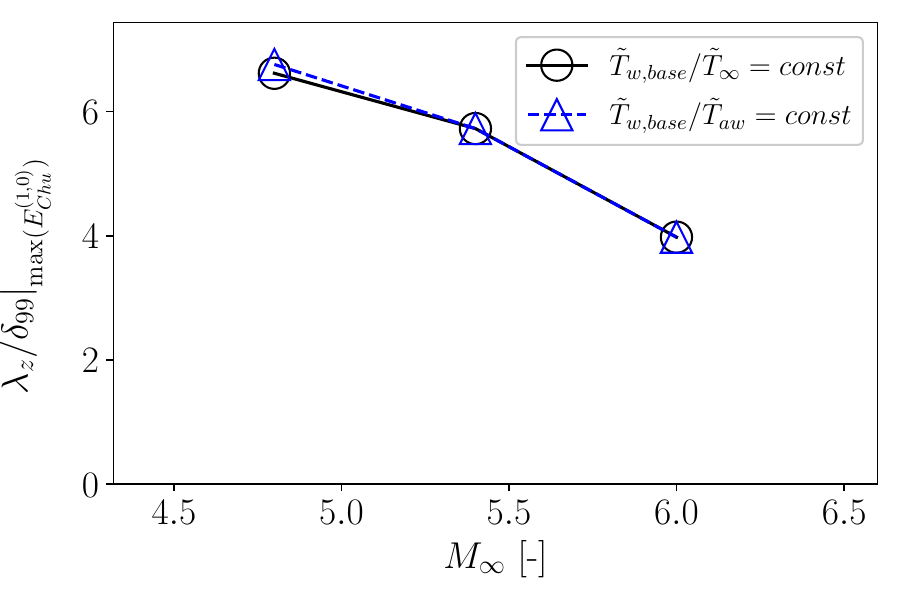}\label{fig:M_effect_streaks_wavelength}}
  \caption{DNS results showing the influence of Mach number on (\textit{a}) streak amplitude and (\textit{b}) ratio of streak wavelength to boundary layer thickness at the streamwise location of maximum amplification of the second Mack mode. There is only one data point at $M_{\infty}=6$.}
\end{figure}

For the uncontrolled configurations, the effect of base flow wall temperature on the streamwise distribution of the amplitude of the second Mack mode, $(f,k)=(1,0)$, static pressure fluctuations is firstly investigated (figure \ref{fig:M_effect_second_mode_uncontrolled}). Relative to the case with $\tilde{T}_w/\tilde{T}_{\infty}=const=3$, for the $M_{\infty}=4.8$ (figure \ref{fig:M4p8_effect_second_mode}) and $M_{\infty}=5.4$ (figure \ref{fig:M5p4_effect_second_mode}), the second Mack mode amplification significantly increases as the wall temperature is reduced to hold the ratio to the adiabatic wall temperature constant. This is consistent with the existing literature about the destabilizing effect of wall cooling on second Mack growth rate \citep{Mack1975}. For the controlled configurations, the effect of wall cooling at a fixed Mach number has only a modest effect on streak amplitude (figure \ref{fig:M_effect_streaks_amplitude}) and spanwise wavelength to boundary layer thickness ratio (figure \ref{fig:M_effect_streaks_wavelength}). On the other hand the effect of Mach number on both quantities is noticeable. This is a result of freestream total enthalpy and Reynolds number being held constant for this analysis. As the Mach number reduces, the freestream static temperature increases and so does the streak amplitude as a result of a greater, dimensional spanwise temperature variation. Similarly, as the Mach number increases the boundary layer thickness increases almost quadratically with Mach number \citep{Anderson1989} and the streak wavelength ratio to the boundary layer thickness at the streamwise location of maximum amplification of second Mack mode also reduces.

\begin{figure}
  \centering 
\includegraphics[width=0.5\textwidth]{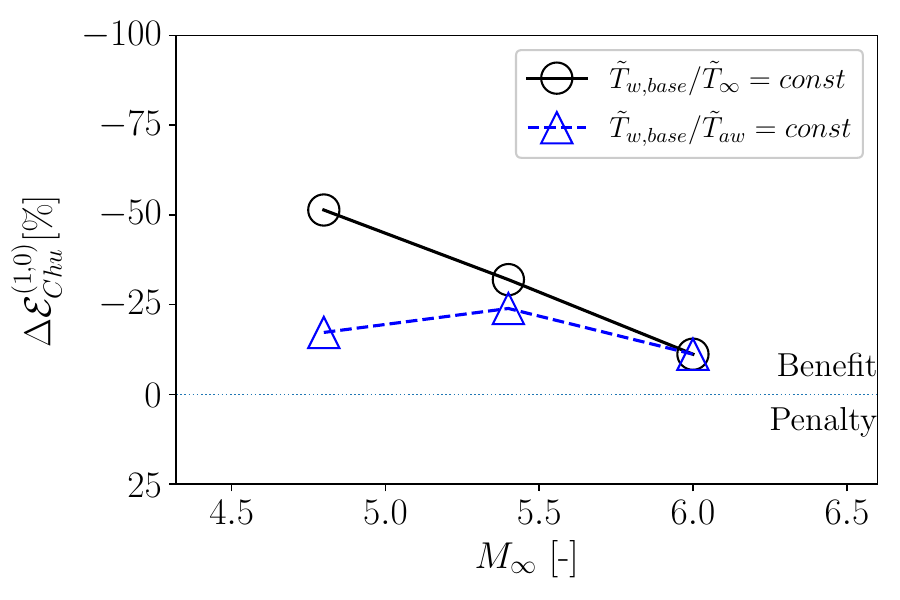}
  \caption{DNS results showing the effect of Mach number on second Mack mode stabilisation. There is only one data point at $M_{\infty}=6$.}
\label{fig:M_effect_stabilisation}   
\end{figure}

The control-streaks have a stabilizing effect on second Mack mode for all the configurations investigated, and the control effectiveness increases with a reduction in Mach number (figure \ref{fig:M_effect_stabilisation}). This is expected as the streak amplitude increases from approximately $2.7\%$ at $M_{\infty}=6$ to $4.8-5.4\%$ at $M_{\infty}=4.8$ (figure \ref{fig:M_effect_streaks_amplitude}), and also the streak wavelength increases from approximately $4\delta_{99}$ to $6.7\delta_{99}$ (figure \ref{fig:M_effect_streaks_wavelength}). This is consistent with the results presented in section \ref{sec:results_controlled}. However, the effect of base flow wall temperature at a fixed Mach number on second Mack mode stabilisation requires further investigations.
For the configurations with $M_{\infty}=5.4$ and $M_{\infty}=4.8$, the stabilisation effect of the streaks reduces with a reduction in $T_{w,base}$ from $T_{w,base}=3$ to $2.5$ and $2.1$, respectively. This is consistent with a stronger amplification of the second Mack mode for the baseline, uncontrolled configuration (case $\tilde{T}_{w,base}/\tilde{T}_{aw}=const$ in figure \ref{fig:M_effect_second_mode_uncontrolled}), and it is somewhat expected given that streak amplitude and wavelength are similar for the two configurations. 

\begin{figure}
  \centering 
  \subfloat[]{\includegraphics[width=0.5\textwidth]{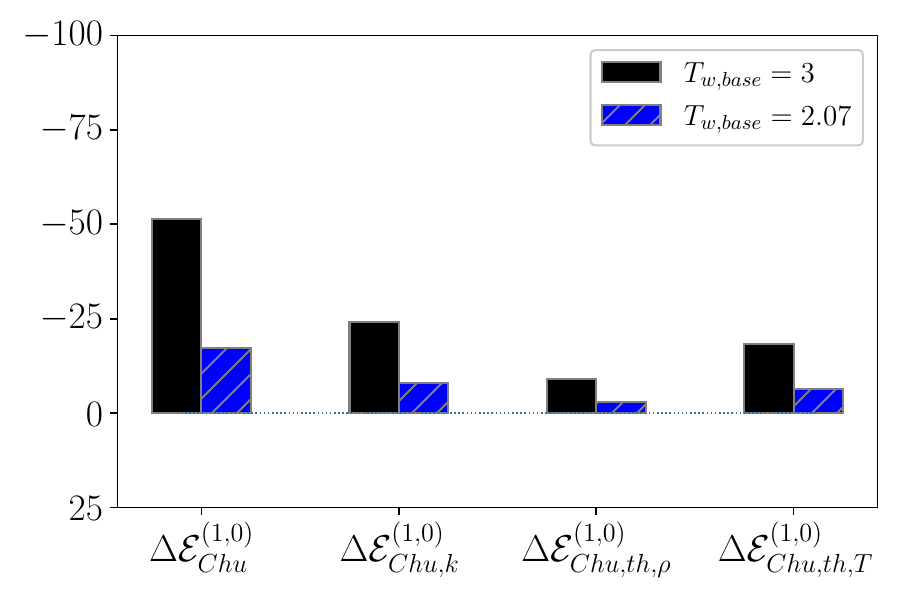}\label{fig:Tw_effect_M4p8_stabilisation}} 
  \hfill 
  \subfloat[]{\includegraphics[width=0.5\textwidth]{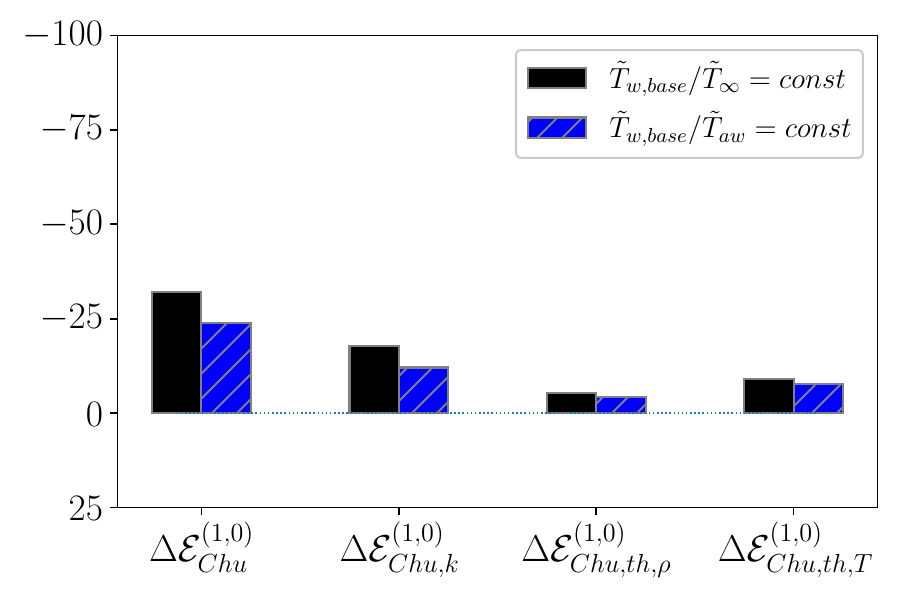}\label{fig:Tw_effect_M5p4_stabilisation}}
  \caption{DNS results showing the influence of base flow wall temperature on the effect of the streaks on second Mack mode stabilisation, and breakdown into the constitutive kinetic and thermodynamic energy components. (\textit{a}) $M_{\infty}=4.8$ and (\textit{b}) $M_{\infty}=5.4$. Negative is benefit, and positive is penalty.}
\end{figure}

The reduction in base flow wall temperature evaluated for the case with $M_{\infty}=4.8$ is greater compared to the case with $M_{\infty}=5.4$, and the effect of base flow wall temperature on the control method effectiveness is also greater. This is further inspected through the analysis of the constitutive component's of the energy for both the second Mack mode and the streaks. The kinetic ($\Delta \mathcal{E}_{Chu,k}^{(1,0)}$) and thermodynamic ($\Delta \mathcal{E}_{Chu,th,\rho}^{(1,0)}$ and $\Delta \mathcal{E}_{Chu,th,T}^{(1,0)}$) energy contributions to the stabilisation effect of the second Mack mode energy are computed as follows,
\begin{equation}
\Delta \mathcal{E}_{Chu,k}^{(1,0)} = \frac{\left( \int_{x_s}^{x_e} E_{Chu,k,c}^{(1,0)}dx - \int_{x_s}^{x_e} E_{Chu,k,nc}^{(1,0)}dx \right)}{\int_{x_s}^{x_e} E_{Chu,nc}^{(1,0)}dx}100
\end{equation}
\begin{equation}
\Delta \mathcal{E}_{Chu,th,\rho}^{(1,0)} = \frac{\left( \int_{x_s}^{x_e} E_{Chu,th,\rho,c}^{(1,0)}dx - \int_{x_s}^{x_e} E_{Chu,th,\rho,nc}^{(1,0)}dx \right)}{\int_{x_s}^{x_e} E_{Chu,nc}^{(1,0)}dx}100
\end{equation}
\begin{equation}
\Delta \mathcal{E}_{Chu,th,T}^{(1,0)} = \frac{\left( \int_{x_s}^{x_e} E_{Chu,th,T,c}^{(1,0)}dx - \int_{x_s}^{x_e} E_{Chu,th,T,nc}^{(1,0)}dx \right)}{\int_{x_s}^{x_e} E_{Chu,nc}^{(1,0)}dx}100
\end{equation}
where $E_{Chu,k}$, $E_{Chu,th,\rho}$ and $E_{Chu,th,T}$ refer to the first, second and third term in equation \ref{eq:Echu}, respectively. For the $M_{\infty}=4.8$ case, an increase in base flow wall temperature from $T_{w,base}=2.1$ to $3$ leads to a significant stabilisation of both the kinetic and thermodynamic energy components (figure \ref{fig:Tw_effect_M4p8_stabilisation}). For the $M_{\infty}=5.4$ case, the increase in the stabilisation through the streaks of the thermodynamic energy of the second Mack mode due to an increase in $T_{w,base}$ from $2.5$ to $3$ is marginal (figure \ref{fig:Tw_effect_M5p4_stabilisation}). 

\begin{figure}
  \centering 
  \subfloat[]{\includegraphics[width=0.5\textwidth]{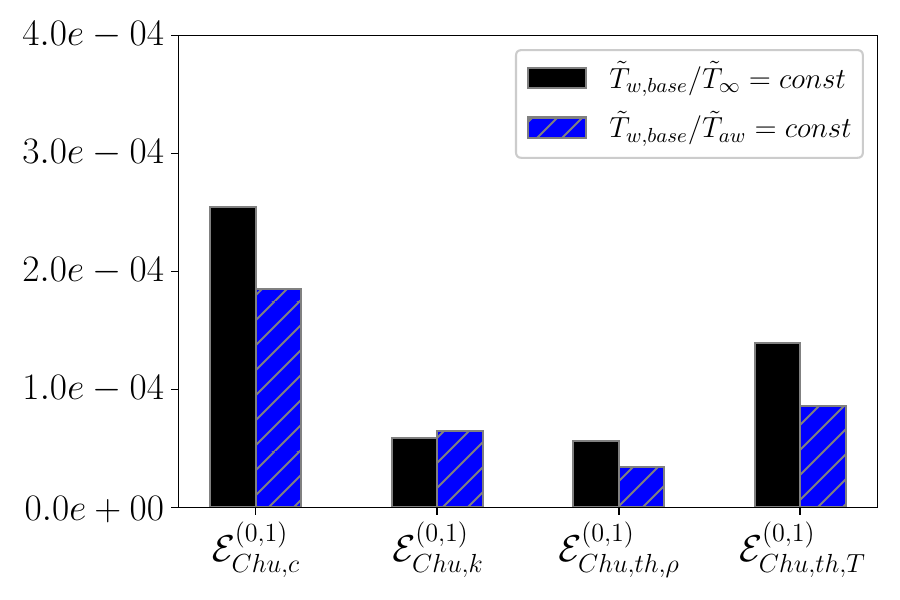}\label{fig:Tw_effect_M4p8_streaks}} 
  \hfill 
  \subfloat[]{\includegraphics[width=0.5\textwidth]{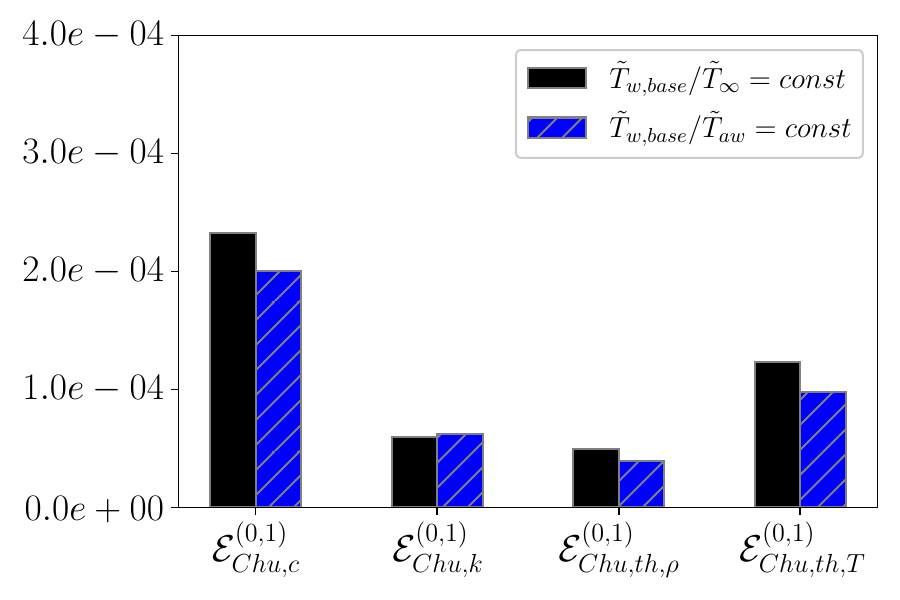}\label{fig:Tw_effect_M5p4_streaks}}
  \caption{DNS results showing the influence of base flow wall temperature on the modal energy of the streaks, and breakdown into the constitutive kinetic and thermodynamic energy components. Controlled configurations, (\textit{a}) $M_{\infty}=4.8$ and (\textit{b}) $M_{\infty}=5.4$.}
   \label{fig:Tw_effect_streaks}
\end{figure}

A similar evaluation of the integral modal energy of the control-streaks ($\mathcal{E}_{Chu,c}^{(0,1)} = \frac{1}{L_x} \int_{x_s}^{x_e} E_{Chu,c}^{(0,1)}dx$) shows that this reduces with the decrease in base flow wall temperature (figure \ref{fig:Tw_effect_streaks}). For both the $M_{\infty}=4.8$ and $M_{\infty}=5.4$ configurations, this reduction is driven by the reduction in thermodynamic energy due to the lower spanwise temperature variation for the colder cases. This is dictated by the control parameter for the amplitude of the spanwise temperature variation ($A_{T_w}$) being held constant, $A_{T_w}=0.3$. In the control law for the temperature boundary condition (equation \ref{eq:Tw_eq}), $A_{T_w}$ acts as a perturbation to the base flow wall temperature. Relative to the $M_{\infty}=5.4$ case, for the $M_{\infty}=4.8$ configuration the reduction of the streaks modal energy due to the decrease in $T_{w,base}$ is greater due to the larger base flow wall temperature variation that was investigated to keep $\tilde{T}_{w,base}/\tilde{T}_{aw}$ constant. Overall, the differences in the effect of the base flow wall temperature between the $M_{\infty}=4.8$ and $M_{\infty}=5.4$ configurations are driven by both a difference in the wall temperature range investigated, as well as by the modal, thermal energy of the streaks ($\mathcal{E}_{Chu,th}^{(0,1)}$). This indicates that for streaks generated through a manipulation of surface temperature the classical streak amplitude metric based on the streamwise velocity perturbation relative to the base flow may be insufficient for a complete characterization of the stabilisation effectiveness.

\subsection{Control method effectiveness under heated conditions}

In a low-enthalpy, wind tunnel test facility, the passive generation of the streaks exploiting the aerothermodynamics of the flow is not viable  due to the low driving potential for heat transfer, $\propto (T_{aw}-T_w)$. However, streaks can be generated through active heating \citep{Ozawa2025}. In this section, the effectiveness of the control method for a more practical wind tunnel implementation is computationally investigated. This provides further guidance for future experimental tests. The active heating system for the generation of the streaks generates a (uniform) perturbation of the base flow temperature, and the wall temperature boundary condition in the DNS is therefore modified as follows,

\begin{equation}
\begin{cases}
    T_w = T_{w,base} \left(A_{T_{w}} + A_{T_{w}} \sin \left( \frac{2 \pi}{\lambda_z} z \right) \right), \, \textnormal{if} \, \, Re_x\ge Re_{x_{T_w,s}}\\
    T_w = T_{w,base}, \, \textnormal{if} \, \, Re_x < Re_{x_{T_w,s}}
\end{cases}
\label{eq:Tw_eq_wt}
\end{equation}
where $Re_{x_{T_w,s}}$ is the start of the (active) control method. For the uncontrolled configurations, the base flow temperature ($T_{w,base}$) is also uniformly increased relative to the initial nearly adiabatic conditions for $Re_x \ge Re_{x_{T_w,s}}$, such that $T_{w,base}$ remains the same for the controlled and uncontrolled cases. Four different configurations are investigated (table \ref{tab:wind_tunnel_case}), and the operating conditions are based on the studies presented in section \ref{sec:h0_effect}, and typical operating range for high-speed blow-down wind tunnels \citep{Rees2020}. The disturbance forcing frequency ($F$) is close to the linearly optimal frequency for $A_{T_{w}}=0$, and the wavelength of the streaks is selected based on the studies presented in section \ref{sec:F_effect}. 

\begin{table}
  \begin{center}
\def~{\hphantom{0}}
  \begin{tabular}{cccccccccc}
      $\textnormal{Case no.}$ & ${M}_{\infty}$ & $\tilde{T}_{\infty}$ & $Re_{unit}$ & $T_{w,base}$ & $F$ & $Re_{x_{T_w,s}}$ & $\lambda_{z}/\delta_{99}|_{\max(E_{Chu}^{(1,0)})}$ & $A_{T_w}$\\ [3pt]
     1 & $6$ & $85.4$K & $2.44\times10^6$1/m & $7.0$ & $7.5\times10^{-5}$ & $1500$ & $\sim 12$ & $0.15$\\  
     2 & $5$ & $83.3$K & $10.7\times10^6$1/m & $6.0$ & $10\times10^{-5}$ & $1000$ & $\sim 10$ & $0.1$\\ 
     3 & $5$ & $50$K & $24.6\times10^6$1/m & $6.0$ & $10\times10^{-5}$ & $1000$ & $\sim 10$ & $0.1$\\ 
     4 & $5$ & $50$K & $24.6\times10^6$1/m & $6.0$ & $10\times10^{-5}$ & $1000$ & $\sim 8$ & $0.3$\\      
  \end{tabular} 
  \caption{Operating and boundary conditions for the heated configurations.}
  \label{tab:wind_tunnel_case}
  \end{center}
\end{table}

\begin{figure}
  \centering 
  \subfloat[]{\includegraphics[width=0.5\textwidth]{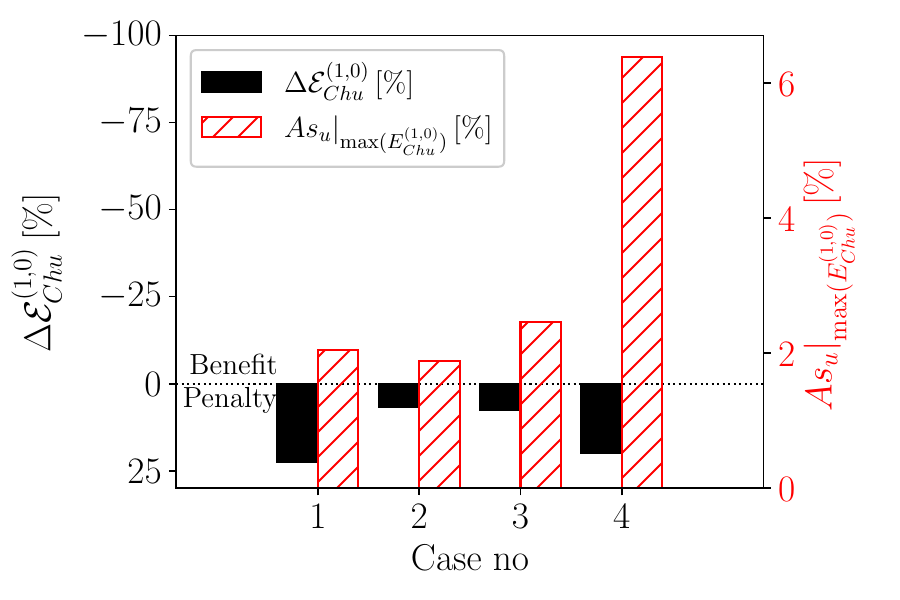}\label{fig:summary_heated}} 
  \hfill 
  \subfloat[]{\includegraphics[width=0.5\textwidth]{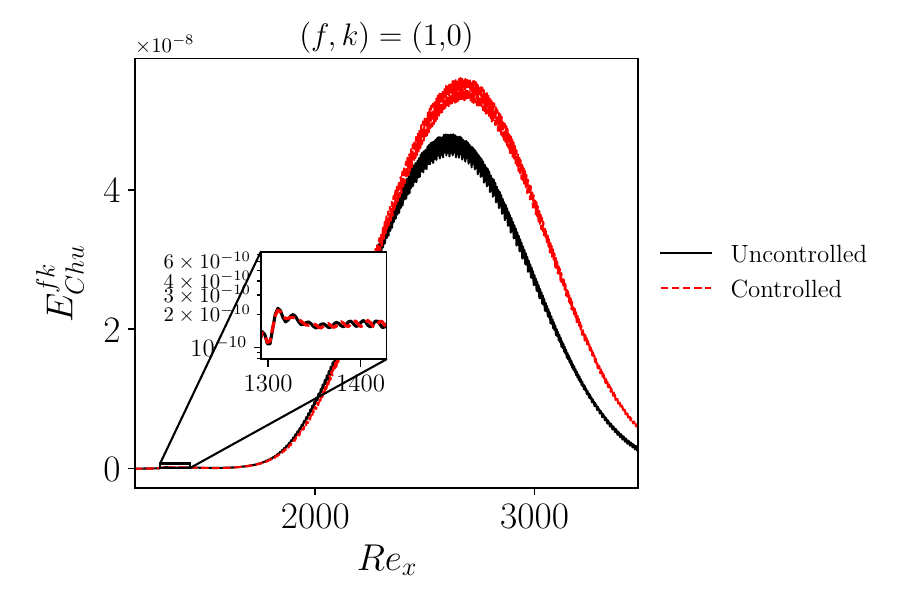}\label{fig:EChu_Case_1}}
  \caption{DNS results showing (\textit{a}) control method effectiveness (left y-axis) and amplitude of the streaks (right y-axis) for the heated configurations; (\textit{b}) streamwise distribution of second Mack mode energy for case 1. The inset in (\textit{b}) depicts the energy of the forcing disturbance.}
   \label{fig:Echu_heated}
\end{figure}

\begin{figure}
  \centering 
\includegraphics[width=\textwidth]{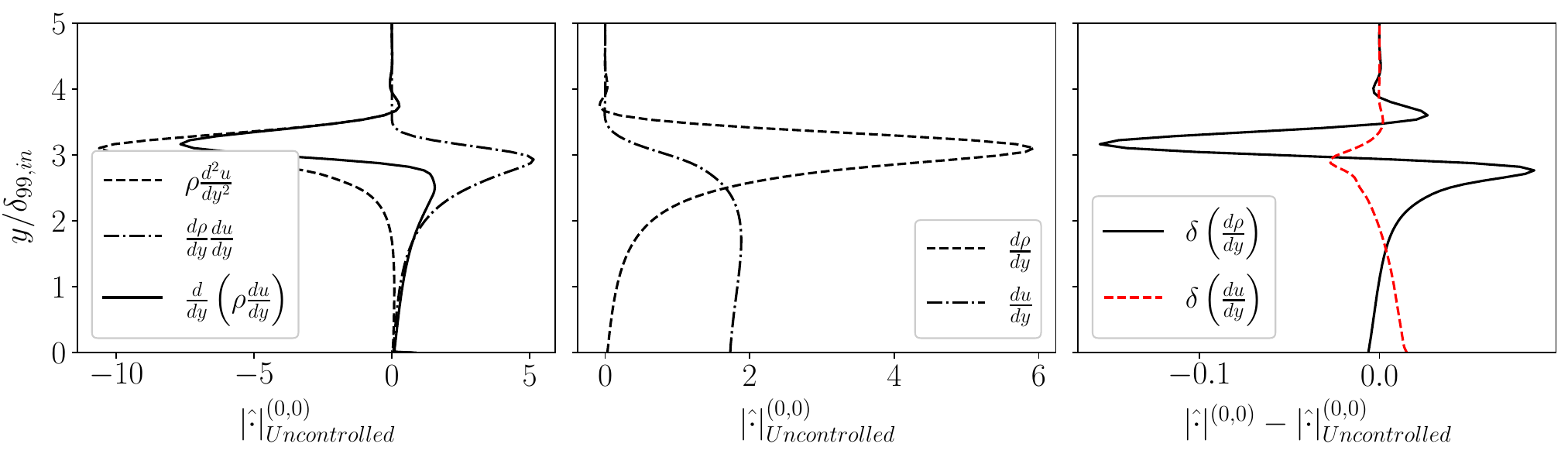}
  \caption{DNS base flow, wall normal profiles for case 1 at $Re_x\approx3100$. (\textit{a}) Generalized inflection point condition and product-rule decomposition, and (\textit{b}) wall-normal gradients of density and streamwise velocity for the uncontrolled configuration. (\textit{c}) Perturbation profiles ($\delta(\cdot)$) for the controlled configuration relative to the uncontrolled case.}
   \label{fig:GIP_heated}
\end{figure}

\begin{figure}
  \centering 
\includegraphics[width=\textwidth]{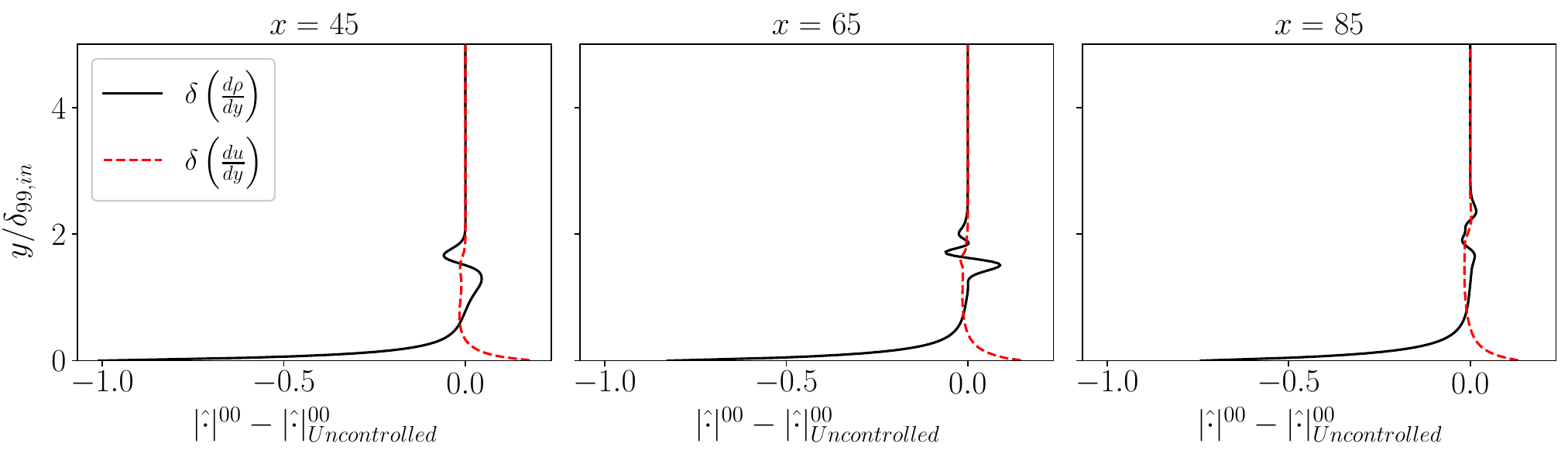}
  \caption{DNS cold base flow ($T_{w,base}=3$) configuration: $M_{\infty}=6$, $\tilde{h}_{0,\infty}=0.7\times10^6 \, \textnormal{J/kg}$ in table \ref{tab:h0_effect}. Perturbation profiles at various streamwise location in the region of second Mack mode amplification for the controlled configuration relative to the uncontrolled case.}
   \label{fig:GIP_cooled}
\end{figure}

For all the conditions investigated, the thermally generated streaks under heated ($T_w>T_{aw}$) conditions destabilise the second Mack mode (figure \ref{fig:summary_heated}). The onset of the second Mack mode amplification remains the same, but the energy peak of the instability is increased by the streaks (figure \ref{fig:EChu_Case_1}). Section \ref{sec:stab_mechanisms} showed that under cold ($T_w<T_{aw}$) conditions, the stabilisation mechanism is driven by the two-dimensional base flow modification due to the streaks. In figure \ref{fig:GIP_heated}, the analysis of the base flow proposed by \citet{Kuehl2018} is closely followed to determine the reason for the destabilisation due to the streaks for case 1, and similar conclusions apply to the other cases. Downstream of the maximum amplification of the second Mack mode ($Re_x\approx3100$, figure \ref{fig:EChu_Case_1}), for the uncontrolled configuration, the presence of the generalized inflection point ($\frac{d}{dy}\left(\rho \frac{du}{dy}\right)=0$, \cite{Lees1946}) closer to the wall ($y/\delta_{99,in}\approx 3$, left graph in figure \ref{fig:GIP_heated}) is driven by a strong positive density gradient (middle graph in figure \ref{fig:GIP_heated}), which is the cause for the instability. This is consistent with \citet{Kuehl2018}, and it further confirms the thermoacoustic nature of the second Mack mode. Relative to the uncontrolled configuration, the thermally generated streaks further increase the wall normal density gradient (right graph in figure \ref{fig:GIP_heated}). Under heated conditions, the penalising effect of the streaks on the density field dominates the beneficial increase in the velocity gradient, $\delta (\frac{du}{dy})$, at the wall. On the other hand, under cold ($T_w<T_{aw}$) conditions the positive increase in $\delta (\frac{du}{dy})$ at the wall due to the streaks dominates (figure \ref{fig:GIP_cooled}), and the second Mack mode is stabilised ($\tilde{h}_{0,\infty}=0.7\times10^6 \, \textnormal{J/kg}$ case in figure \ref{fig:h0_effect}). This indicates that to experimentally assess the beneficial effect of the novel control method on the stabilisation of the second Mack mode, active cooling methodologies \citep{Paquin2023} for the generation of the streaks should also be investigated. 

\section{Conclusions}\label{sec:conclusions}
The influence of spanwise non-uniform surface temperature distribution on second Mack mode stability has been investigated under deterministic forcing. The effectiveness of the novel control method has been determined and quantified based on an energy norm. The spanwise non-uniform surface temperature generates streaks whose amplitude can be controlled either through a change in the amplitude of the spanwise temperature distribution, axial position and extent of the hot and cold patches, as well as spanwise wavelength. For the latter, it is shown that a near-optimal solution is achieved for a wavelength of the streaks which is approximately $10$ times the boundary layer thickness at the location of maximum amplification of the second Mack mode. This provides initial guidance for future experimental investigations.

A set of parametric studies has been used to assess the robustness of the control method under different operating conditions. A range of freestream total enthalpies representative of both ground testing and flight conditions has been studied. The control effectiveness of the method increases at flight conditions. It is shown that this is driven by the wall temperature difference among the three scenarios investigated. Compared to flight conditions, during ground testing the wall is colder  and the stronger amplification of the second Mack mode may require stronger amplitude streaks. These are difficult to attain through this passive, non-intrusive flow control strategy due to the lower freestream total enthalpies and, consequently, heat transfer rate. More intrusive, passive (e.g., vortex generators or roughness elements) or active (e.g., local blowing and suction or heating and cooling strips) flow control strategies should be considered to achieve greater amplitude streaks. A range of Mach numbers where, based on linear stability theory, the second Mack mode dominates the initial, linear stage of laminar to turbulent transition has been assessed. The effectiveness of the method increases with a reduction in Mach number, due to the greater amplitude of the control-streaks that have been generated. It is shown that this effect is driven by an increase of the spanwise wavelength of the streaks relative to the local boundary layer thickness. This is further confirmation of the notable effect of the streak wavelength on the second Mack mode stabilisation, which is in this case achieved via a change in freestream Mach number. Finally, the effect of base flow wall temperature is independently investigated for two Mach number configurations. As a result of wall cooling, the stabilisation effect of the streaks is reduced due to the combined effect of a greater amplification of the second Mack mode and lower modal thermal energy injected through the streaks.

Finally, for a practical low-enthalpy, wind tunnel implementation of the control method with the streaks generated through an active heating system, it is found that the streaks destabilise the second Mack mode regardless of amplitude, wavelength and operating conditions. The destabilisation is driven by the streaks leading to an increase in the positive density gradient off the wall, which dominates the beneficial increase in the streamwise velocity gradient at the wall. This indicates the need to also investigate active cooling strategies for the generation of the streaks in blow-down high-speed wind tunnel facilities to further confirm the role of surface temperature on the control method effectiveness.

Overall, the results indicate that streaks can be generated through a spanwise non-uniform surface temperature, and second Mack mode energy can be significantly reduced. Compared to other control strategies, this method appears sub-optimal to attain large amplitude control streaks. However, the novel mechanism of streak generation is a potentially promising non-intrusive and passive flow control strategy, and therefore an evaluation on transition to turbulence remains of interest. Suitable scaling parameters to provide stabilisation of the second Mack mode under deterministic forcing and small-amplitude disturbances have been identified. This provides initial guidance for future studies to assess the effectiveness of the method on transition to turbulence. This may require further optimisation of the control parameters to ultimately achieve aero-thermal-structural efficiency benefits.    

\backsection[Acknowledgements]{The authors gratefully acknowledge EPSRC for the computational time made available on the UK supercomputing facility ARCHER2 via the UK Turbulence Consortium (EP/R029326/1).}

\backsection[Funding]{This research received financial support of Dstl through the WSRF program (task number 0105).}

\backsection[Declaration of interests]{The authors report no conflict of interest.}

\backsection[Author ORCIDs]{L. Boscagli, \url{https://orcid.org/0000-0002-8121-4208}; G. Rigas, \url{https://orcid.org/0000-0001-6692-6437}; O. Marxen, \url{https://orcid.org/0000-0002-5746-1962}; P. J. K. Bruce, \url{https://orcid.org/0000-0002-1361-8737}}

\appendix

\section{Governing equations}\label{appendix:gov_eq}
The non-dimensional equations for the conservation of mass, balance of momentum and energy conservation are as follows:

\begin{equation}
\frac{\partial \rho}{\partial t} + \frac{\partial}{\partial x_j} \left(\rho u_j \right) = 0
\end{equation}

\begin{equation}
\frac{\partial \rho u_i}{\partial t} + \frac{\partial}{\partial x_j} \left(\rho u_i u_j + p\delta_{ij} \right) = \frac{\partial \sigma_{ij}}{\partial x_j}
\end{equation}

\begin{equation}
\frac{\partial E}{\partial t} + \frac{\partial}{\partial x_j} \left[ \left(E + p \right)u_j \right] = -\frac{\partial q_j}{\partial x_j} + \frac{\partial}{\partial x_k} \left(u_j \sigma_{jk} \right)
\end{equation}
In the preceding equations, Einstein notation is used, and $\sigma_{ij}$, $E$ and $q_j$ are the viscous stress tensor, the total energy per unit mass, and the heat flux vector, respectively, and these are defined as follows:
\begin{equation}
\sigma_{ij} = \frac{\mu}{Re_{\infty}} \left( \frac{\partial u_i}{\partial x_j} +\frac{\partial u_j}{\partial x_i} -\frac{2}{3} \frac{\partial u_k}{\partial x_k}\delta_k \right)
\end{equation}

\begin{equation}
E = \rho e + \frac{1}{2}\rho u_i^2
\end{equation}

\begin{equation}
q_i = - \frac{1}{Re_{\infty} Pr_{\infty}} k_e \frac{\partial T}{\partial x_i}
\end{equation}

\section{Spanwise grid refinement studies}\label{appendix:grid_refinement}
\begin{figure}
  \centering 
  \subfloat[]{\includegraphics[width=0.48\textwidth]{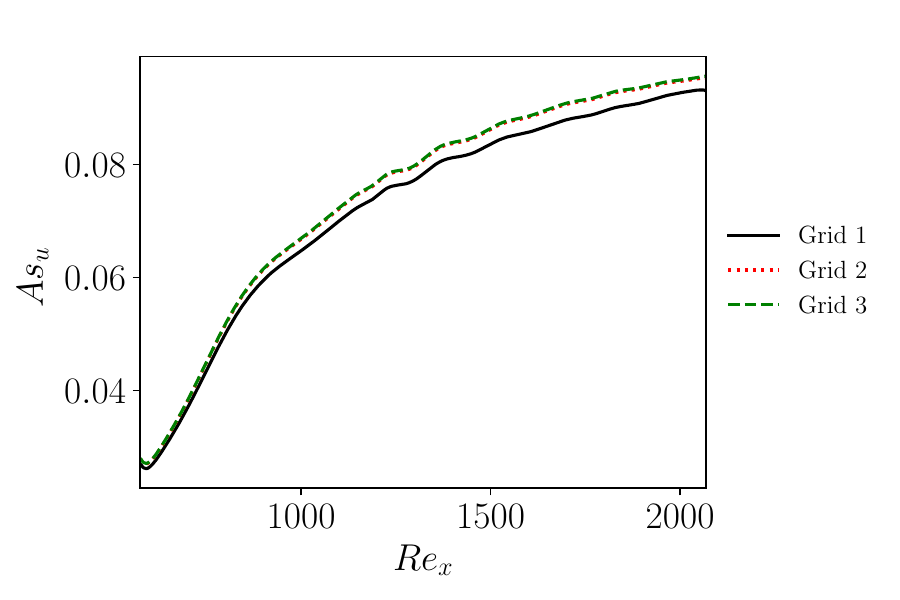}\label{fig:Asu_grid_ref}} 
  \hfill 
  \subfloat[]{\includegraphics[width=0.48\textwidth]{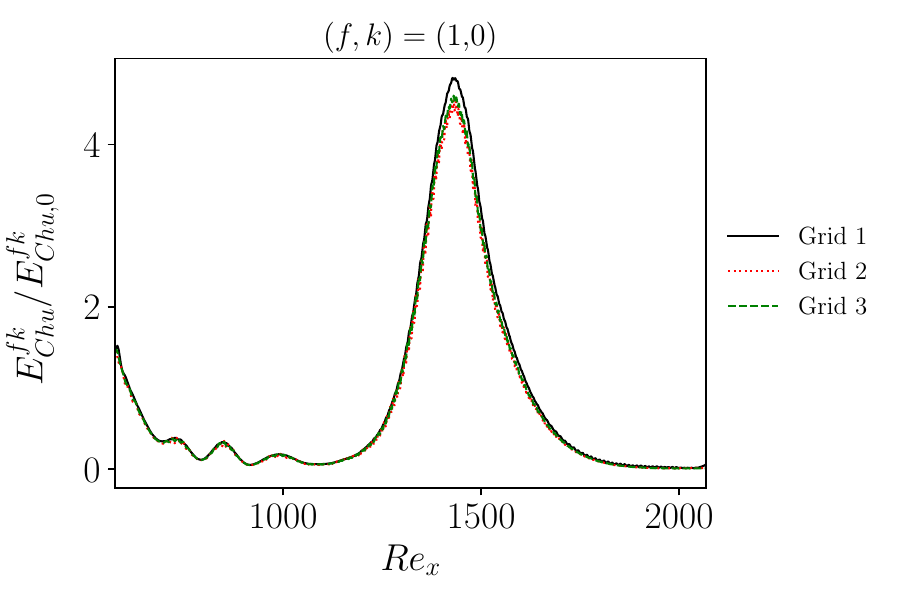}\label{fig:Echu_grid_ref}} 
  \caption{Effect of spanwise grid refinement on the streamwise distribution of (\textit{a}) streak amplitude and (\textit{b}) second Mack mode linear amplification; x-axis cropped downstream of the blowing and suction strip at $Re_x\approx500$ .}
\end{figure}

The influence of the number of nodes per fundamental spanwise wavelength ($n_z$) of the streaks on streak amplitude and second Mack mode amplification is investigated. Three grid refinement levels are used with $n_z=13$, $45$ and $75$, and these are named 1, 2 and 3, respectively (table \ref{tab:grid_refinement}). The finest grid level is based upon previous grid convergence studies for the assessment of the hypersonic boundary layer transition with surface roughness \citep{Lefieux2019}. Overall, for the three grid levels the maximum amplitude of the streaks (figure \ref{fig:Asu_grid_ref}) and of the linear amplification of the second Mack mode energy (figure \ref{fig:Echu_grid_ref}) is within approximately $0.8\%$ and $5\%$, respectively.
 
\begin{table}
  \begin{center}
\def~{\hphantom{0}}
  \begin{tabular}{ccccc}
      Grid level & No. nodes ($n_x \times n_y \times n_z$) & $\lambda_z$ & $T_{w,base}$ & $A_{T_w}$\\[3pt]
       1  & $1200\times211\times13$ & 1.2 & 3.6 & 0.3\\
       2  & $1200\times211\times45$ & 1.2 & 3.6 & 0.3\\
       3  & $1200\times211\times75$ & 1.2 & 3.6 & 0.3\\
  \end{tabular}
  \caption{Summary of the spanwise grid refinement studies; $M_{\infty}=4.8$, $(Re_{\infty}M_{\infty})=1.0\times 10^5$, $\tilde{h}_{0,\infty}=0.3\times 10^6$J/kg.}
  \label{tab:grid_refinement}
  \end{center}
\end{table}
 
\section{Assessment of overlap between spanwise non-uniform surface temperature and disturbance forcing region}\label{appendix:overlap_effect}

\begin{figure}
  \centering 
  \subfloat[]{\includegraphics[width=0.48\textwidth]{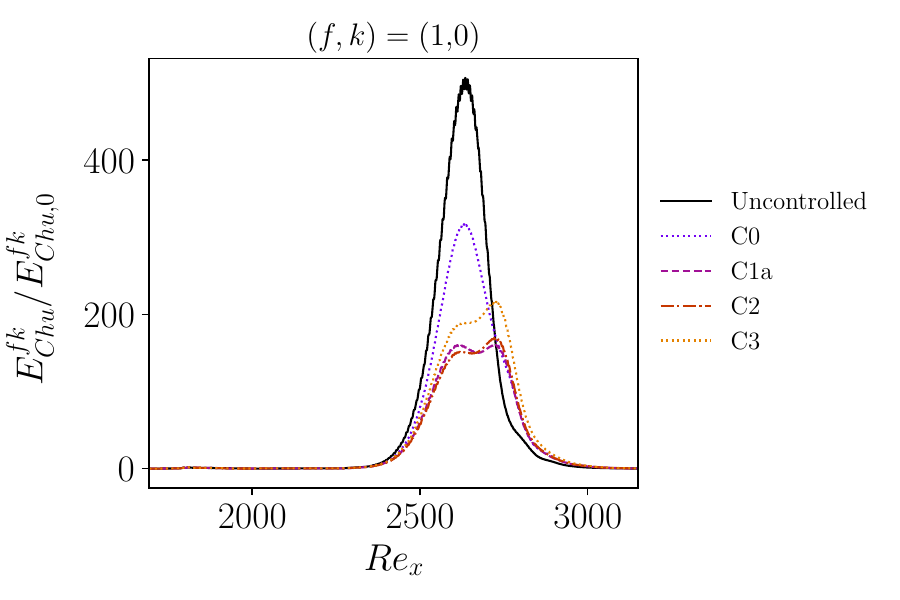}\label{fig:EChu_amplification_overlap_second_Mack_mode}}
  \hfill  
  \subfloat[]{\includegraphics[width=0.48\textwidth]{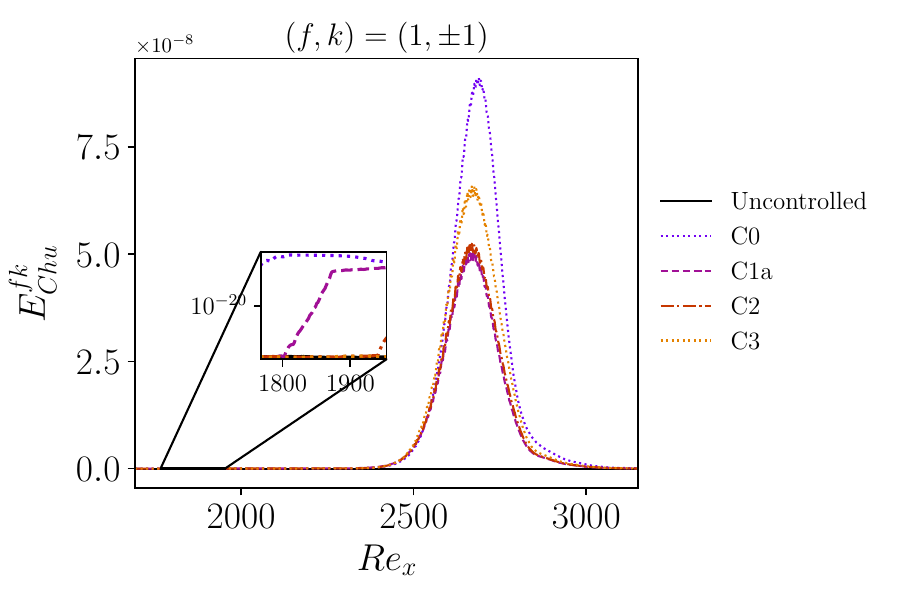}\label{fig:EChu_overlap_non_linear}} 
  \hfill  
  \subfloat[]{\includegraphics[width=0.48\textwidth]{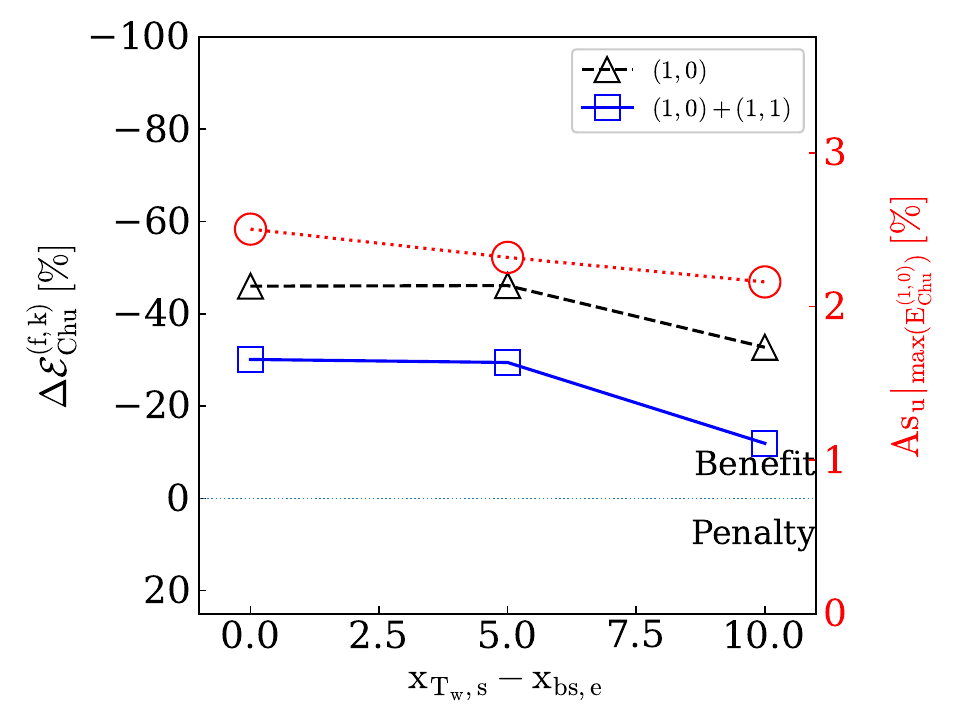}\label{fig:EChu_no_overlap_fk11}}   
  \caption{Effect of the overlap between the disturbance forcing region and the control on (\textit{a}) linear amplification of the second Mack mode, $(f,k)=(1,0)$, and (\textit{b})  energy due to non-linear interaction between streaks and second Mack mode, $(f,k)=(1,\pm 1)$. (\textit{c}) Effect of the contribution of the non-linear terms to the control method effectiveness.}
\end{figure}

For the case with overlap between the disturbance forcing region and the control method, the amplitude of the wall normal momentum perturbations introduced by the actuator is no longer of the same magnitude as for the uncontrolled case (inset in figure \ref{fig:EChu_xtw_effect}), due to the effect of the surface temperature on the density of the flow. This prevents a direct comparison of the modal energies for the controlled and uncontrolled case. To attempt to remove this spurious effect, the Chu's energy is normalised with the value downstream of the actuator region ($E_{Chu,0}^{fk}$) to determine the amplification factor. Relative to the uncontrolled (uniform surface temperature) case, the linear amplification of the second Mack mode is reduced for all the configurations investigated (figure \ref{fig:EChu_amplification_overlap_second_Mack_mode}). The controlled configurations with no overlap (C1a, C2 and C3) have similar streamwise distribution of the linear amplification of the second Mack mode, and this increases as the streak amplitude reduces. However, the amplification factor for the overlap configuration (case C0) follows a different streamwise distribution and trend. Despite a greater amplitude of the streaks, the control method is less effective for C0 (figure \ref{fig:EChu_amplification_overlap_second_Mack_mode}). This may be associated to non-linear interactions, $(f,k)=(1,\pm 1)$, between the actuator and the spanwise non-uniform surface temperature distributions, which generate three dimensional static pressure disturbances at the actuator region (inset in figure \ref{fig:EChu_overlap_non_linear}). Although the actuator law is two dimensional for this case study, the overlap region between the actuator and the control method generates spurious non-linear terms, due to the quadratic nature of the Navier Stokes equations. As such, a simple rescaling of the modal energy with the energy downstream of the actuator is not sufficient to remove the spurious effects, that result from the overlap between the actuator and the control method. For the configuration further investigated in this work (C1a, $x_{T_w,s}-x_{bs,e}=0$ in figure \ref{fig:EChu_no_overlap_fk11}), the contribution of the non-linear terms to the effectiveness of the control method remains below $\sim15\%$.

\section{Influence of streaks sub-harmonics on second Mack mode linear amplification}\label{appendix:sub_harmonic}
\begin{figure}
  \centering 
  \subfloat[]{\includegraphics[width=0.48\textwidth]{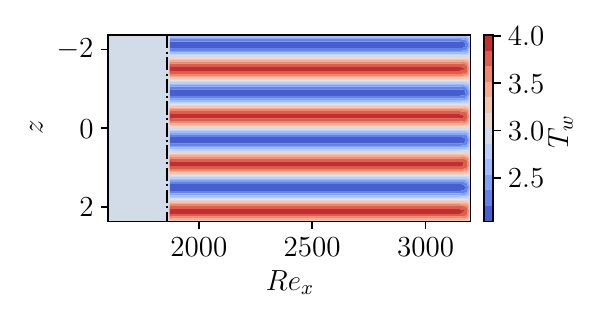}\label{fig:twall_sub_harmonic}} 
  \hfill 
  \subfloat[]{\includegraphics[width=0.48\textwidth]{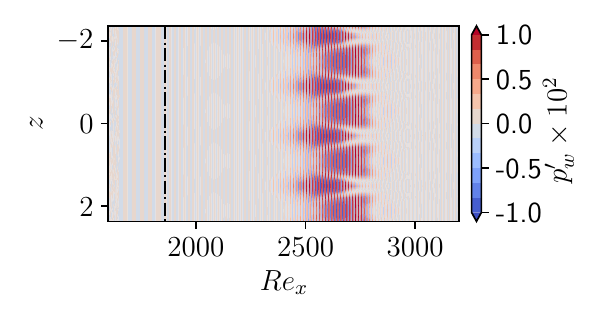}\label{fig:pwall_sub_harmonic}} 
  \caption{Distribution of wall (\textit{a}) temperature and (\textit{b}) instantaneous static pressure fluctuations for the case with $\lambda_{z,domain}=4\lambda_z$. The black-dashed line marks the end of the region of blowing and suction (actuator).}
\end{figure}

Previous research has identified modal energy transfer mechanism from the higher to lower wavenumbers that can have a significant influence on  both first and second Mack mode stability \citep{Paredes2017,Caillaud2025}. For wall-bounded turbulent flows, this local mechanism of energy transfer and production is referred to as backscatter \citep{Piomelli1996}, and it is associated to an energy bifurcation in the buffer layer towards both the wall (direct cascade) and the core flow (reverse cascade, \citep{Cimarelli2013}). Nevertheless, within the context of this research where a single (fundamental) harmonic of the second Mack mode is triggered through small-amplitude (linear) disturbances, the flow does not exhibit any chaotic behaviour, and it is envisaged that there is no energy transfer mechanism from the smaller ($\leq \lambda_z$) to the larger length scales ($> \lambda_z$). To verify this, the spanwise extent of the computational domain ($\lambda_{z,domain}$) is increased by a factor of 4 relative to the streaks size ($\lambda_{z,domain}=4\lambda_z=4.8$, figure \ref{fig:twall_sub_harmonic}). This enabled an assessment of a possible influence of streaks sub-harmonics ($\lambda \geq 2\lambda_z$) on the linear amplification of the second Mack mode. The number of grid nodes in the spanwise direction was similarly increased by a factor of $4$ to keep the spanwise resolution per fundamental wavelength constant. The instantaneous distribution of wall static pressure fluctuations (figure \ref{fig:pwall_sub_harmonic}) is similar to the configuration C1a, despite the fact that the spectral resolution has increased for the configuration with the larger domain ($\lambda_{z,domain}=4\lambda_z$). 

\begin{figure}
  \centering 
  \subfloat[]{\includegraphics[width=0.48\textwidth]{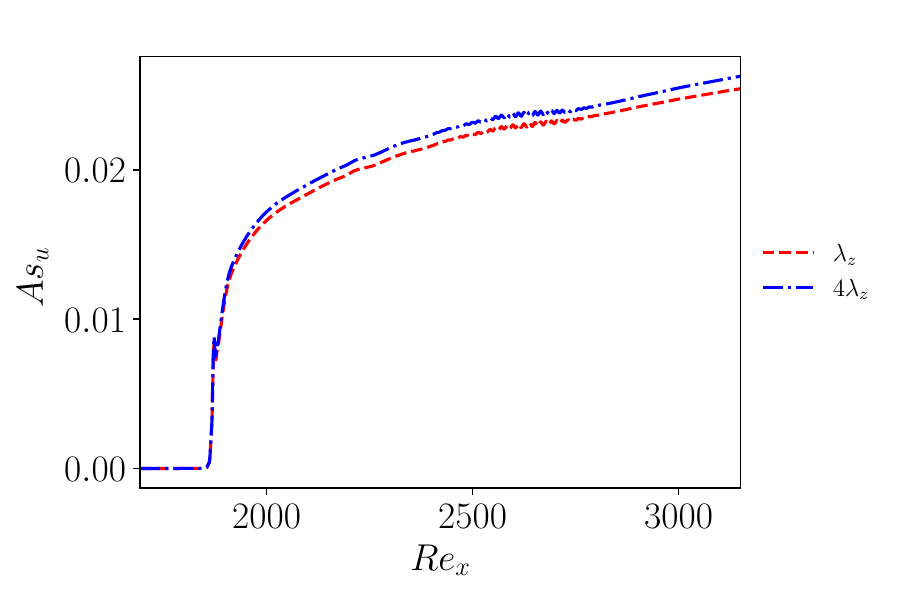}\label{fig:Asu_sub_harmonic}} 
  \hfill 
  \subfloat[]{\includegraphics[width=0.48\textwidth]{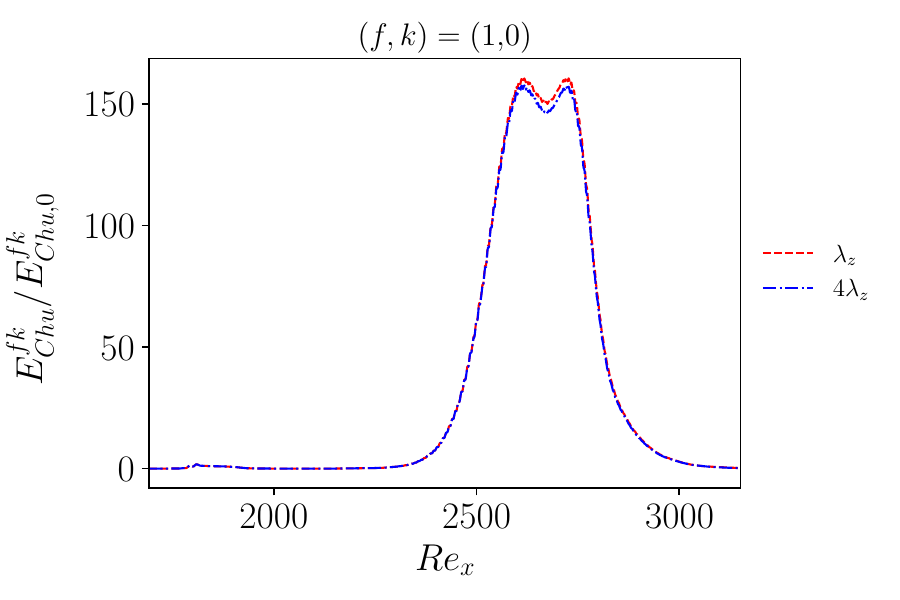}\label{fig:Echu_sub_harmonic}} 
  \caption{Streamwise distribution of (\textit{a}) streak amplitude and (\textit{b}) second Mack mode linear amplification for the configurations with $\lambda_{z,domain}=\lambda_z$ (red) and $4\lambda_z$ (blue).}
\end{figure}

For a more quantitative assessment both the streak amplitude (figure \ref{fig:Asu_sub_harmonic}) and the linear amplification of the second Mack mode (figure \ref{fig:Echu_sub_harmonic}) are determined for both the configuration with $\lambda_{z,domain}=\lambda_z$ and $4\lambda_z$. Overall, it is showed that for this case study where the amplitude of the perturbations introduced by the actuator is sufficiently small and deterministic, there is no influence of the streaks sub-harmonics on the amplification of the second Mack mode (figure \ref{fig:Echu_sub_harmonic}). As such, a computational domain with $\lambda_{z,domain}=\lambda_z$ is sufficient for the investigations presented in this work.

\section{Benchmark of current LST results}\label{appendix:LST_verification}
\begin{figure}
\centering
\includegraphics[width=0.5\textwidth]{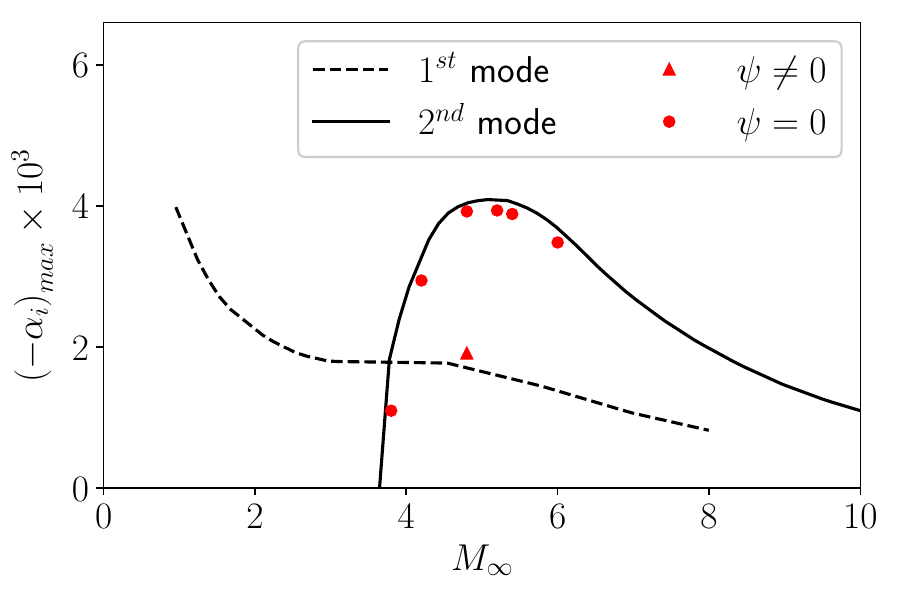}
\caption{Effect of Mach number on maximum spatial growth rate of first and second Mack mode instability within the laminar (self-similar) boundary layer over an adiabatic flat plate for $Re_x=1500$. Red markers represent the current study; solid and dashed lines are reproduced from \cite{Mack1975}.}
\label{fig:LST_benchmark}
\end{figure}

The linear stability theory (LST) code used within this study is benchmarked with previous data in the literature \citep{Mack1975}. The effect of Mach number on maximum spatial growth rate (figure \ref{fig:LST_benchmark}) is assessed at a fixed freestream specific total enthalpy ($\tilde{h}_{0,\infty}\approx 0.31\times 10^6$ J/kg). The analysis was carried out at a fixed $Re_x$ ($=1500$) and by varying the non-dimensional frequency ($F=\omega/Re_x$) of the perturbation. Overall, the agreement in maximum spatial growth rate for the second Mack mode was deemed satisfactory for the purpose of this work. 
   
\bibliographystyle{jfm}
\bibliography{lb_icl}

\end{document}